\documentclass{aa}
\usepackage{natbib}
\usepackage{graphicx}
\usepackage{float}
\usepackage{txfonts}
\usepackage{amsfonts}
\usepackage{longtable}
\usepackage{afterpage}
\usepackage{changepage}
\usepackage{geometry}
\usepackage{hhline}
\usepackage{tablefootnote}
\usepackage{subcaption}
\usepackage{graphicx}
\usepackage{lscape}
\usepackage{comment}
\usepackage{hyperref}
\bibpunct{(}{)}{;}{a}{}{,}
\begin{document}

\title{Searching for substellar companion candidates with \textit{Gaia}}
\subtitle{IV. New planetary-mass companions found in debris disks}
\author{A.-M. Lagrange\inst{1,2}  \and S. Toucheboeuf\inst{1} \and  F. Kiefer\inst{1} \and  P. Rubini  \and J. Milli\inst{2} \and L. Matr\`a\inst{3} \and H. Beust\inst{2} \and C. Perrot\inst{1} \and  M. Mâlin\inst{4}   \and A. Boccaletti\inst{1} \and C. Wilkinson\inst{1} \and P. Delorme\inst{2}}

\institute{
\label{inst:1}LIRA, CNRS, Observatoire de Paris, Universit\'e PSL,  Sorbonne Universit\'e, Universit\'e de Paris, 5 place Jules Janssen, 92195 Meudon, France\thanks{Please send any request to anne-marie.lagrange@obspm.fr} \and \label{inst:2}Univ. Grenoble Alpes, CNRS, IPAG, F-38000 Grenoble, France  \and
\label{inst:3}School of Physics, Trinity College Dublin, the University of Dublin,
College Green, Dublin 2, Ireland
\and \label{inst:4}STScI, Steven Muller Building, 3700 San Martin Drive, Baltimore, MD 21218, USA
}
        
   \date{Received ; accepted }
  
  \abstract
{
 {When spatially resolved, debris disks often exhibit structures (gaps, rings, and spirals) that may indicate the presence of unseen planetary-mass companions.} However, such planets have not been detected in most cases. \textit{Gaia} opens a new mass-separation parameter space for detecting unseen planets. A tool named GaiaPMEX has been developed to detect and characterize companions to stars. It is based on two binarity indicators derived from the DR3: the \textit{Gaia} astrometric signature, and, when the sources were also observed by \textsc{Hipparcos}, the \textit{Gaia}-\textsc{Hipparcos} proper motion anomaly.}
{We  search for companions in resolved debris disks using absolute astrometry.}
{We used GaiaPMEX on 94 stars hosting resolved debris disks to search for companions. When companions were identified,  we combined GaiaPMEX results with high-contrast imaging and radial velocity data to further constrain their mass and orbital properties. }
{GaiaPMEX found stellar or substellar companions around five targets (HD 54341, HD 146897, HD 206893, HD 107146, and HD 106906) and indications (2$\sigma $) for companions around three  additional targets (BD+45° 598, TWA 7, and HD 120326). The companion to the $150$ Myr star HD 107146, which harbors a ring-like dust disk, is found to be a 2-6$M_J$ planet orbiting between 2-10 au. A massive 3-20$M_J$ companion is found  3-10 au from the close binary HD 106906, which already hosts a remote planet orbiting at more than 700 au, and an edge-on disk. The tentative companions to TWA 7 and HD 120326 are planetary-mass companions at 1-3 au. Both stars host a multiple-ring system. We finally provide meaningful upper limits on the possible presence of companions in the range of 0-100 au around a few emblematic stars (49 Cet, AU Mic, \ensuremath{\beta} Pic, and \ensuremath{\epsilon} Eri) and stars hosting well-defined ring-like structures.}
{These newly detected planets or candidates belong to complex systems and give us the opportunity for detailed dynamical modeling of planet-disk and planet-planet interactions. Current or upcoming facilities, VLTI/Gravity, JWST/AMI, or ELT, will provide an improved orbital and possibly spectral characterization of these companions. From a methodological point of view, these results demonstrate the synergy between absolute astrometry, radial velocity, and direct imaging for detecting new bodies with planetary masses orbiting a few astronomical units away from their stars, or for determining significant constraints on unseen planets.  }


   \maketitle

\section{Introduction}

Debris disks are valuable tracers of planetary formation. Most are found around stars in young (10-150 Myr) nearby associations, but a few have also been resolved around more mature stars with ages of up to 5 Gyr. The young stars trace phases when giant planets have formed and terrestrial planets may be forming (for a detailed introduction, see \citet{Matra2025}, hereafter Ma25).
In many instances, the disks appear to be highly structured, with warps, gaps, inner voids, {and} rings. These structures are often interpreted as indirect signatures of planets \citep{Pearce2024}{. However,} it has been {shown} that alternative mechanisms might exist that do not involve planets \citep{Lyra13} (examples of such features were reported in Ma25).
To date, definite evidence of planets shaping disks is limited to two systems, however: $\beta$ Pictoris \citep{Mouillet97,Augereau01, Lagrange09bpic,Lagrange10}, and TWA7 \citep{Lagrange2025b}. In some cases, such as HR 8799, planets and disk rings are observed  \citep{Booth16,Faramaz21,Boccaletti24}, but they are not directly linked. It is therefore crucial to search for planets within these disks that might account for the observed structures.

The identification of low-mass companions in disks has mostly relied on ground-based high-contrast imaging (HCI) so far. The number of detected planets {remains} limited because the achievable contrasts only allow us to detect young, massive ($\geq $ 3$M_J$) giants that in most cases orbit at separations of 10 au or more from the central stars. Such massive planets are rare, as demonstrated by surveys of hundreds of stars that are {part of} young, nearby moving groups \citep{Nielsen19, Vigan21}. However, lighter planets, such as the Saturn-mass planet TWA 7b, are now detectable with JWST \citep{Lagrange2025b, Crotts2025}. {Short-period  planets have also been found in the AU Mic system with transit photometry \citep{Plavchan20,Wittrock23}}.

\textit{Gaia} absolute astrometry is expected to detect thousands of giant planets in the future. In particular, the \textit{Gaia} DR4, expected in 2026, is expected to  deliver the full orbital and mass characterization of {hundreds of planets  with periods typically shorter than 60 months, which is the baseline of DR4 data}. It is already possible with the DR3 data to identify and constrain the mass and orbital properties of the planets, however, as demonstrated by \citet[hereafter Ki25a]{Kiefer2025a}, \citet{Kiefer2025b}, and \citet[hereafter La25a]{Lagrange2025a}. 

We used our tool GaiaPMEX to search for and constrain low-mass companions to stars hosting spatially resolved debris disks. Our sample and approach are presented in Sect. \ref{sec:sample_approach}. Detections of companions are presented in Sects. \ref{sec:stellar_substellar_companions} and \ref{sec:hints_for_companions}, covering GaiaPMEX companions and indications for companions, respectively. Detection limits are provided for a subsample of systems in Sect. \ref{sec:limdets}.

\section{Sample, approach and outputs}\label{sec:sample_approach}
{We detail the disk sample in Sect. \ref{sec:sample} and the approach we used to detect \textit{Gaia} excesses and the use of additional data to further constrain the companion candidate properties (Sect. \ref{sec:approach}), and we describe the outputs of this approach (Sect. \ref{sec:GaiaPMEX_outputs}).}

\subsection{Sample} \label{sec:sample}
Our sample consists of the 74 systems with resolved belts that were studied in Ma25 and the 20 additional systems with resolved disks that were not included in Ma25, but were found in the literature. The main disk and stellar properties of this sample are presented in Table \ref{tab:all_systems}, and Table \ref{tab:planet_catalog} provides more extensive information on some emblematic  systems, including AU Microscopii, $\beta$ Pictoris, $\epsilon$ Eridani, HD 106906, HD 206893, and TWA 7. Our sample covers a wide range of stars, both young and old (6-5000Myr), hotter and cooler (B8-M2Ve), near and far (3-152pc), abd with thin and wide belts (width of 4-185au).

\subsection{Method} \label{sec:approach}

\subsubsection{Identification of planet candidates in debris disk systems}

GaiaPMEX explores the \textit{Gaia}-DR3 database to identify targets with \textit{Gaia} "excess", that is, with a significant excess in the residuals of the astrometric displacement of the source, as measured by \textit{Gaia}, with respect to a single-star motion model. These targets are quantified  by the renormalized unit weight error (RUWE) {and} the astrometric excess noise (AEN). {Underlying instrumental $\sigma_{\rm calib}$ and measurements $\sigma_{\rm AL}$ noises are determined as a function of the $G$ magnitude and the $Bp-Rp$ color of sources (see Ki25). When these noises are properly corrected for, RUWE and AEN enable us to infer the presence of a companion. } When available, GaiaPMEX also uses the  proper motion anomaly (PMa)  from~\citet{Kervella2022}. For each flagged target, GaiaPMEX maps the family of companion mass – semimajor axis (sma) solutions that are consistent with the observed signal. {At each possible value of mass and sma, it draws 100 random Keplerian orbits within the parameter distributions shown in Table~\ref{tab:distro}. It then simulates \textit{Gaia} observations using the epoch and scan-angles retrieved from the \textit{Gaia} scanning law available in the \textit{Gaia} observation forecast tool}\footnote{\texttt{https://gaia.esac.esa.int/gost/}}{ (GOST), and randomly generates measurements following a combination of normal distributions centered on 0 and scaled by $\sigma_{\rm calib}$ and $\sigma_{\rm AL}$. It also simulates \textsc{Hipparcos} observations using the epoch, scan angles, and data scatter retrieved from the ESA/\textsc{Hipparcos} intermediate astrometric data. A five-parameter model, including position, proper motion, and parallax, is fitted to the simulated observations of both \textit{Gaia} and \textsc{Hipparcos}, leading to  simulated PMa, RUWE, and AEN. These simulated values are compared to the  observational data to determine the consistent models of mass and sma.}  

\textit{Gaia} and PMa information can be combined to derive final (sma and mass) solutions. These solutions are associated with properly quantified confidence levels (for a detailed description of this last step, see Ki25, and for examples of use, see La25a).

{As a first step, we ran GaiaPMEX on the whole sample, adopting a 50$\times$50 grid of sma and mass, drawing 100 simulations per grid bin. We recall the distribution of parameters sampled at each tested bin of the sma-mass 50$\times$50 grid in Table~\ref{tab:distro}.} {At this stage, no assumption was made on the inclination of the possible companion. We note, however, that the inclination has a very limited effect on the GaiaPMEX outputs.}

{For the targets showing evidence of the presence of a companion, we then ran GaiaPMEX again, restricting the inclination range to that of the disk to force coplanar solutions (second step in Table \ref{tab:distro}) because for these targets, we used complementary data, as described in Sect. \ref{sec:compldata}, and in particular, RV data, which are sensitive to inclination.}
Moreover, we assumed eccentricities in the range 0-0.9, but we verified that this had little effect on the results at this stage.

\begin{table}[hbt]
    \centering
    \caption{Distribution of parameters sampled in each tested bin of the mass-sma 50$\times$50 grid.}
    \label{tab:distro}
    \begin{tabular}{lccc}
        Parameter &  Type & Bounds or law & Bounds or law\\
         &   & 1st step & 2nd step\\
        \hline
        $\log(M_c /M_{\rm sun})$ & Un. & $\pm$$\frac{1+\log M_\star}{100}$  & $\pm$$\frac{1+\log M_\star}{100}$\\
        $\log({\rm sma}_{\rm au})$ &  &  $\pm$$0.05$ &  $\pm$$0.05$\\ 
        $e$ & Un &  $0$--$0.9$ & $0$--$0.9$\\
        $\omega$ & Un. &  $0$--$\pi$ &  $0$--$\pi$\\
        $\Omega$ & Un. &  $0$--$2\pi$ &  $0$--$2\pi$\\
        $\phi$ & Un. &  $0$--$1$ &  $0$--$1$\\
        $I_c$ & Un. or $\sin i$ &  $0$--$\pi/2$ & $i_{disk}$\\
        $\varpi$ & Norm. & $\mathcal N$(PLX,$\sigma_{\rm PLX}^2$)  & $\mathcal N$(PLX,$\sigma_{\rm PLX}^2$)\\
        $M_\star$ & Norm. & $\mathcal N$($M_\star$,$\sigma_{M_\star}^2$)  & $\mathcal N$($M_\star$,$\sigma_{M_\star}^2$)\\
        \hline
    \end{tabular}
    \tablefoot{"Un." stands for uniform and "Norm." stands for normal. {$M_c$ was explored from 0.1 to $M_\star$ and sma from 0.01 to 1000\,au. A uniform distribution of $\log M_c$ and $\log\text{sma}$ was adopted in each bin whose width is indicated here.}}
\end{table}

\subsubsection{Use of complementary data to further constrain the candidates} \label{sec:compldata}
When {companions} were identified, we used complementary radial velocity (hereafter RV) data and/or high-contrast imaging (hereafter HCI) data to further  constrain  the properties of the possible candidates whenever possible. Otherwise, the RV data came from previous surveys we made with the European Southern Observatory HARPS or the Observatoire de Haute Provence SOPHIE in the past decades, and were reduced with the Spectroscopic data via Analysis of the Fourier Interspectrum Radial velocities (SAFIR), a software dedicated to the measurements of RV of rapid rotators such as early-type stars and young stars  \citep{Galland05,Lagrange09,Borgniet17, Borgniet19,Grandjean20,Grandjean21,Grandjean23}. The HCI data were taken from our library of SPHERE data that were analyzed with the PAtch COvariances (PACO) algorithm \citep{flasseur_paco_2020,Chomez23,Chomez25} as part of the COBREX ERC project. The contrast maps derived from direct imaging were converted into detection limit maps using the AMES-Cond \citep{Baraffe2003} or the HADES model \citep{Wilkinson2024} for the SPHERE and JWST data, respectively.

To place tighter constraints on the companion properties, we first identified the intersection between the GaiaPMEX solutions and the detection limits from RV and HCI. Furthermore, when the GaiaPMEX detection was greater than 3$\sigma$, we applied an MCMC analysis considering all these data. Details and examples of such an approach can be found in La25a.

\subsubsection{Detection limits} \label{sec:PyMESS}

{We computed the GaiaPMEX detection limits for all our sample stars without astrometric excess. In addition, we combined the \textit{Gaia} upper limits with RV and/or HCI detection limits to tighten the constraints on any unseen companions for stars of particular interest (those with known companions and those with well-structured rings; Ma25). The detection limits were computed using PyMESS,} a Monte Carlo framework originally designed to fuse RV and HCI constraints \citep{Lannier16} {that was later upgraded to use GaiaPMEX constraints.} Illustrative applications are provided in La25a.

\subsection{GaiaPMEX outputs}
\label{sec:GaiaPMEX_outputs}
All our targets but Fomalhaut and $\alpha $  Lyrae (these targets are too bright for \textit{Gaia}) were analyzed by GaiaPMEX. In most cases, GaiaPMEX did not indicate any companions.  The sensitivity of GaiaPMEX depends on the target distances and masses, as analyzed in La25a. GaiaPMEX maps for non-detections are shown in Fig. \ref{fig:nondetections} for the sample studied in Sect. \ref{sec:limdets}.

Companions were detected with a confidence level $\geq3\sigma$ around 5 of the 94 sample stars based on \textit{Gaia} and \textit{Gaia}-\textsc{Hipparcos} PMa. We additionally found indications for new companions around 3 stars, 2 of which were based on \textit{Gaia} alone because no \textsc{Hipparcos} data were available. Finally, HR 8799 is studied in detail in a separate paper to address the complexity of the system, which harbors four known planets, some of which {contribute} to the \textit{Gaia} signal.

We show in  Fig.~\ref{fig:unambiguous_detections} the (sma and mass) solutions for the five  unambiguous detections (two stellar companions and three substellar companions) when using \textit{Gaia} data alone, PMa alone, and \textit{Gaia} data combined with PMa, and we discuss each target hereafter. {At this stage, we made no specific assumptions on the inclination or eccentricity of the planet (hence, they varied between 0 and 0.9).}

\section{Results: GaiaPMEX companions}\label{sec:stellar_substellar_companions}
We report unambiguous ($\geq 3\sigma$ significance) GaiaPMEX detections of one  stellar companion, one companion in the stellar or brown dwarf (hereafter BD) regime, and three unambiguous detections of planets or BDs.

\subsection{A stellar companion to HD 54341}
A stellar companion is clearly detected for 
HD 54341 (HIP 34276, frac\footnote{Frac indicates the fraction of exposures for which multiple modes have been detected, published in \textit{Gaia} DR3.}=0, gof\footnote{ Gof measures the amplitude of the sinusoidal variation of the goodness-of-fit of the PSF fit, published in \textit{Gaia} DR3.}=0.016). \cite{Chen2014} reported that the disk is divided into two parts: a hot component at 4.9au, and a cold component at 128.1au. \cite{Vican2016} fit a double belt to be consistent with \cite{Chen2014}, but the associated $\chi^2$ values for single- and double-belt assumptions were similar, and Ma25 indicated a ring size of 170 au. 
The GaiaPMEX map (row 1 of Fig. \ref{fig:unambiguous_detections}) clearly indicates ($>10\sigma$) the presence of a $\geq 200 M_J$ stellar companion with an sma of between 0.6 and 3 au. To our knowledge, this star has never been reported as a close binary. Dedicated dynamical simulations are needed to explore the impact of the stellar companion on the disk structure.

\subsection{A stellar/BD companion to HD 146897}
The F2 star HD 146897 (HIP 79977; frac=0, gof=0.054)  is a member of the Sco-cen association. It harbors a highly inclined debris disk that has been resolved with SPHERE, GPI, and SCExAO on the Subaru telescope \citep{Thalmann2013, Engler2017, Goebel2018, Crotts2024}.

The GaiaPMEX map clearly indicates ($>4.1\sigma$) the presence of a $>15M_J$ stellar or brown dwarf companion (row 4 of Fig. \ref{fig:unambiguous_detections}), with a semimajor axis between 0.05 and 3 au ($\sim$30 mas). To our knowledge, this star has never been reported as a close binary, and it was not reported as a \textit{Gaia} non-single star in \citet{Vallenari23}.
RV data over about 500 days revealed no significant variations. They exclude an equal-mass binary companion. However, their poor temporal sampling prevented us from further constraining the companion between a low-mass companion star and a brown dwarf. Based on the ring size, 82 au (Ma25), the  binary companion, orbiting within 3 au, is not expected to contribute significantly to the disk shaping. 

\subsection{A {(previously known)} planet and a BD around HD 206893}\label{sec:HD206893} This $\simeq$ 160 Myr star (HIP 107412; frac=0, gof=0.05) is surrounded by a wide (100 au) disk at about 100 au that is resolved at millimeter wavelengths (Ma25). According to \cite{Pearce2024}, the flat inner-edge slope of this disk is consistent with a pure collisional evolution. A first substellar companion, HD 206893B, was imaged at a projected separation of 10 au by \citet{Milli2017b}. Various estimates of the companion mass have been produced, with an overlap around 20-30 $M_J$. An inner companion with a mass of about 13 $M_J$ to HD 206893c was found  orbiting at separations between 2 and 3 au using RV and absolute astrometry data  \citep{Grandjean2019}, and it was later confirmed with Gravity measurements \citep{Kammerer2021,Hinkley2023}. {We ran GaiaPMEX, taking the two known companions into account in the modeling. The output was no longer indicative of the presence of a companion. We conclude that \textit{Gaia} does not indicate the presence of an additional companion to HD 206893.}

\subsection{A planet around HD 107146}\label{sec:HD107146}

\begin{figure}[hbt]
    \centering
    \begin{subfigure}{0.49\textwidth}
        \includegraphics[width=\textwidth]{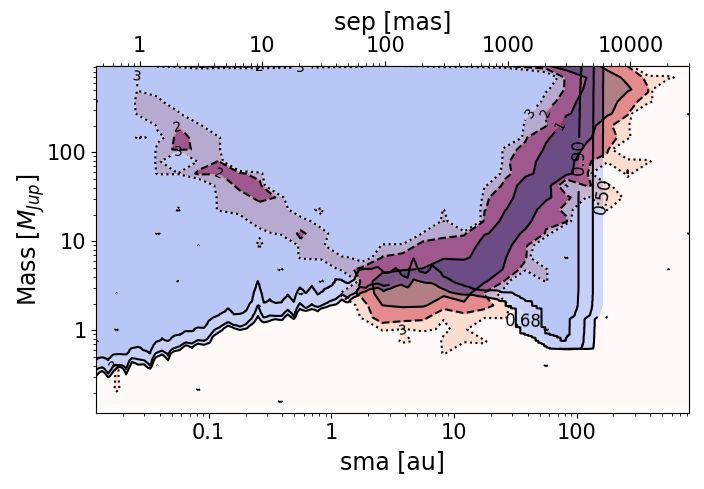}
        \caption{}
        \label{fig:HD107146_gpmex_with_limdet}
    \end{subfigure}
    \begin{subfigure}{0.49\textwidth}
        \includegraphics[width=\textwidth]{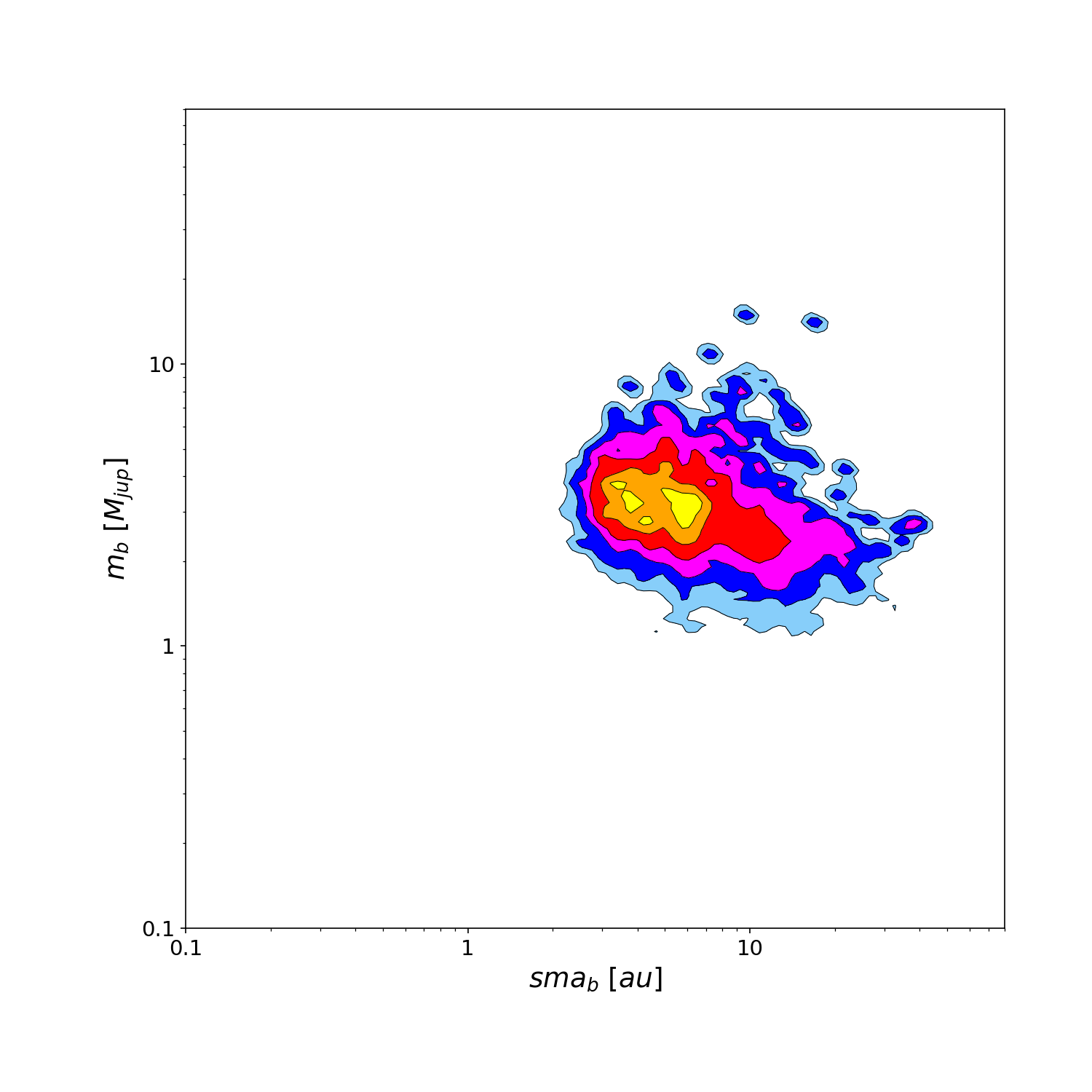}
        \caption{}
        \label{fig:HD107146_MCMC}
    \end{subfigure}
    \caption{Planet around HD 107146.
    Panel a: Superimposition of the GaiaPMEX solutions with limits set by RV and HCI data (SPHERE and JWST). The GaiaPMEX (sma and mass) solutions  are shown in red/brown, with darker shades showing the likeliest solutions. The regions shaded in blue above the black lines correspond to the (sma and mass) excluded by detection limits from RV and HCI data. Altogether, the possible (sma and mass) left are those in red/brown below the blue region. The sma is expressed in au (bottom) and mas (top).
    Panel b: Solutions (sma and mass) found by the MCMC analysis  using RV, \textit{Gaia}, PMa, and HCI data as inputs. The colors represent increasing likelihood. Light blue shows the likelihood at $3 \sigma$,  blue at $2.5 \sigma$, magenta at $2 \sigma$, red at $1.5 \sigma$, orange at $1 \sigma$, and yellow at $0.5 \sigma$.}
    \label{fig:Fig_HD107146}
\end{figure}

The $150$ Myr star with a mass of $1.03M_{\odot}$, HD 107146 (HIP 60074; frac=0, gof=0.028), hosts a wide ($\simeq $ 100 au) ring-like disk at 130 au that was first resolved by HST/ACS \citep{Ardila2004}. {Early millimeter images revealed extended emission up to 150 au \citep{Corder2009,Roccatagliata2009}. Higher angular resolution millimeter images fully resolved the disk, showing an inner edge at about 50 au} \citep[Ma25]{Marino2018}. According to \citet{Pearce2024}, the inner edge is steep enough to be compatible with planet sculpting. Neither direct imaging nor radial velocity {has} identified a companion so far, while GaiaPMEX clearly shows a companion.

Combining the RV and HCI detection limits with GaiaPMEX either by simple super-imposition or within an MCMC, as done in La25a, allowed {us} to conclude that the companion is a 2-6 $M_J$ planet orbiting between 2 and 10 au, with the bulk of the sma solutions between 2-5 au (Fig.~\ref{fig:Fig_HD107146}). { We assumed that the planet has the same inclination as the disk, that is, between 18 and 22 $\deg$, and an eccentricity lower than 0.3.  As a check, we proceeded similarly without any assumptions on the inclination and on the eccentricity. The solutions are not significantly different, but are very slightly spread on a wider (sma and mass) range.}

In addition to the SPHERE HCI data (see Sect. \ref{sec:compldata}), we used recent JWST/MIRI coronagraphic images (ID 1668). Because of its short separation, HD 107146 b alone cannot cause the inner disk edge at 50 au. However, it might dynamically excite smaller, more remote planets that might sculpt the inner disk. 
Finally, a direct detection of this 150 Myr planet orbiting less than 100 mas from its parent star {is not within Gravity's detection capabilities and} requires the spatial resolution and sensitivity of the ELT.

\subsection{An additional planet/BD around HD 106906}\label{sec:HD106906}

\begin{figure}[hbt]
 \centering
    \begin{subfigure}{0.49\textwidth}
        \includegraphics[width=\textwidth]{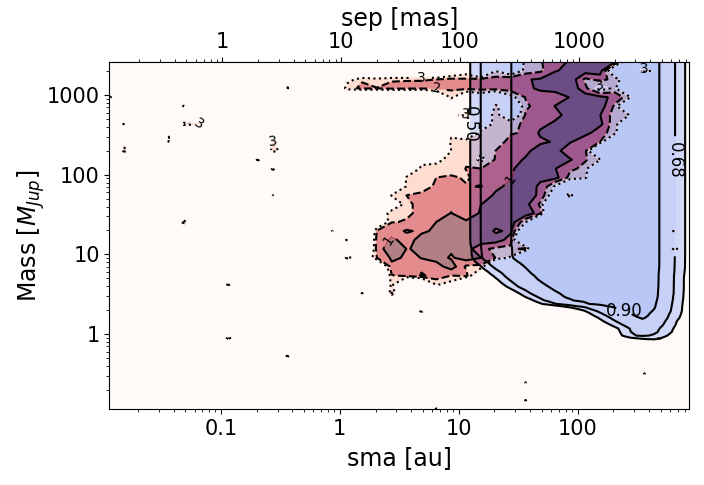}
        \caption{}
        \label{fig:HD106906_GPMEX_wcompanion}
    \end{subfigure}
    \begin{subfigure}{0.49\textwidth}
        \includegraphics[width=\textwidth]{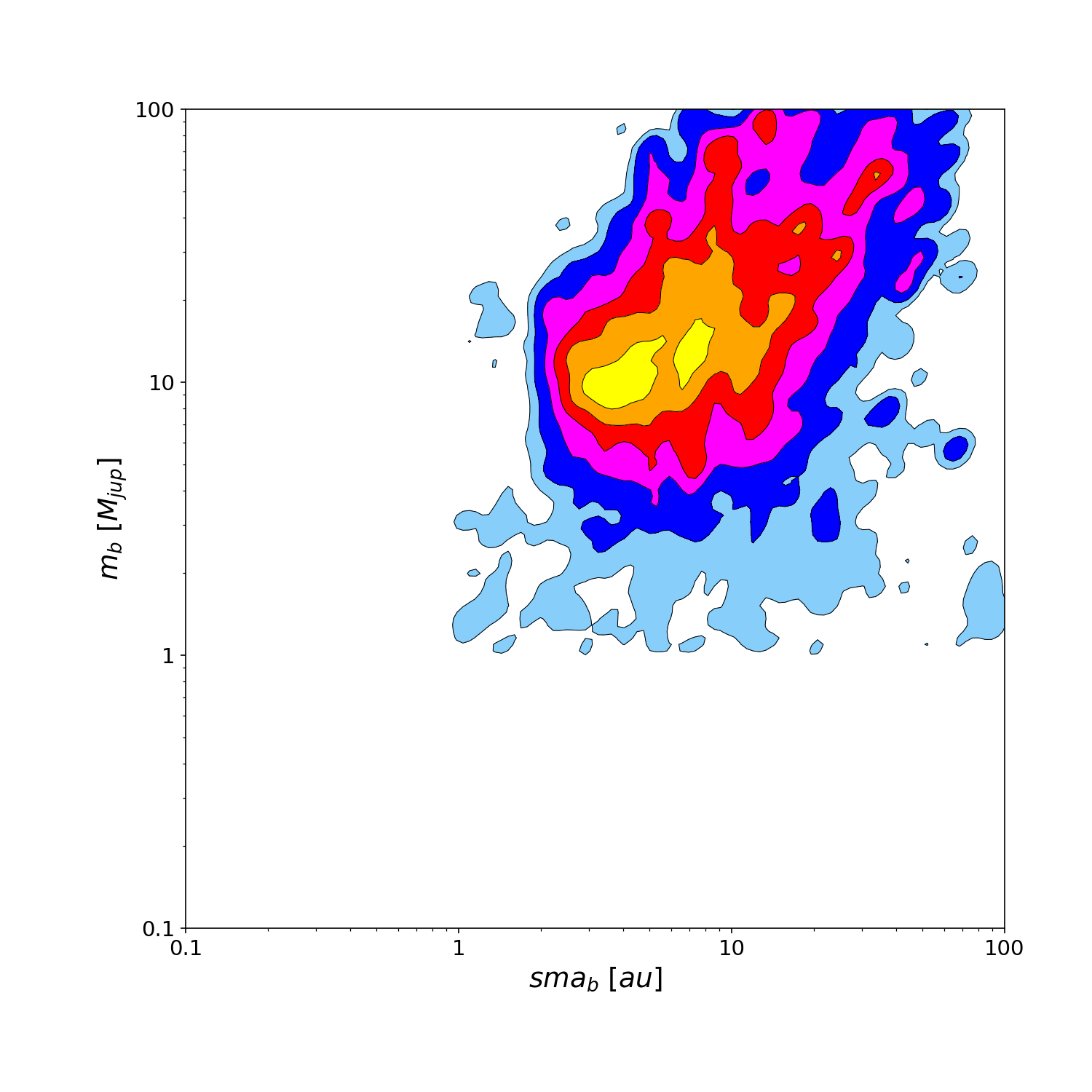}
        \caption{}
        \label{fig:HD106906_emcee}
    \end{subfigure}
    \caption{Inner substellar companion to HD 106906. Panel a: Solutions (sma and mass) for HD 106906, taking the known binary companion into account and using detection limits from HCI data overplotted with PyMESS. The possible (sma and mass) solutions are those in red/brown and outside the blue region excluded by HCI. Panel b: Solutions (sma and mass) found by the MCMC analysis for HD 106906 c using \textit{Gaia}, PMa, and HCI data as inputs. The colors are explained in Fig. \ref{fig:Fig_HD107146}}
    \label{fig:HD106906}
\end{figure}

HD 106906 (HIP 59960; frac=0, gof=0.007) is a young ({approximately} $13 $ Myr), close, equal-mass spectroscopic binary with a 50-day period (corresponding to an sma of 0.4 au) and an RV amplitude of $\sim$6 km/s. It hosts a wide debris disk between 35 and 110 au inclined at 85.3°, as observed by SPHERE \citep{Lagrange2016} and recently by JWST/MIRI \citep{Rouan2025}, as well as a remote super-Jupiter orbiting at more than 600 au \citep{Bailey2014}. Recent observations of HD 106906 b reported an eccentric orbit, inclined with respect to the disk plane \citep{Nguyen2021}. The origin of the current position of HD 106906 b is {unclear}. In situ formation would have required a very large protoplanetary disk.  \cite{Rodet2017,Rodet2019} proposed a  formation a few astronomical units from the central binary. In this framework, the planet migrated inward and was scattered onto a high-eccentricity orbit through an unstable mean-motion resonance with the binary. Through the action of an external stellar perturber, it was then trapped in a stable region. An alternative scenario by \cite{Moore2023} {suggests} that the planet formed in the disk around the binary or from a portion of a self-gravitating molecular cloud {that} fragmented into a binary and companion. Through a recent close encounter with a free-floating planet, it was then scattered to its current position. 
 
We ran GaiaPMEX in a two-component mode that accounts for the spectroscopic pair. The result clearly indicates a companion (Fig. \ref{fig:HD106906_GPMEX_wcompanion}). High-contrast imaging with SPHERE and JWST further excludes solutions at separations beyond 30 au, as shown in the overplotted GaiaPMEX map. An MCMC analysis shows that the most likely solutions are located at separations 3-30 au (with a bulk in the 3-10 au region, however), and with masses between 3-70$M_J$ (bulk in the 7-20$M_J$), as shown in Fig. \ref{fig:HD106906_emcee}. {The results are not significantly affected by the  assumptions on the companion inclination.}

The identification of this additional companion  necessitates revisiting earlier dynamical models to account for the observed properties of HD 106906 c. In particular, scenarios invoking formation near the star followed by outward migration and ejection through interactions with the binary \citep{Rodet2019} should be reassessed. Crucial issues remain unresolved, such as whether HD 106906 b formed inside or beyond HD 106906 c, and {determining} their respective masses. 
{Additional HCI observations, possibly with VLTI/Gravity, JWST/AMI, or forthcoming ELT facilities with improved spatial resolution and contrasts are needed } to further constrain its mass and orbital parameters, while spectroscopic data might provide crucial insights into its atmospheric properties. 

\section{Candidate detections in debris disks}\label{sec:hints_for_companions}
Evidence of companions was found around three additional stars, each at a confidence level of $\geq 2\sigma$.
For two of them, BD+45° 598 and TWA 7, the detections are based on \textit{Gaia} data alone because no \textsc{Hipparcos} data are available. For these systems, we present solutions based on AEN data, as in both cases the significance is higher than from RUWE data. For HD 120326, our conclusion is based on \textit{Gaia} and \textit{Gaia}-\textsc{Hipparcos} PMa. We show in  Fig.~\ref{fig:ambiguous_detections_all} the (sma and mass) solutions when using the relevant data, and we discuss each target hereafter. 

\subsection{BD+45° 598}
The system presents a debris disk between 50.5 and 85.5 au, inclined at $86.1$°, as seen at $2.2\mu m$ \citep{Hinkley2021}. For BD+45° 598 (frac=0, gof=0.034), the GaiaPMEX maps indicate a companion at a level of 2.1 $\sigma $ for the AEN map. The possible solutions are degenerate, with inferred companion masses spanning the planetary to stellar regime. No RV or HCI data are available. BD+45° 598 was not reported as a \textit{Gaia} non-single star in \citet{Vallenari23}. RV measurements and HCI data are needed to conclude on the nature of this candidate companion. If it is of planetary nature, the possible solutions are at separations smaller than 100 mas, which excludes any direct detection with current instruments, except possibly with VLTI/Gravity or JWST/AMI. 

\subsection{HD 120326}
\begin{figure}[!ht]
    \centering
    \begin{subfigure}{0.48\textwidth}
        \includegraphics[width=75.mm]{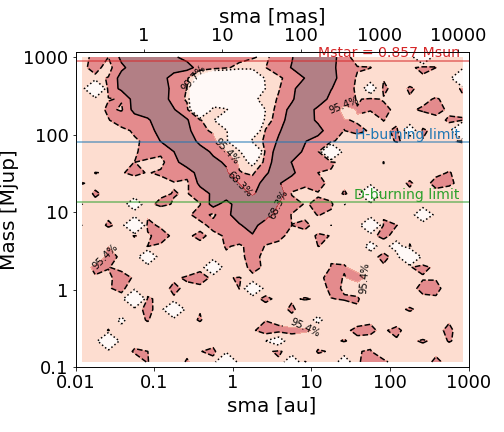}
        \caption{}
    \end{subfigure}
    \begin{subfigure}{0.48\textwidth}
        \includegraphics[width=75.mm]{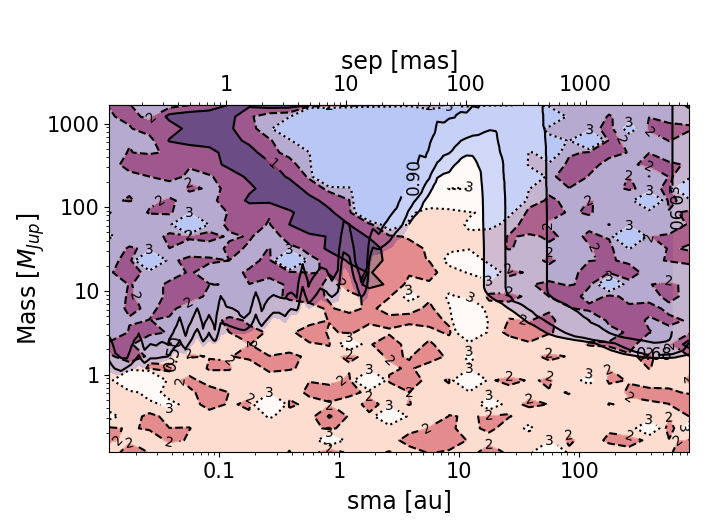}
        \caption{}
        \label{fig:HD120326_limdet_overplot}
    \end{subfigure}
     \begin{subfigure}{0.48\textwidth}
     \includegraphics[width=75.mm]{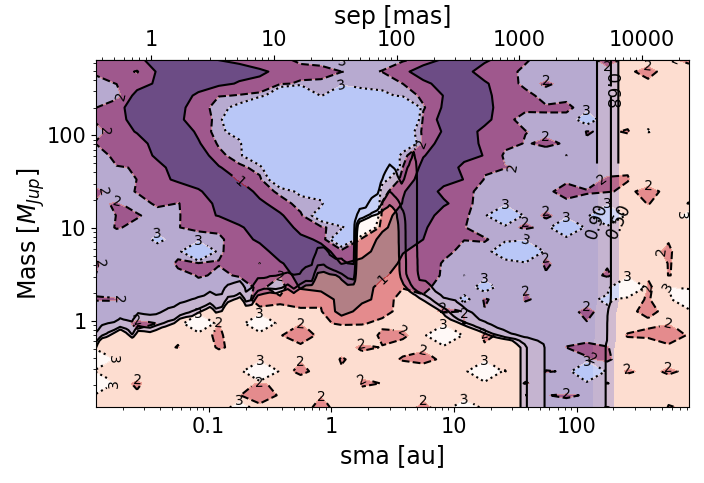}
        \caption{}
         \end{subfigure}
    \caption{Indications of companions around BD+45° 598, HD 120326, and TWA 7. Panel a: GaiaPMEX map for BD+45° 598 based on AEN data, assuming the planet to be coplanar with the disk.
    Panel b: RV and HCI detection limits plotted over the GaiaPMEX map (combination of RUWE and PMa) for HD 120326.
    Panel c: RV and HCI detection limits plotted over the GaiaPMEX map for TWA 7. The  step in mass beyond 30 au is due to limits in the \cite{Baraffe2003} model used to convert contrast maps into mass detection limits.}
    \label{fig:ambiguous_detections_all}
\end{figure}

HD 120326 (HIP 67497; frac=0, gof=0.039) harbors a double ring observed with SPHERE at 59 au and 130 au and inclined by $\simeq $ 78° \citep{Bonnefoy2017,Desgrange2025}. The GaiaPMEX map obtained from \textit{Gaia} RUWE data indicates a close companion (2.5 $\sigma $ level) with a mass down to about 10 $M_J$ and a separation smaller than 10 au. Based on RUWE and PMa, the GaiaPMEX map indicates a companion at less than 5 au and a mass higher than about 8 $M_J$. RV data taken from our reduction of HARPS data over 800 days do not show variations, but they cannot exclude periods of 400 days or shorter, even though the probability is low. The associated detection limits are shown overplotted on the GaiaPMEX map in Fig. \ref{fig:HD120326_limdet_overplot}. If the planet is coplanar with the disk, a small overlap is left between the GaiaPMEX solutions and the RV data, which indicates a 10-20 $M_J$ companion orbiting at about 2 au. Lower inclinations would lead to more massive solutions closer to the star. Simulations, as {those} done by \citet{Lee2016}, are needed to show whether this putative inner companion, if confirmed, might explain the double disk ring.

\subsection{TWA 7}
 TWA 7 (frac=1, gof=0.012) harbors three rings seen in scattered light with SPHERE, STIS, and NICMOS \citep{Olofsson2018,Ren2021,Lagrange2025b}: the first centered at about 25 au and extending beyond 100 au, the second at 52 au of width $<7$ au, and the third at 93 au of width $>40$ au \citep{Ren2021}. The rings are seen almost pole-on at $13$°{, with tentative detections of a spiral arm and a dusty clump} \citep{Olofsson2018}. A 0.3 $M_J$ companion, TWA 7b, was recently detected around this star at a separation of 52 au using JWST (ID 3662) direct imaging \citep{Lagrange2025b}. A third candidate companion, with a mass of 0.17 $M_J$ and at a separation of 0.09 au, was suggested through radial velocity analysis \citep{Donati2025_TWA7}. This third candidate companion is denoted TWA 7c.
 
 GaiaPMEX indicates a companion at a level of 2.1 $\sigma $ for the AEN around TWA 7. We denote this candidate companion TWA 7d. TWA 7b is too far away and too light to contribute to the \textit{Gaia} signal. TWA 7c is too close and too light to contribute to the \textit{Gaia} signal. When we incorporated the available HARPS RV, SPHERE, and JWST HCI data, most degeneracies were removed, and we found a mass of about 2–8 $M_J$ at a separation of about 0.7–4 au for TWA 7d.
 Additional constraints can be derived from dynamical simulations. Depending on its mass and orbital properties, the inner planet might affect the evolution of TWA 7b and the stability of the narrow ring on which this planet sits. For instance, preliminary simulations similar to those performed in \citet{Lagrange2025b} showed that a 4 $M_J$ {planet} at a {separation} of 4 au is not compatible with the observations. {Depending on its precise mass and orbital properties, simulations can also show whether this additional planet can explain the entire void, up to the inner edge of the inner ring, or if additional planets are needed.} In any case, TWA 7 now represents a key benchmark for models of planet–disk coupling.
    {
\section{Upper limits of low-mass companions in debris disks}\label{sec:limdets}
We estimated upper limits for possible potential planets that are not yet known using \textit{Gaia} and PMa (Fig.~\ref{fig:detection limits all}, \ref{fig:detection limits all long}). These upper limits fall in the brown dwarf domain for orbital distances of a few tenths to 10-20 au and in the planetary-mass domain in the range of 1-5 au, depending on the stellar properties.  In addition, we combined the \textit{Gaia} (and PMa) data with available RV and HCI data  to extend the upper limits to a wider range of orbital distances for a few iconic systems (47 Cet, AU Mic, \ensuremath{\beta} Pic, and \ensuremath{\epsilon} Eri), as well as for  all disks showing well-defined ring-like structures. }
The detection limits in (sma and mass) were computed between 0.01 and a few tens of astronomical units, bridging the gap between RV and HCI detection limits. They are shown in Fig.~\ref{fig:detection limits all}, together with the locations of the ring(s).

\subsection{49 Cet}
The star 49 Cet (HD 9672; 1.89\ensuremath{M_{\odot}}/A1V/45Myr) hosts a wide  ($\sim185$ au) outer disk with an inner edge at about 65 au \citep[][{near-IR images}]{Choquet2017}, and it is inclined by 79.1° (Ma25, {millimeter images}). It also harbors an inner warm dust disk from $\sim4$ to 60 au \citep[][{mm images}]{MacGregor2016}. 
The inner disk is devoid of gas, while the outer disk contains a substantial amount of gas \citep{Hughes2008} that cannot be primordial and might instead come from the destruction of comet-like ices \citep{Roberge2013}.  No companion more massive than 4-40 $M_J$ is present in the system.

\subsection{AU Mic}
AU Microscopii (HD 197481, 0.64\ensuremath{M_{\odot}}/M1VeBa1/23Myr) has an edge-on millimeter disk peaking at 35 au (Ma25), also seen in scattered light, with multiple fast-moving features and no single explanation regarding their origin \citep[][{near-IR images}]{Bocca18}. It also hosts three confirmed transiting planets and one possible candidate companion, all close ($<1$ au from the star) and fairly light ($M_c<20M_{\oplus} \approx 0.06M_j$; \citealt{Donati25,Martioli21,Plavchan20,Wittrock23}).
The {young age (23 Myr) and }proximity {(9.7 pc)} of this star system explain the very low detection limits obtained from HCI and PMa.
The presence of planets more massive than Jupiter at separations greater than 8 au has been excluded by JWST coronography \citep{Lawson2023}. GaiaPMEX provides tighter  constraints closer to the star. Planets more massive than about 2$M_J$ can be excluded between 0.1 and 1 au, and sub-Jupiter mass planets can be excluded beyond and up to the disk.

\subsection{\ensuremath{\beta} Pic}
$\beta$ Pictoris (HD 39060, 0.78\ensuremath{M_{\odot}}/A6V/23Myr) is surrounded by a broad (92 au) millimeter disk at R = 100 au seen edge-on (86.6°) (Ma25). The disk has been extensively studied at shorter wavelengths \citep{Golimowski2006, Boccaletti2009, Apai2015, Millar-Blanchaer2015, Hughes2018, Rebollido2024, Worthen2024}, showing various asymmetries and an inner warp. The inner disk radius is 23.6 au in scattered light and 63.2 au at millimeter wavelengths.
The system hosts two planetary companions, $\beta$ Pic b (8$M_J$, 8 au) and $\beta$ Pic c (9$M_J$, 2.7 au; \citealt{Lagrange09bpic, Lagrange2019}). $\beta$ Pic b was shown to explain the observed inner disk warp.
Detection limits from GaiaPMEX, RV, and HCI allowed us to exclude additional companions with a mass greater than 1-8 $M_J$, depending on the separations. Hence, $\beta$ Pic b or c are the most massive planets in the system. 

\subsection{\ensuremath{\epsilon} Eri }
$\epsilon$ Eri (HD 22049, 0.81\ensuremath{M_{\odot}}/K2V/600Myr) harbors a narrow ($\sim 12$ au) millimeter debris disk at 68.3 au seen close to face-on at 30° (Ma25). An inner disk has also been detected between 12 and 16 au in the far-infrared (FIR; \citealt{Greaves2014}). The system has one confirmed $0.66M_J$ planet at 3.53 au \citep{Llop2021, Hatzes2000} that is {unlikely to contribute} to the shaping of the disk due to its low mass and distance from the inner disk edge.
The RV data we used were taken from \cite{Rosenthal2021, Benedict2006, Fischer2014, Butler2006, HIRES_Pub2017, Campbell1988} and ESPRESSO data retrieved from the Geneva Data \& Analysis Center for Exoplanets (DACE). The available HCI detection limits are not as good as for other stars because the system is much older; they could be significantly improved with JWST/MIRI data. Below 10 au, we find no additional companions more massive than about 2 $M_J$. The suggested carving of the inner edge of the disk in \cite{Booth2017} by a 0.4-1.2$M_J$ companion at 48 au remains plausible and compatible with our present detection limits. 

\subsection{GJ 14}
GJ 14 (0.65\ensuremath{M_{\odot}}/K7V/500Myr) hosts a 33 au wide ring at 99 au as seen in millimeter wavelengths (Ma25), or 37.2 au wide at 54.5 au, as seen in FIR \citep{Marshall2021}. The RV data we used were taken from \citet{HIRES_Pub2017, Rosenthal2021}, and HCI from JWST. We exclude planetary companions with masses $>1 M_J$ at separations up to about 30 au, that is, within the main disk.

\subsection{HD 105} HD 105 (1.04\ensuremath{M_{\odot}}/G0V/50Myr) hosts a debris disk seen from 55 to 85 au in ALMA {millimeter data} and inclined by 50° \citep{Marshall2018}. The lack of dust interior to the ring, evidenced by the lack of mid-IR excess in the SED, might be due to one or more planets \citep{Donaldson2012}. The RV data we used were taken from \cite{HIRES_Pub2017} and our reduction of HARPS data. The detection limits exclude companions more massive than about $3M_J$ at separations up to 200 au.

\subsection{HD 53143}
HD 53143 (0.91\ensuremath{M_{\odot}}/G9V/1010Myr) hosts a narrow (23 au) millimeter disk at 88.4 au inclined by 56.1° (Ma25), with a notable eccentricity between 0.19 and 0.23 \citep{MacGregor2022}. The millimeter disk shows significant differences compared to the HST scattered-light data. In particular, an inner disk ranging from 5.5 au to 55 au was  observed almost face-on {with optical HST/STIS} \citep{Schneider2014}. To explain the non-detection of the inner disk at millimeter wavelengths, \cite{MacGregor2022} suggested a previous scattering event or dynamical instability in the disk. The RV data we used were taken from \cite{Perdelwitz2024}. We exclude companions more massive than $20M_J$ within the disk and its gap and between 1-7 $M_J$ for separations smaller than 5 au, that is, interior to the inner ring.

\subsection{HD 92945 } HD 92945 (0.82\ensuremath{M_{\odot}}/K1V/29Myr) hosts a disk imaged at millimeter wavelengths, with sharp inner and outer edges at 52 au and 121 au, respectively, as well as a gap at 73.4 au with a width of 20 au \citep{Marino2019}.  \citet{Marino2019} suggested a maximum mass of $0.6M_J$ for a companion located within the gap. The RV data we used were taken from \cite{HIRES_Pub2017} and HCI data from SPHERE and JWST (ID 1668). Our detection limits exclude companions of mass $>2M_J$ up to 100 au. This is still compatible with the maximum mass of the companion predicted by \citet{Marino2019}.  

\subsection{\ensuremath{\eta} Cru} $\eta$ Cru (HD 105211; 1.56\ensuremath{M_{\odot}}/F2V/1300-3990Myr) is a billion-year-old system with a narrow (23 au) millimeter ring at 131.5 au (Ma25). \cite{Yang2024} noted that the system is depleted of small grains and is potentially dynamically cold. They also pointed out a dust disk at 13 au, likely caused by an in situ asteroid belt and comet delivery, possibly indicating the presence of planets. They reported that the ratio of the true disk radius to the blackbody radius is lower than expected, implying a disk composed primarily of larger dust grains.
The GaiaPMEX map does not constrain the presence of planets in the system. RV and HCI data are needed to further the search for exoplanets in this system.

\subsection{\ensuremath{\eta} Crv}$\eta$ Crv (HD 109085; 1.46\ensuremath{M_{\odot}}/F2V/1400Myr) presents a 53 au wide (millimeter) debris ring at 153 au and an anomalously bright inner disk component at 8.5 au \citep{Matra2025, Duchene2014}.  \citet{Marino2017} found that if a planet is carving the inner edge of the disk, it should be at separations 75-100 au and of mass 3-30$M_{\oplus}$, while \citet{Pearce2022} found a  minimum mass of 0.34$M_J$. RV data were taken from \citet{Perdelwitz2024}. The  current detection limits do not allow us to test the putative planets proposed by \cite{Marino2017} and \citet{Pearce2022}. At these separations, our detection limits only exclude stellar companions with a mass greater than $\sim 300M_J$.

\subsection{HD 109573}
HD 109573 (HR 4796; 1.99\ensuremath{M_{\odot}}/A0V/10Myr) is part of a wide binary system. It is  surrounded by a narrow (14.8 au) belt at 77.8 au (Ma25) in millimeter wavelengths. The center of the scattered-light ring is offset from the star by 4.76 au{, as observed in near-IR} \citep{Rodigas2015}. \citet{Kennedy2018} estimated that a planet that caused the ring eccentricity would be more massive than Neptune and orbit beyond 40 au. \citet{Lagrange2012} suggested that a giant planet might be at the origin of the sharp outer edge by orbiting outside the ring. \citet{Schneider2018} also noted that the particle depletion within the ring{, observed in optical wavelengths,} might be indicative of planets. No planet has been found in the system so far, however, with previous detection limits excluding masses greater than 2$M_J$ beyond 100 au and 4$M_J$ in the 30-100 au region \cite{Milli2017}. We exclude brown dwarf companions down to 1 au and massive giants down to 5 au.

\subsection{HD 170773} HD 170773 (1.32\ensuremath{M_{\odot}}/F5V/1500Myr) is a billion-year-old system hosting a wide debris disk ranging from 81 au to 265 au inclined by 31.3° as seen by Herschel \citep{Moor2015}, or centered at 194 au of width 68 au as seen in millimeter wavelengths (Ma25). \cite{Sepulveda2019} explained that a hypothetical long-period giant planet clearing material inward of the inner disk edge would have to lie beyond 91 au to remain below direct-imaging detection limits. Our combined detection limits exclude a massive planet ($>10M_J$) at separations 3-30 au and up to 20$M_J$ within the disk.

\subsection{HD 181327}
This young solar-type star (1.26$M_\odot$/F6V/18.5Myr) hosts a narrow ring, seen by STIS between 82.3 au and 90.5 au \citep{Stark2014} and at millimeter wavelengths between 73.3 and 89.3 au, close to pole-on at an inclination of 29.9° (Ma25). 
\citet{Xie2025} reported water ice in the debris disk of this system, highlighting that the disk is dynamic from the gradient in the water-ice distribution. Recently, \cite{Fox2025} attributed the disk brightness asymmetry seen in optical STIS images to a 2-5 $M_J$ planet orbiting at 61-64 au (circular orbit). Gemini/NICI detection limits impose a mass lower than $3.5M_J$ \citep{Wahhaj2013, Fox2025}. The HCI data we used were from SPHERE and JWST (ID 3662). We exclude companions more massive than 1-2$M_J$ beyond about 3 au and 0.3-0.8$M_J$ companions between about 0.2 and 1 au. 

\subsection{HD 202628} HD 202628 (0.99\ensuremath{M_{\odot}}/G5V/1100Myr) harbors an eccentric narrow (16 au) debris ring at 155.4 au inclined at 57.9° (Ma25, millimeter images).  \citet{Krist2012} suggested a plausible planetary companion around 120 au with a  $<10M_J$ mass on a similarly eccentric orbit (e=0.18). \citet{Thilliez2016} suggested a $3M_J$ companion at 100 au on an  eccentric (e=0.20) orbit. The RV and HCI data we used were taken from \cite{Perdelwitz2024} and JWST, respectively. Our detection limits exclude companions more massive than 1$M_J$ in the 1-10 au range; the detection limits increase to $\sim 2.5M_J$ at 100-120 au, which is hardly compatible with the \citet{Thilliez2016} predictions. Gravity or ELT observations or additional astrometric data are needed to {either confirm or reject} this potential companion.

\subsection{HD 202917} HD 202917 (0.9\ensuremath{M_{\odot}}/G7V/45Myr) harbors a 13.2 au wide ring with inner and outer edges at 55.4 au and 68.6 au, respectively, seen at an inclination of 68.6° in optical wavelengths \citep{Schneider2016}. \citep{Schneider2016} noted the brighter east side, as seen by HST/STIS, and suggested a recent dust enhancement event as its origin. The RV data we used were taken from \cite{Perdelwitz2024, HIRES_Pub2017}. Our detection limits exclude companions more massive than 5$M_J$ at separations of 0-200 au.

\section{Conclusion}\label{sec:conclusion}
Our search for companions in systems with resolved debris disks using astrometric data has yielded the unambiguous detection of five companions. Except for those around HD 206893, they are new: one stellar companion around HD 54341 ($>200M_J, ~0.6-3$ au) and one stellar or brown dwarf companion around HD 146897 ($>15M_J, ~0.05-2$ au). We found one $2-6M_J$ giant planet orbiting $\sim2-10$ au around HD 107146, and one $3-20M_J$ planet/BD orbiting $\sim3-10$ au from the close binary HD 106906. This newly discovered companion calls for a revision of past dynamical modeling of the system. 

We found indications of companions around BD+45° 598, HD 120326, and TWA 7. The radial velocity and high-contrast imaging data available for HD 120326 and TWA 7 led to 10-20 (2-10) $M_J$ companions orbiting less than 5 au from their parent stars. The \textit{Gaia} DR4 is expected to confirm and fully characterize the candidate companions. 

Direct imaging with JWST and ground-based VLTI/Gravity might image some of the detected companions and also provide a spectral characterization. The companions that are closest to stars and faintest will be exquisite targets for the ELT.

While a statistical study was not the objective of this paper, we note that now at least 8 of the 94 stars with structured dusty rings host known giant planets. The occurrence, 8.5$\%$, is slightly higher than  that found by \cite{Meshkat2017} for giant planets with masses 5-20 $M_J$ and separations greater than 10 au around stars with Spitzer IR excess attributed to dusty disks. Their occurrence was 6.3$\%$. If the tentative planets around HD 120326 and  TWA 7 were confirmed, this rate would increase to 10.6$\%$. {We explain the higher rate by the improved sensitivity of our analysis, which combined data and results from absolute astrometry, direct imaging, and radial velocity. }

Finally, for systems without detections, we computed detection limits from \textit{Gaia}, PMa, RV, and HCI data when available for a selection of iconic systems and systems with disks presenting notable ring-like structures.

\begin{acknowledgements}
This project has received funding from the European Research Council (ERC) under the European Union's Horizon 2020 research and innovation programme (COBREX; grant agreement 885593).\\
This publication makes use of SPHERE data analysed within the COBREX data Center, and JWST data (ID 1277, P.I. PO. Lagage; ID 1668, P.I. S. Marino; ID 2538, P.I. S. Hinkley; and ID 3662, P.I. AM. Lagrange).\\ 
The Data \& Analysis Center for Exoplanets (DACE), which is a facility based at the University of Geneva (CH) dedicated to extrasolar planets data visualisation, exchange and analysis. DACE is a platform of the Swiss National Centre of Competence in Research (NCCR) PlanetS, federating the Swiss expertise in Exoplanet research. The DACE platform is available at \url{https://dace.unige.ch}.\\
This research has made use of the NASA Exoplanet Archive, which is operated by the California Institute of Technology, under contract with the National Aeronautics and Space Administration under the Exoplanet Exploration Program.\\
This research has made use of the SIMBAD database, CDS, Strasbourg Astronomical Observatory, France \citep{Wenger2000}. This research has made use of the VizieR catalogue access tool, CDS, Strasbourg Astronomical Observatory, France (DOI : 10.26093/cds/vizier) \citep{Ochsenbein2000}.

\end{acknowledgements}

\bibliographystyle{aa}
\bibliography{main_biblio}

\begin{appendix} %AM
%\appendix AM
\onecolumn
\section{Long figures}
\vspace{-0.7cm}
\begin{figure}[H]
    \centering
    \setlength{\abovecaptionskip}{0.5pt}
    \setlength{\belowcaptionskip}{0.1pt}

    \begin{subfigure}[t]{0.25\textwidth}
        \caption*{\small \hspace{5pt} GJ 14}
        \includegraphics[width=\linewidth,clip]{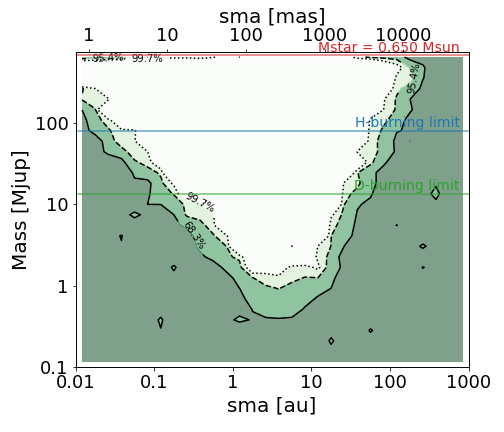}
    \end{subfigure}
    \begin{subfigure}[t]{0.25\textwidth}
        \caption*{\small \hspace{5pt} HD 105}
        \includegraphics[width=\linewidth,clip]{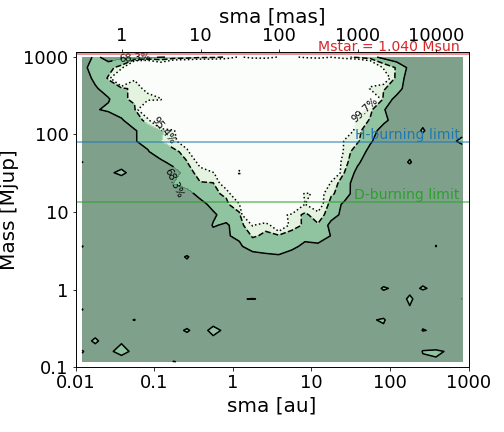}
    \end{subfigure}
    \begin{subfigure}[t]{0.25\textwidth}
        \caption*{\small \hspace{5pt} 49 Ceti}
        \includegraphics[width=\linewidth,clip]{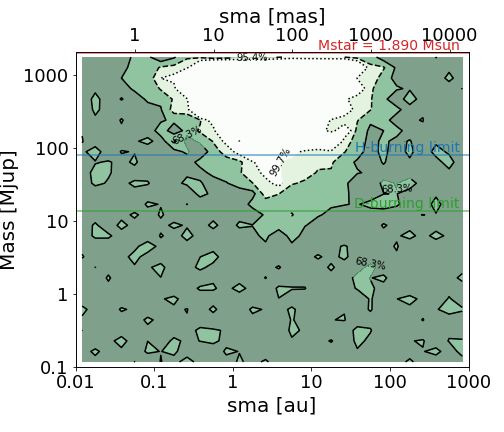}
    \end{subfigure}
    
    \begin{subfigure}[t]{0.25\textwidth}
        \caption*{\small \hspace{5pt} $\epsilon$ Eridani}
        \includegraphics[width=\linewidth,clip]{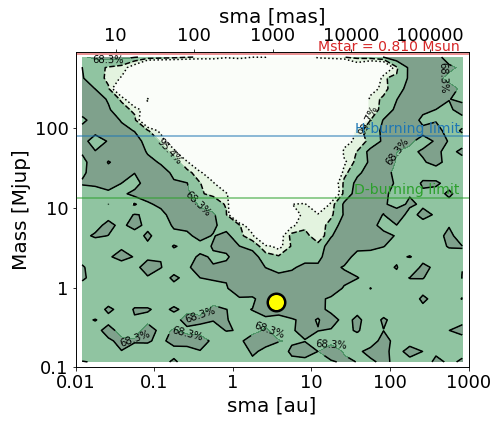}
    \end{subfigure}
    \begin{subfigure}[t]{0.25\textwidth}
        \caption*{\small \hspace{5pt} $\beta$ Pictoris}
        \includegraphics[width=\linewidth,clip]{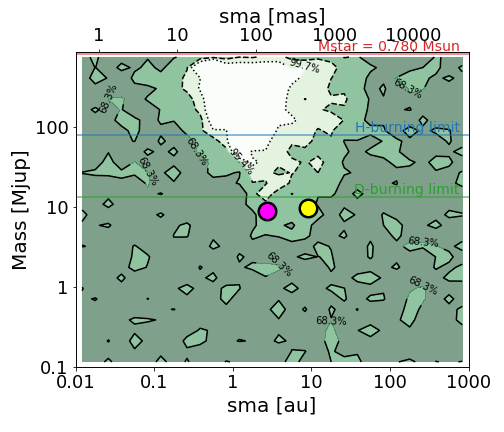}
    \end{subfigure}
    \begin{subfigure}[t]{0.25\textwidth}
        \caption*{\small \hspace{5pt} HD 53143}
        \includegraphics[width=\linewidth,clip]{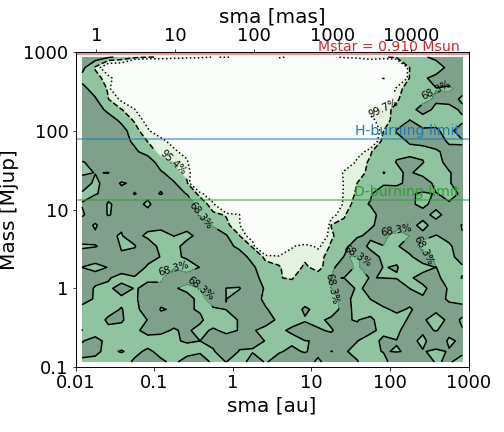}
    \end{subfigure}
    
    \begin{subfigure}[t]{0.25\textwidth}
        \caption*{\small \hspace{5pt} HD 92945}
        \includegraphics[width=\linewidth,clip]{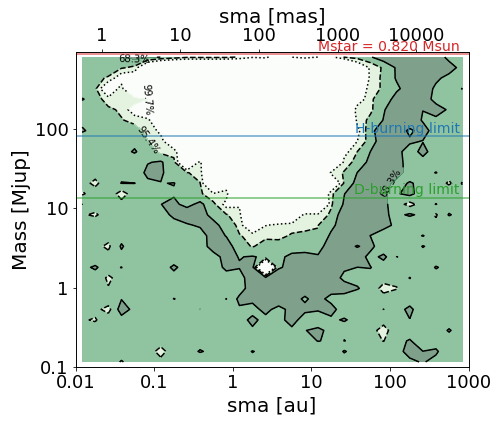}
    \end{subfigure}
    \begin{subfigure}[t]{0.25\textwidth}
        \caption*{\small \hspace{5pt} HD 105211}
        \includegraphics[width=\linewidth,clip]{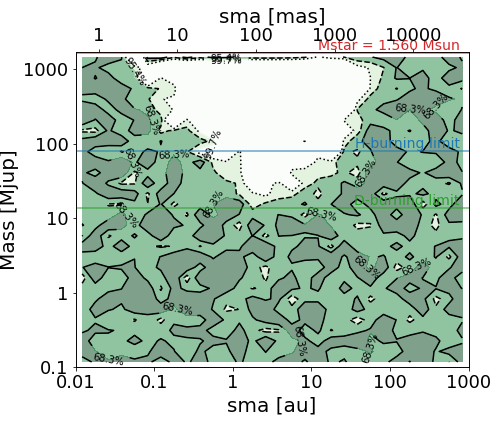}
    \end{subfigure}
    \begin{subfigure}[t]{0.25\textwidth}
        \caption*{\small \hspace{5pt} HD 109085}
        \includegraphics[width=\linewidth,clip]{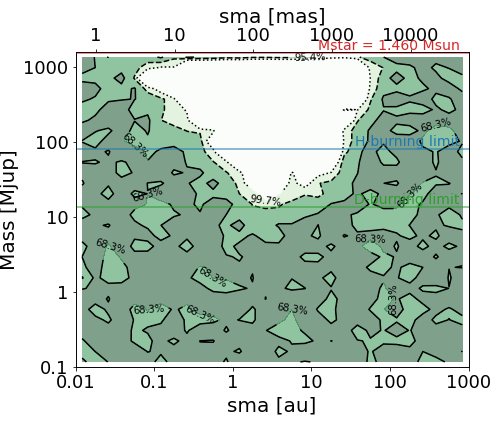}
    \end{subfigure}
    
    \begin{subfigure}[t]{0.25\textwidth}
        \caption*{\small \hspace{5pt} HD 109573}
        \includegraphics[width=\linewidth,clip]{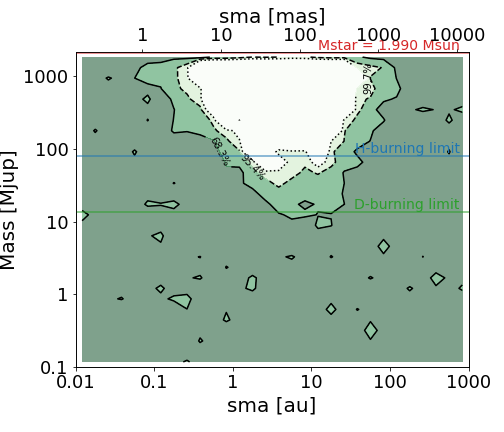}
    \end{subfigure}
    \begin{subfigure}[t]{0.25\textwidth}
        \caption*{\small \hspace{5pt} HD 170773}
        \includegraphics[width=\linewidth,clip]{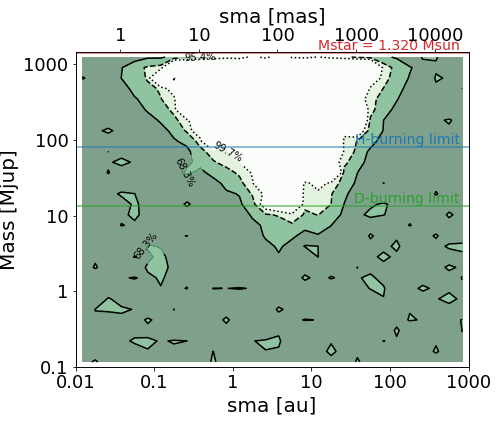}
    \end{subfigure}
    \begin{subfigure}[t]{0.25\textwidth}
        \caption*{\small \hspace{5pt} HD 181327}
        \includegraphics[width=\linewidth,clip]{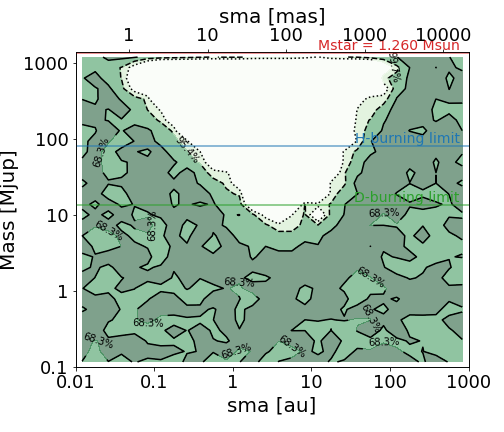}
    \end{subfigure}

    \begin{subfigure}[t]{0.25\textwidth}
        \caption*{\small \hspace{5pt} AU Microscopii}
        \includegraphics[width=\linewidth,clip]{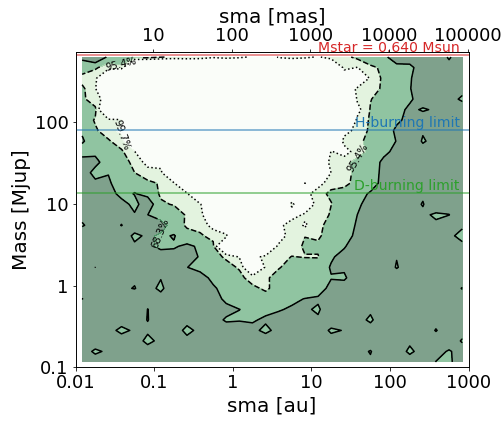}
    \end{subfigure}
    \begin{subfigure}[t]{0.25\textwidth}
        \caption*{\small \hspace{5pt} HD 202628}
        \includegraphics[width=\linewidth,clip]{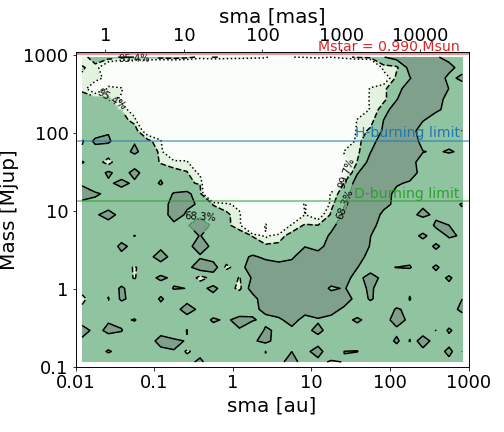}
    \end{subfigure}
    \begin{subfigure}[t]{0.25\textwidth}
        \caption*{\small \hspace{5pt} HD 202917}
        \includegraphics[width=\linewidth,clip]{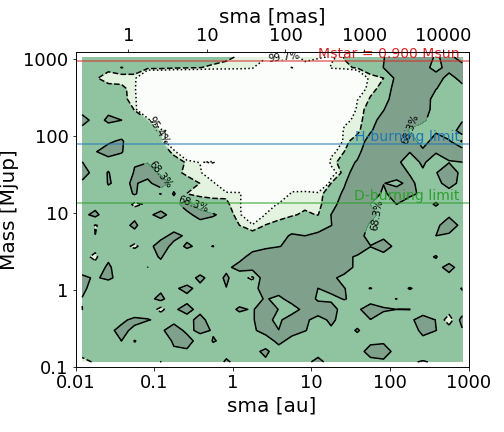}
    \end{subfigure}

    \caption{(sma,mass) GaiaPMEX maps for systems of our sub-sample without detections, taking the debris disk's inclination. From left to right, top to bottom: GJ 14, HD 105, 49 Cet, $\epsilon$ Eri, $\beta$ Pic, HD 53143, HD 92945, HD 105211, HD 109085, HD 109573, HD 170773, HD 181327, AU Mic, HD 202628, and HD 202917. The maps combine RUWE and PMa. {The different shades of color represent different confidence levels, with the darkest shade being $1\sigma$, the intermediate shade $2\sigma$, and the lightest shade $3\sigma$.}}
    \label{fig:nondetections}
\end{figure}

\begin{figure}[H]
    \centering
    \newcommand{\rowwithlabel}[4]{
        \begin{minipage}[c]{6mm}
            \centering
            \rotatebox{90}{\footnotesize #1}
        \end{minipage}
        \begin{minipage}[c]{50mm}
            \includegraphics[width=\linewidth,clip]{#2}
        \end{minipage}
        \begin{minipage}[c]{50mm}
            \includegraphics[width=\linewidth,clip]{#3}
        \end{minipage}
        \begin{minipage}[c]{50mm}
            \includegraphics[width=\linewidth,clip]{#4}
        \end{minipage}\\[3pt]
    }

    \rowwithlabel{HD 54341}
        {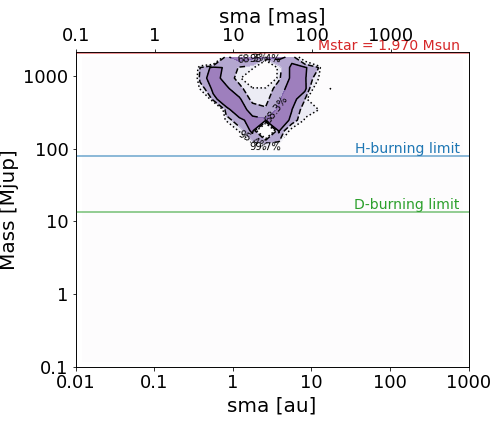}
        {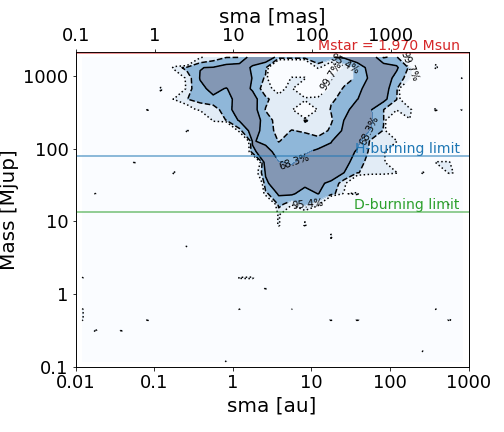}
        {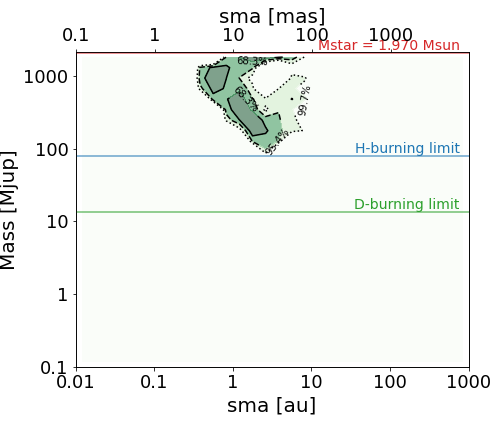}
    
    \vspace{-6pt}
    \rowwithlabel{HD 106906}
        {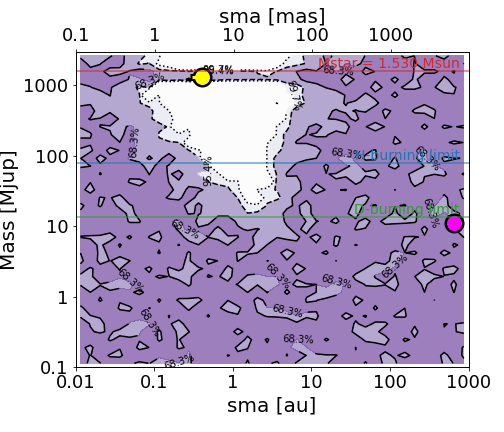}
        {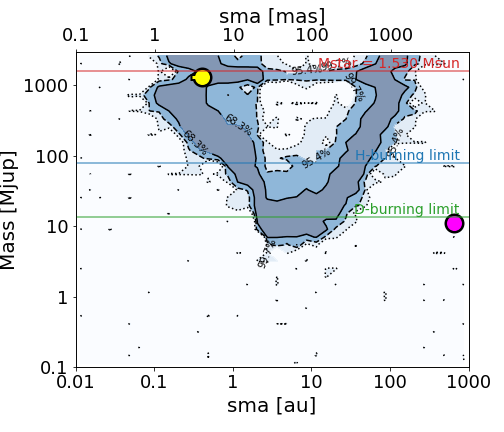}
        {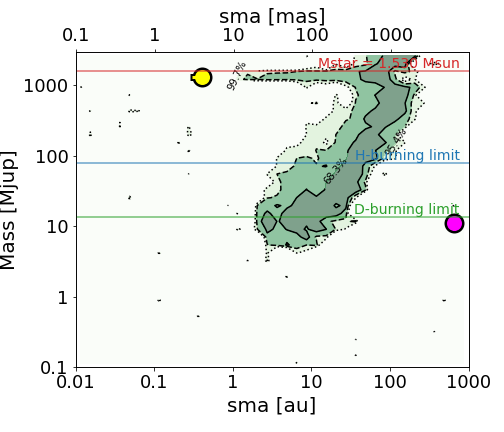}

    \vspace{-6pt}
    \rowwithlabel{HD 107146}
        {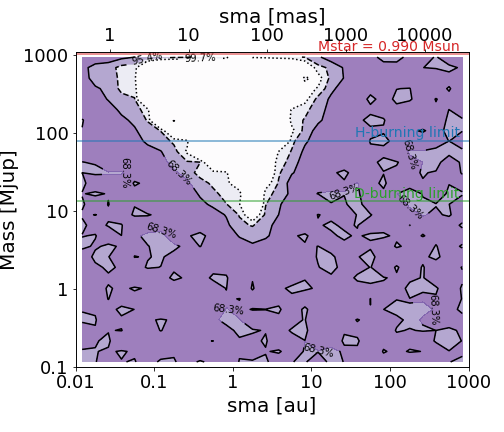}
        {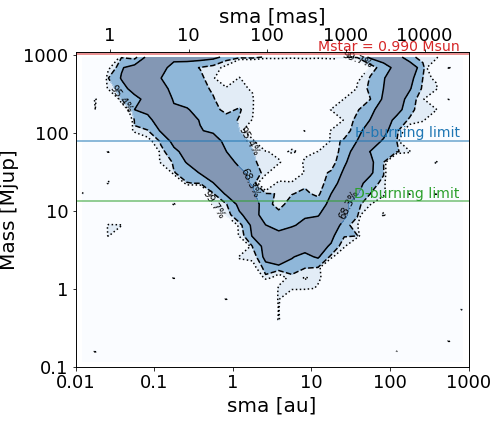}
        {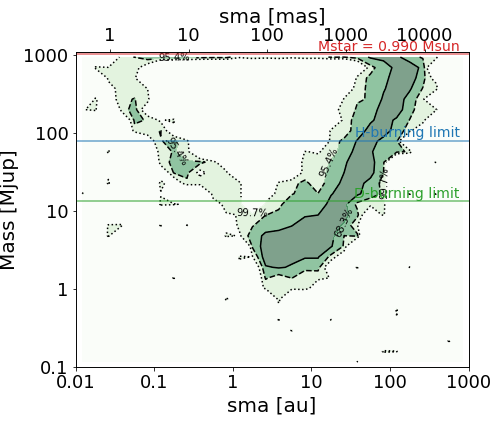}

    \vspace{-6pt}
    \rowwithlabel{HD 146897}
        {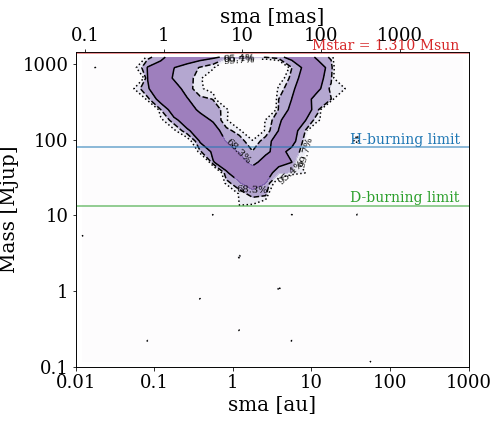}
        {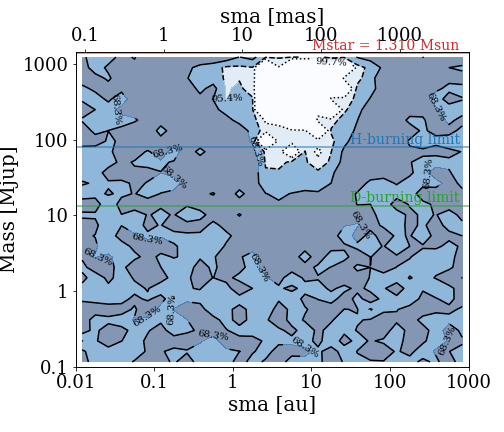}
        {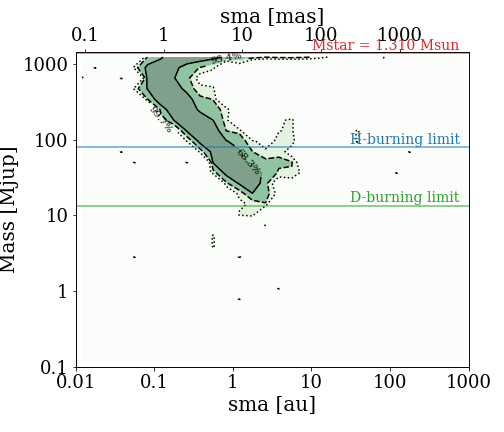}
    \vspace{-6pt}
    \rowwithlabel{HD 206893}
        {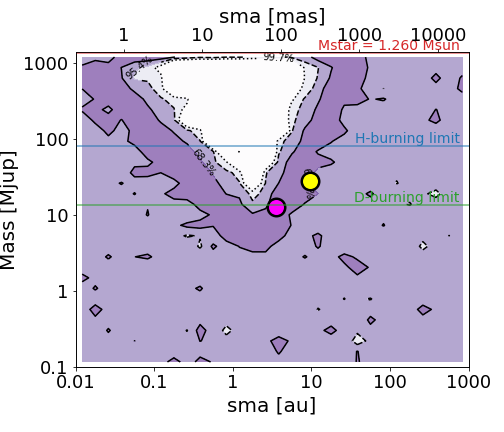}
        {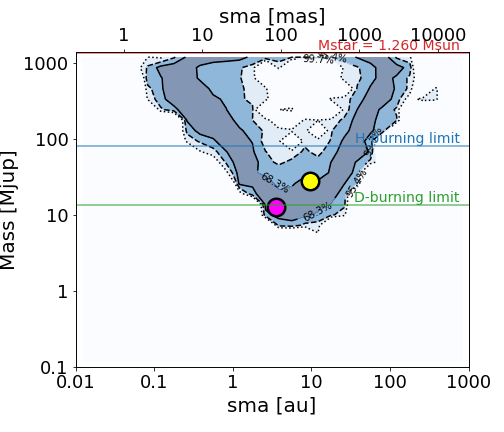}
        {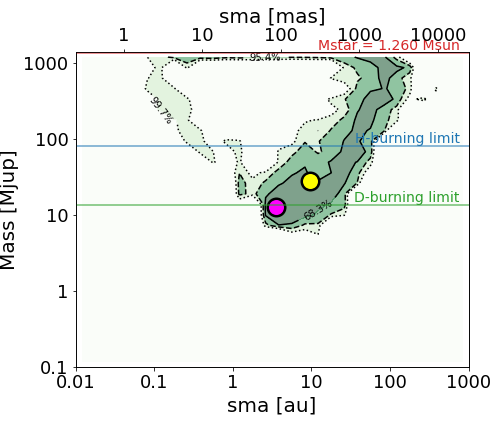}
    \caption{(sma,mass) GaiaPMEX solutions for systems with unambiguous detections. Known companions are plotted as filled circles with the associated uncertainties as vertical/horizontal bars. From Left to right: RUWE (purple), PMa (blue), combination of RUWE and PMa (green). From top to bottom: HD 54341, HD 106906, HD 107146, HD 146897, HD 206893. For HD 106906, we consider the known binary and use a better adapted y-axis. The different shades of color represent different confidence levels, with the darkest shade being $1\sigma$, the intermediate shade $2\sigma$, and the lightest shade $3\sigma$.}
   \label{fig:unambiguous_detections}
\end{figure}

\begin{figure}[H]
    \centering
    \includegraphics[width=55.mm,clip=True]{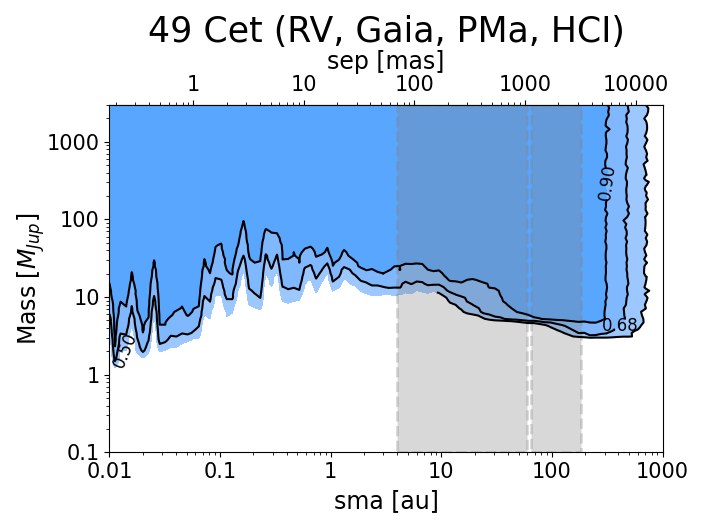}
    \includegraphics[width=55.mm,clip=True]{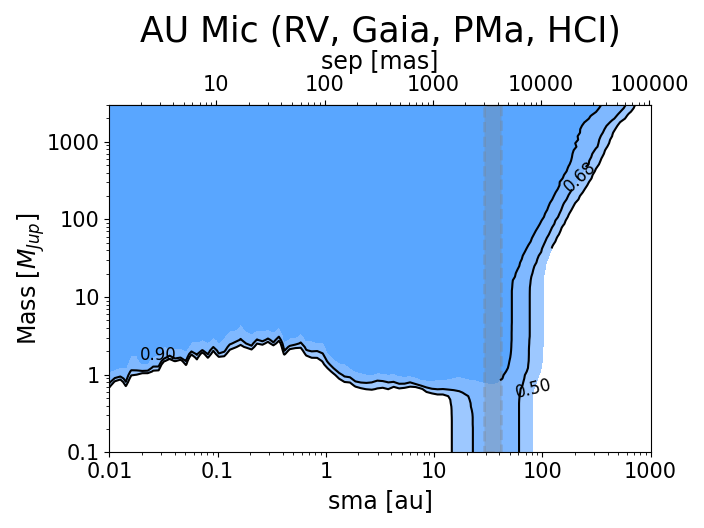}
    \includegraphics[width=55.mm,clip=True]{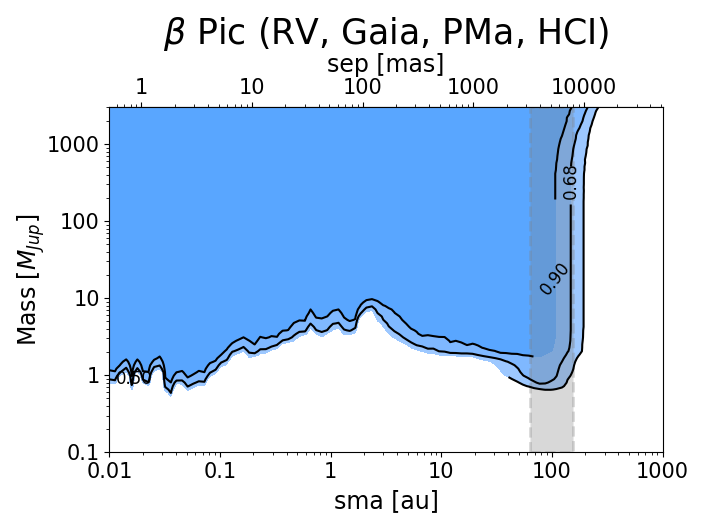}
    \includegraphics[width=55.mm,clip=True]{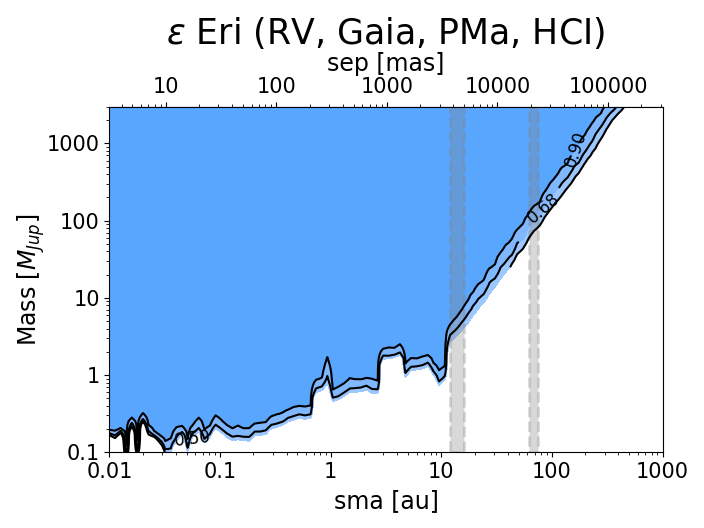}
    \includegraphics[width=55.mm,clip=True]{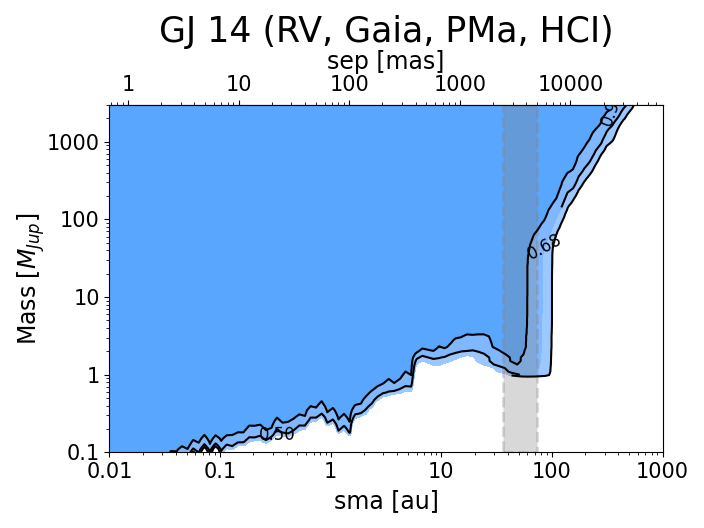}
    \includegraphics[width=55.mm,clip=True]{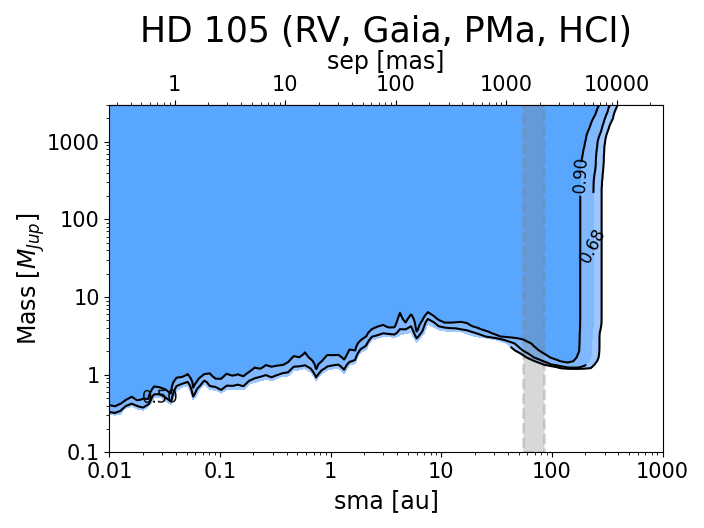}
    \includegraphics[width=55.mm,clip=True]{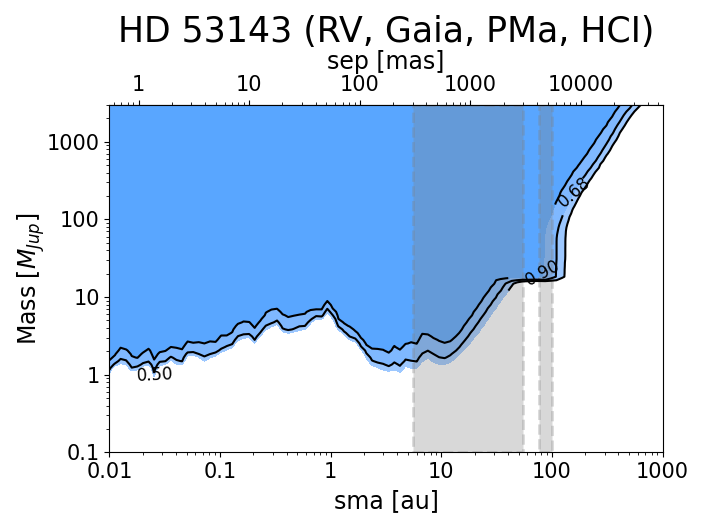}
    \includegraphics[width=55.mm,clip=True]{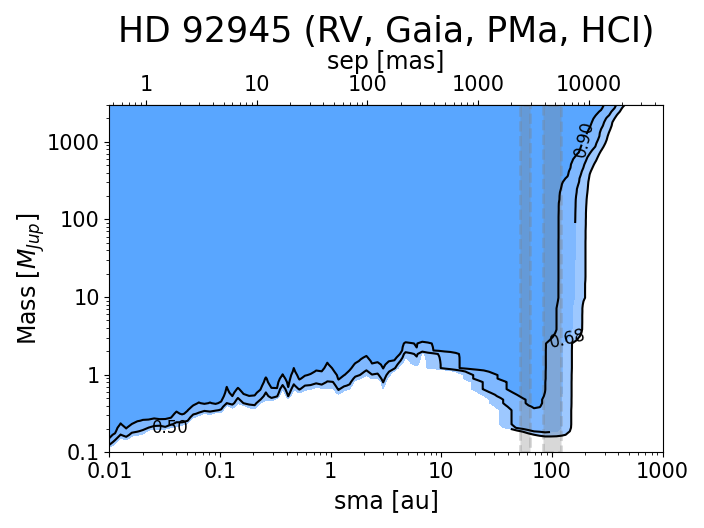}
    \includegraphics[width=55.mm,clip=True]{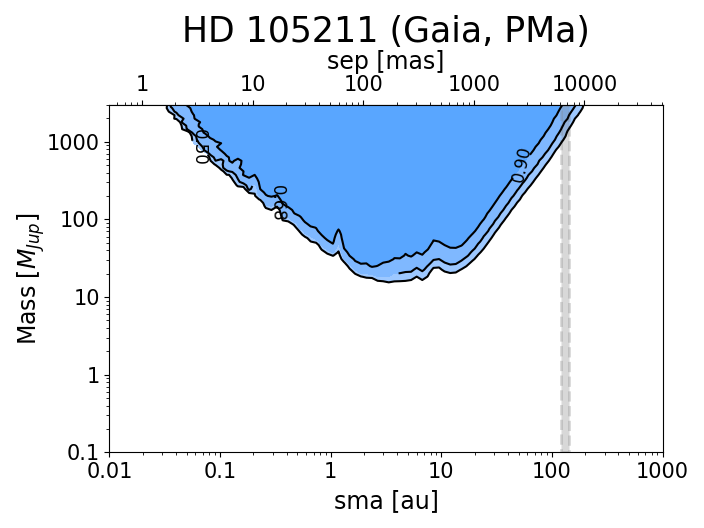}
    \includegraphics[width=55.mm,clip=True]{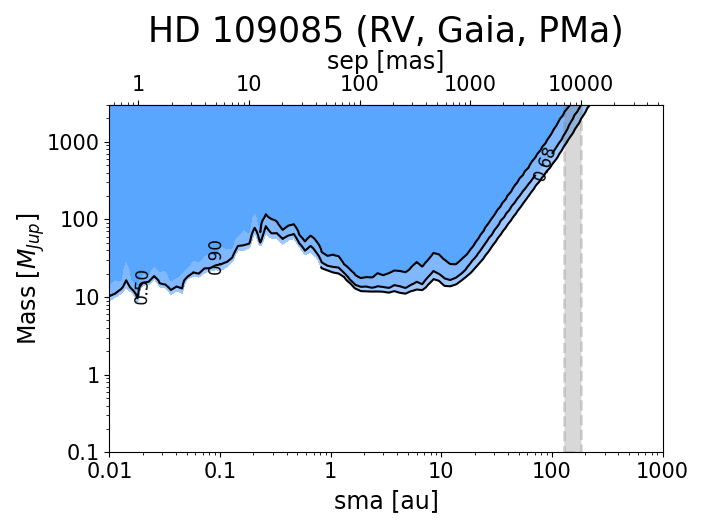}
    \includegraphics[width=55.mm,clip=True]{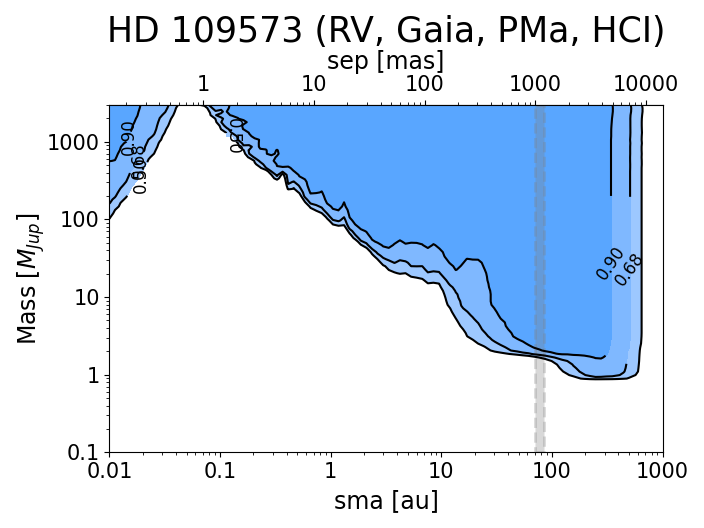}
    \includegraphics[width=55.mm,clip=True]{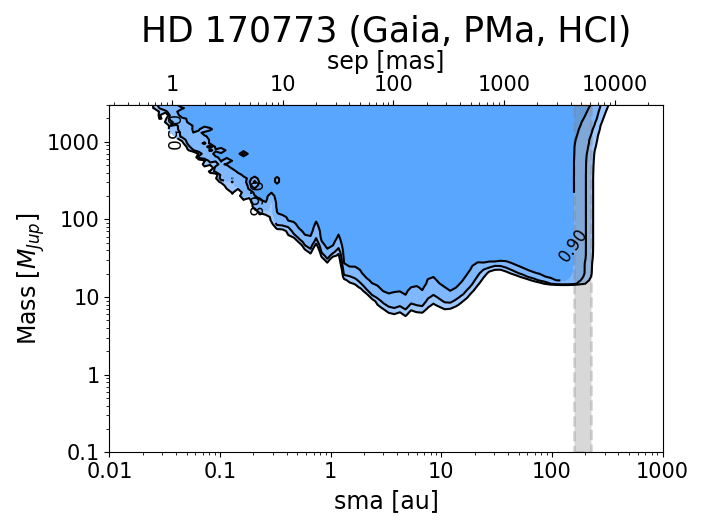}
    \includegraphics[width=55.mm,clip=True]{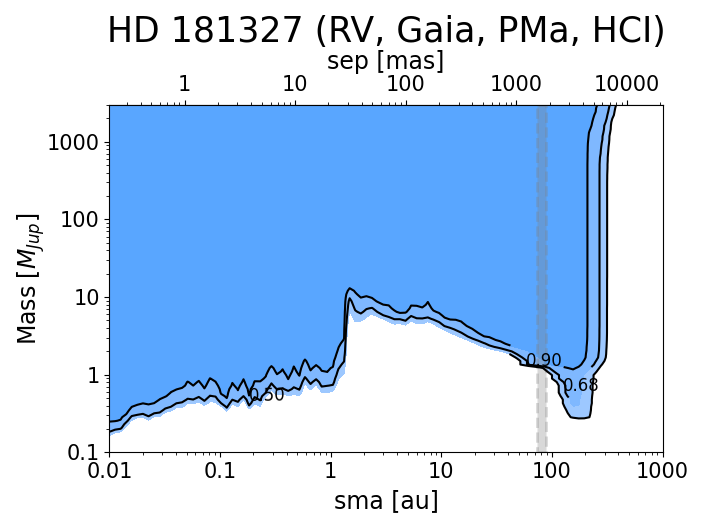}
    \includegraphics[width=55.mm,clip=True]{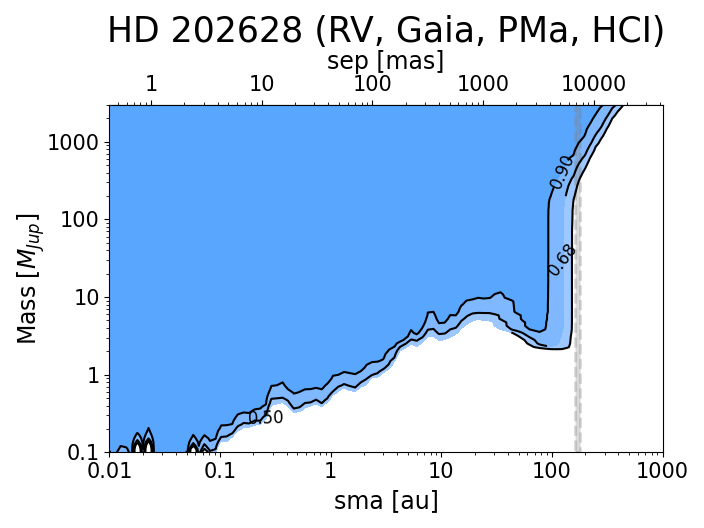}
    \includegraphics[width=55.mm,clip=True]{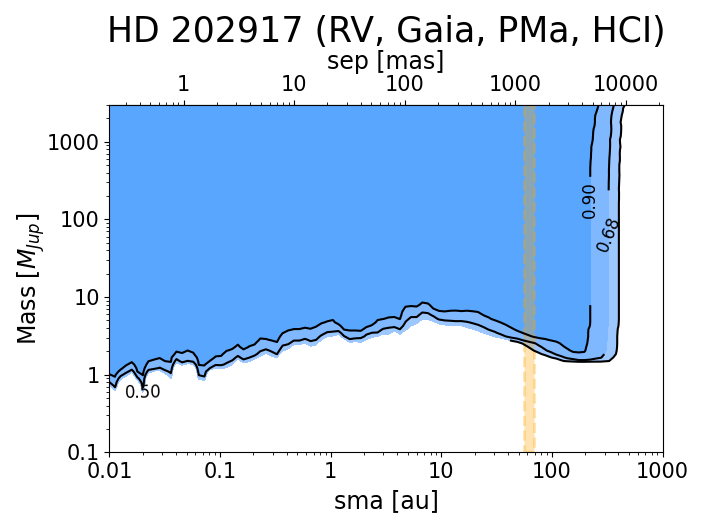}
     \caption{(sma,mass) detection limits obtained with  PyMESS for systems with no detection and part of our sub-sample. The shades of blue correspond to different confidence levels annotated on the graph: 50\%, 68\%, 90\%. The limits of the millimeter debris disks are shown in gray. For HD 202917, since the system was not observed in millimeter wavelengths, we use STIS observations to plot the disk location, shown in orange to distinguish it. The type of data used and the target planet are specified in the title of each graph.}
    \label{fig:detection limits all}
\end{figure}

\begin{figure*}[p]
    \centering
    \begin{subfigure}{0.32\textwidth}
        \includegraphics[width=\linewidth,clip=True]{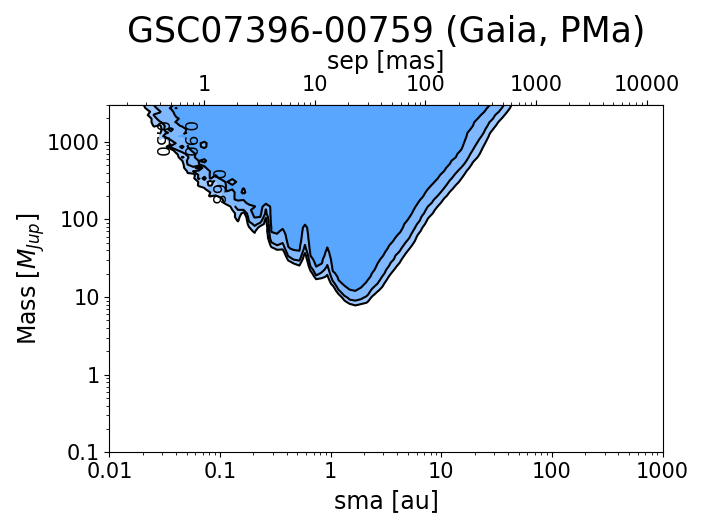}
    \end{subfigure}\hfill
    \begin{subfigure}{0.32\textwidth}
        \includegraphics[width=\linewidth,clip=True]{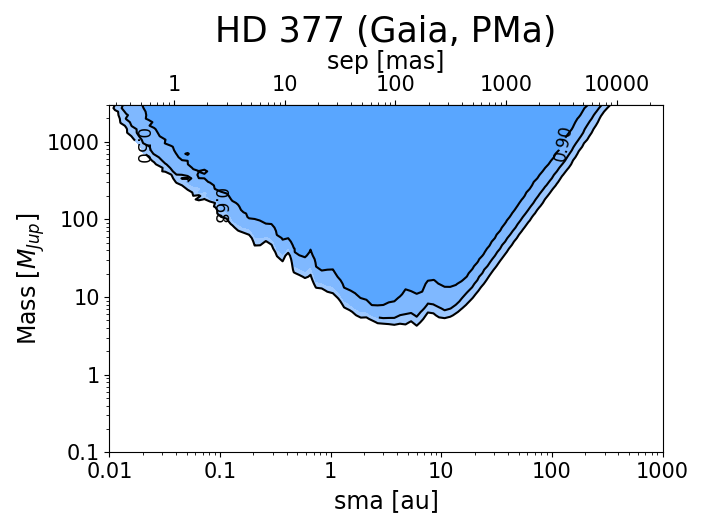}
    \end{subfigure}\hfill
    \begin{subfigure}{0.32\textwidth}
        \includegraphics[width=\linewidth,clip=True]{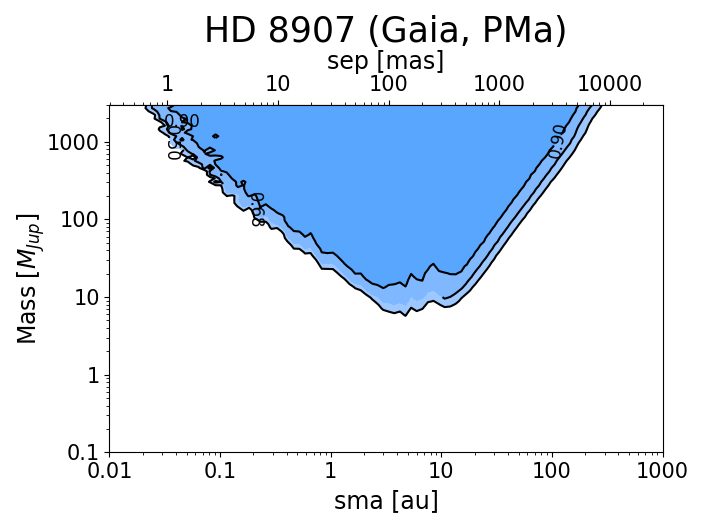}
    \end{subfigure}
    \par\medskip
    \begin{subfigure}{0.32\textwidth}
        \includegraphics[width=\linewidth,clip=True]{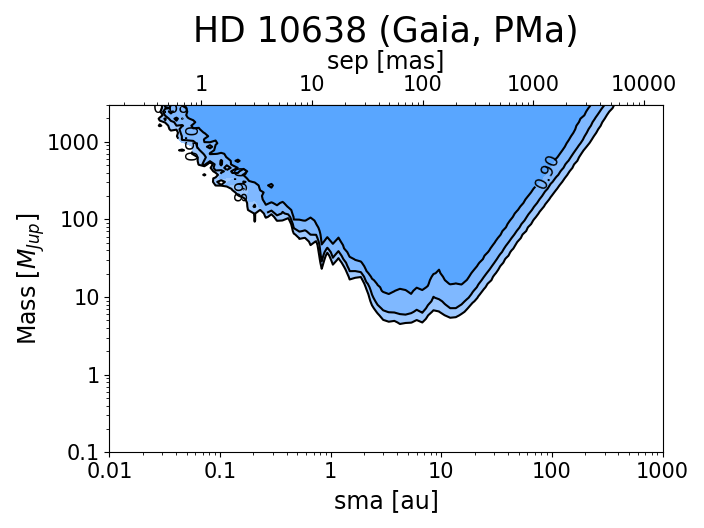}
    \end{subfigure}\hfill
    \begin{subfigure}{0.32\textwidth}
        \includegraphics[width=\linewidth,clip=True]{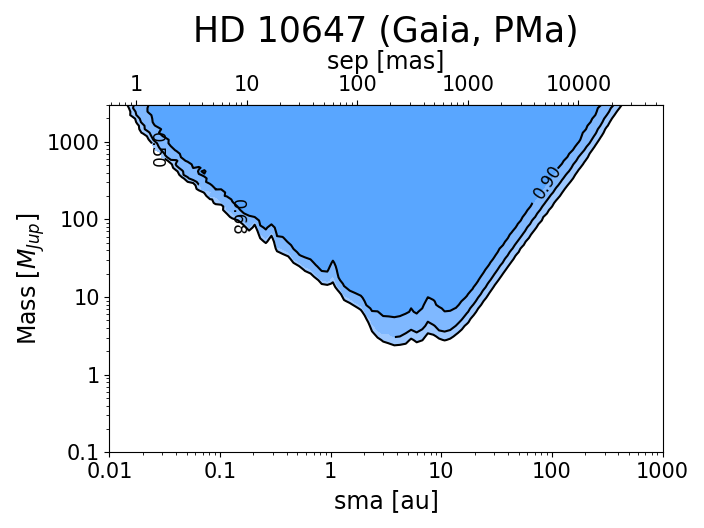}
    \end{subfigure}\hfill
    \begin{subfigure}{0.32\textwidth}
        \includegraphics[width=\linewidth,clip=True]{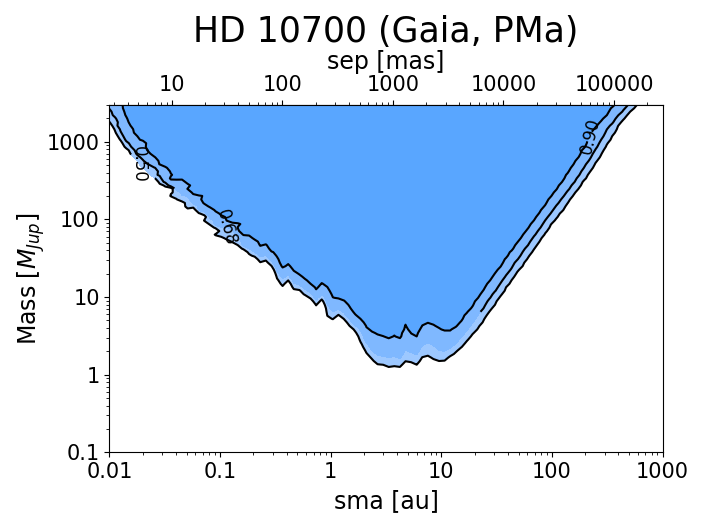}
    \end{subfigure}
    \par\medskip
    \begin{subfigure}{0.32\textwidth}
        \includegraphics[width=\linewidth,clip=True]{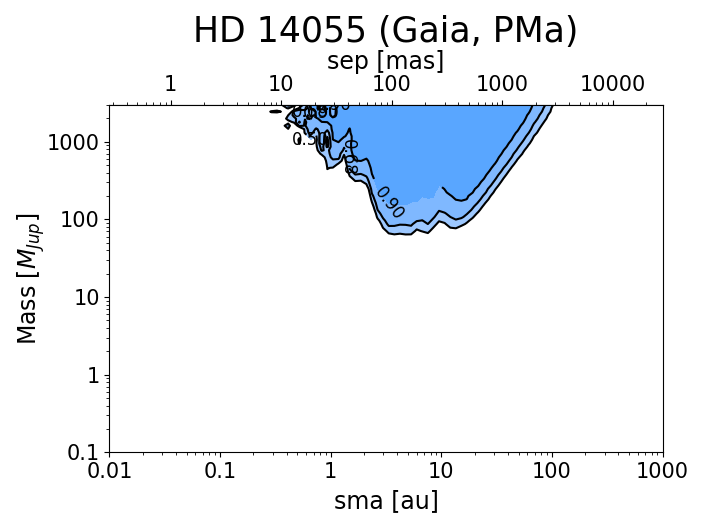}
    \end{subfigure}\hfill
    \begin{subfigure}{0.32\textwidth}
        \includegraphics[width=\linewidth,clip=True]{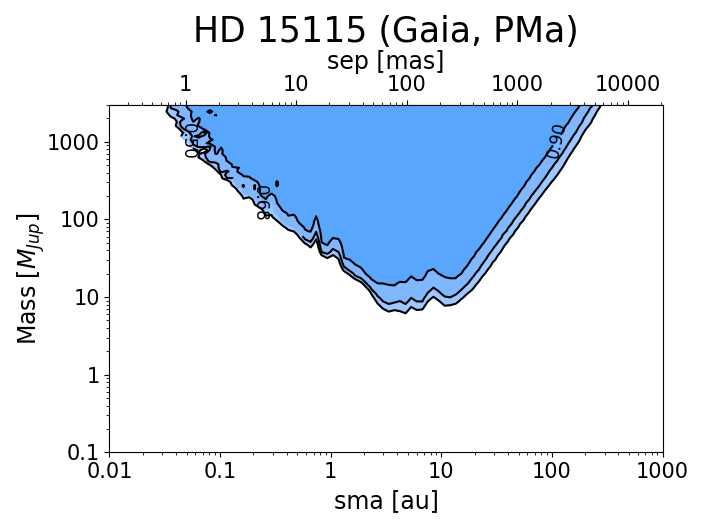}
    \end{subfigure}\hfill
    \begin{subfigure}{0.32\textwidth}
        \includegraphics[width=\linewidth,clip=True]{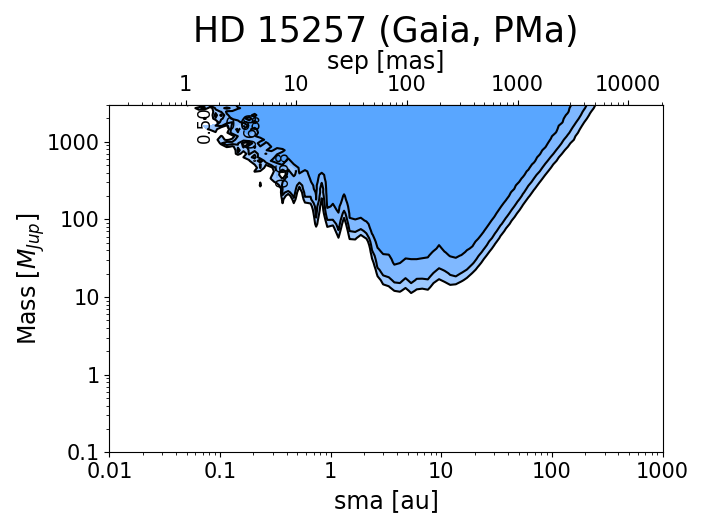}
    \end{subfigure}
    \par\medskip
    \begin{subfigure}{0.32\textwidth}
        \includegraphics[width=\linewidth,clip=True]{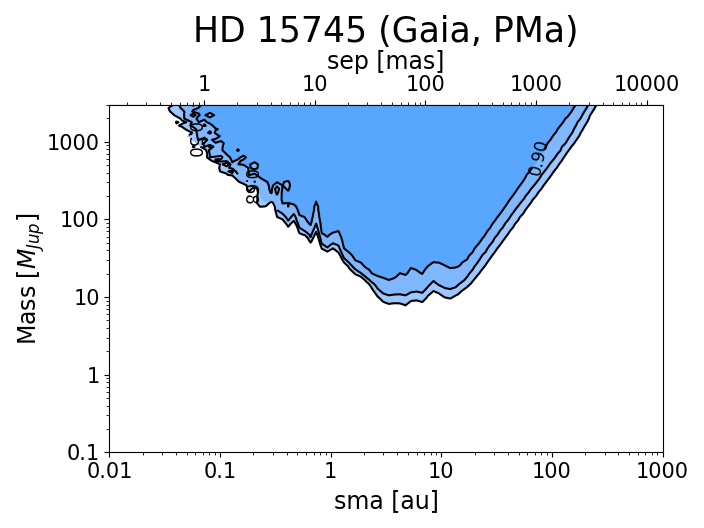}
    \end{subfigure}\hfill
    \begin{subfigure}{0.32\textwidth}
        \includegraphics[width=\linewidth,clip=True]{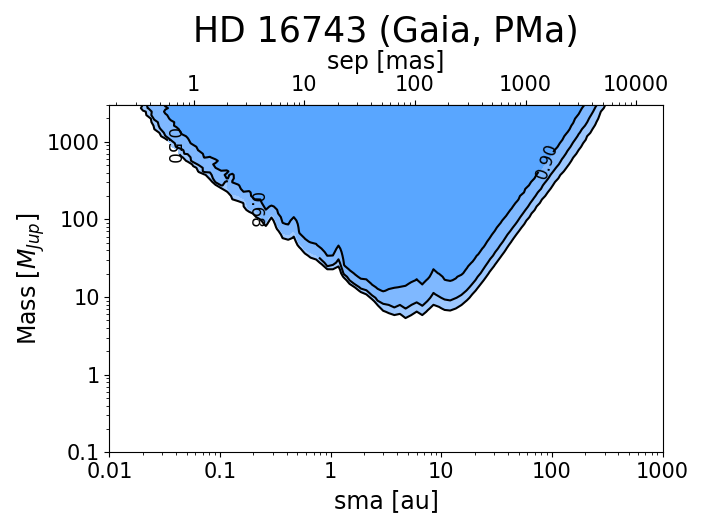}
    \end{subfigure}\hfill
    \begin{subfigure}{0.32\textwidth}
        \includegraphics[width=\linewidth,clip=True]{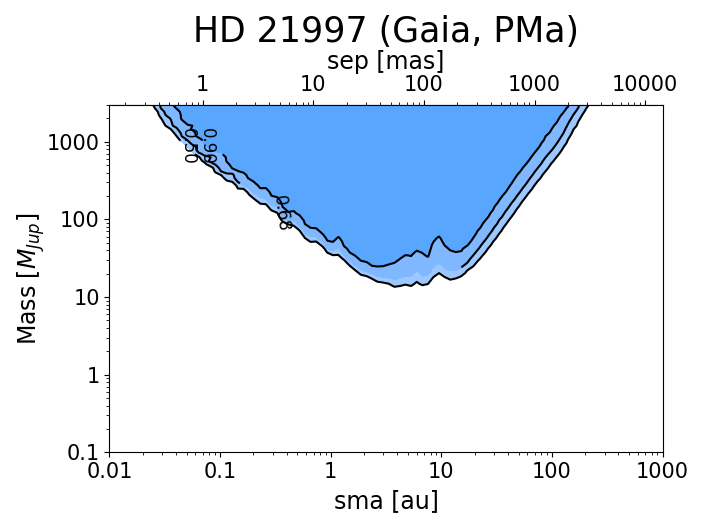}
    \end{subfigure}
    \par\medskip
    \begin{subfigure}{0.32\textwidth}
        \includegraphics[width=\linewidth,clip=True]{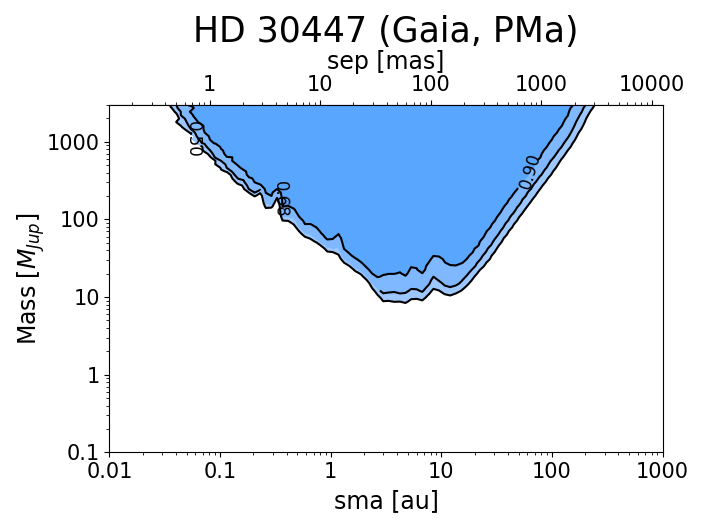}
    \end{subfigure}\hfill
    \begin{subfigure}{0.32\textwidth}
        \includegraphics[width=\linewidth,clip=True]{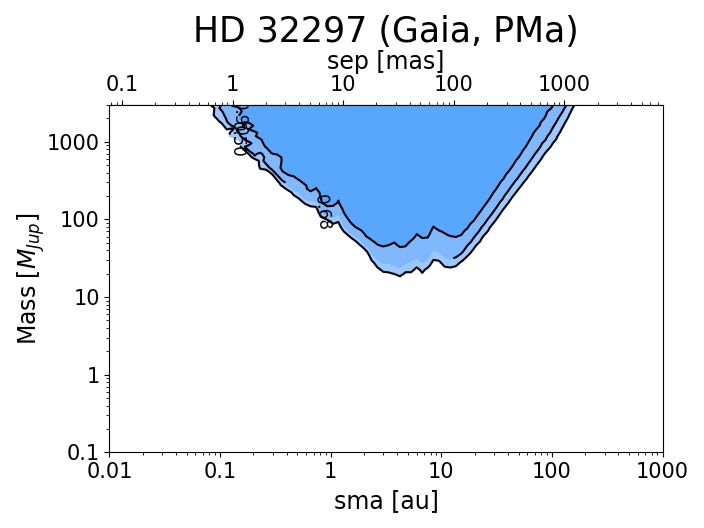}
    \end{subfigure}\hfill
    \begin{subfigure}{0.32\textwidth}
        \includegraphics[width=\linewidth,clip=True]{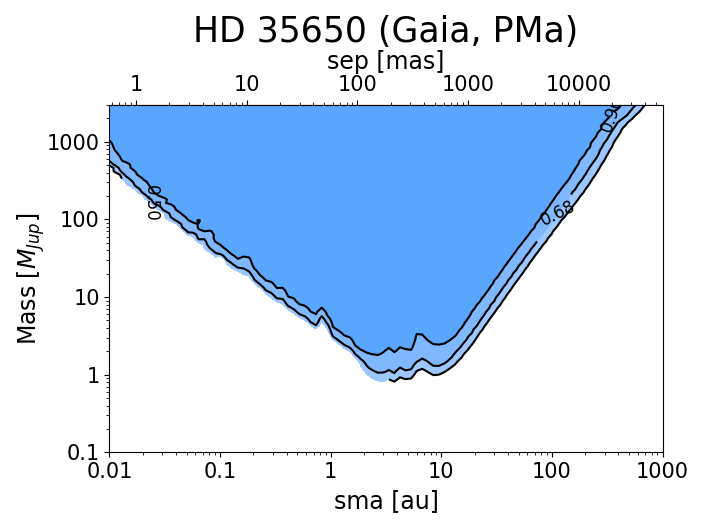}
    \end{subfigure}
    
    \caption{{(sma,mass) detection limits obtained with  PyMESS using AEN and PMa for systems with no detection and not part of our sub-sample. The shades of blue correspond to different confidence levels annotated on the graph: 50\%, 68\%, 90\%. Plots are not given for HD 172167 (Vega) and HD 216956 (Fomalhaut) since they are too bright for \textit{Gaia}.}}
    \label{fig:detection limits all long}
\end{figure*}

\begin{figure*}[p]\ContinuedFloat
    \centering
    \begin{subfigure}{0.32\textwidth}
        \includegraphics[width=\linewidth,clip=True]{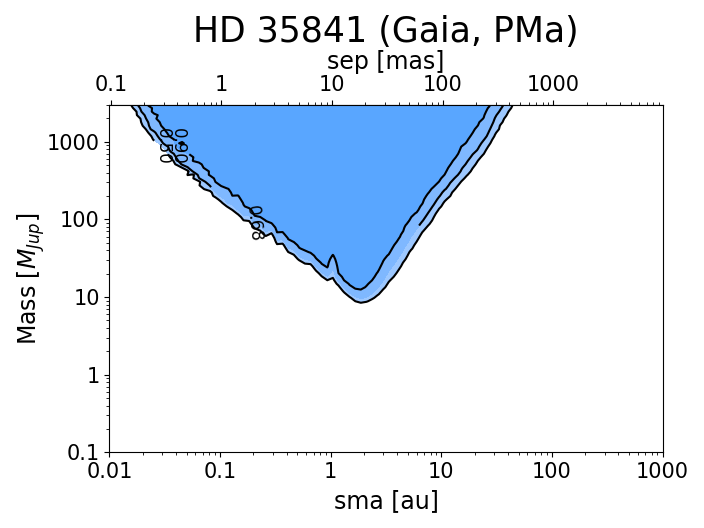}
    \end{subfigure}\hfill
    \begin{subfigure}{0.32\textwidth}
        \includegraphics[width=\linewidth,clip=True]{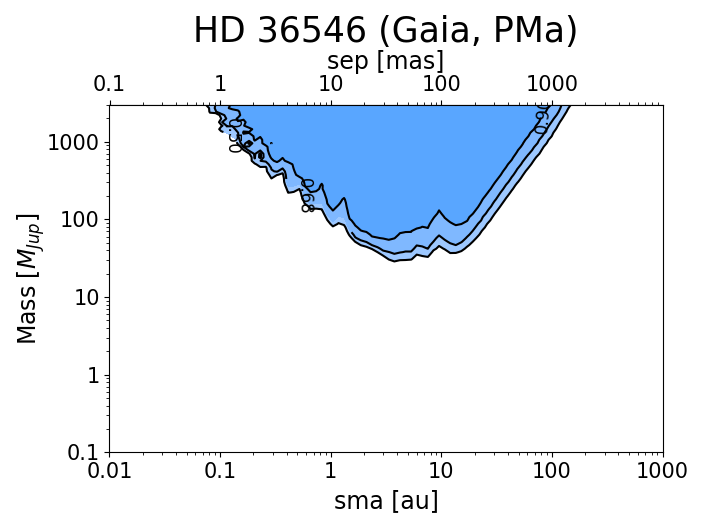}
    \end{subfigure}\hfill
    \begin{subfigure}{0.32\textwidth}
        \includegraphics[width=\linewidth,clip=True]{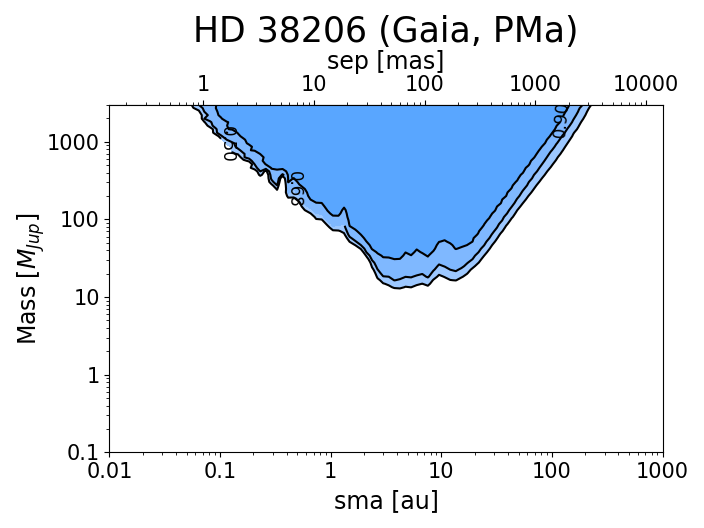}
    \end{subfigure}
    \par\medskip
    \begin{subfigure}{0.32\textwidth}
        \includegraphics[width=\linewidth,clip=True]{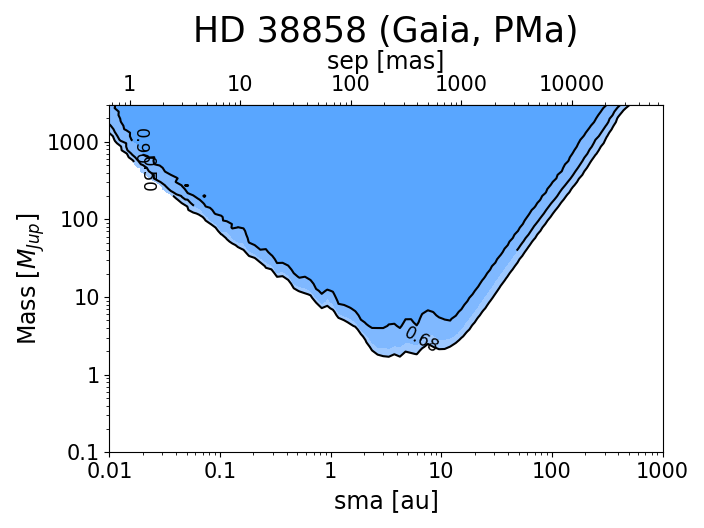}
    \end{subfigure}\hfill
    \begin{subfigure}{0.32\textwidth}
        \includegraphics[width=\linewidth,clip=True]{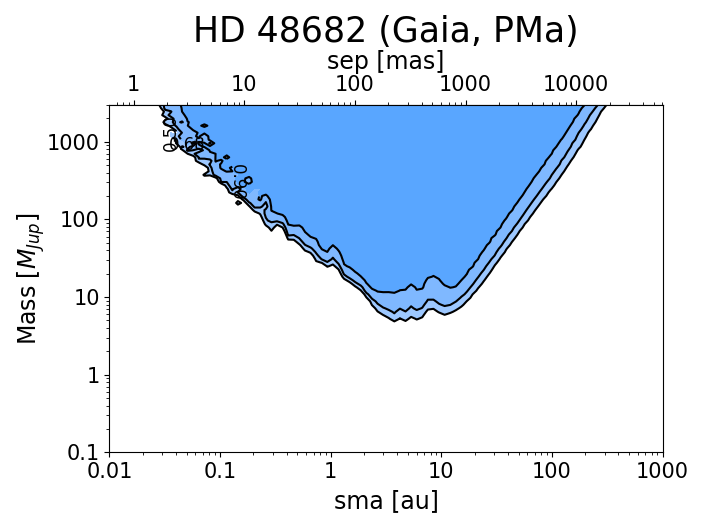}
    \end{subfigure}\hfill
    \begin{subfigure}{0.32\textwidth}
        \includegraphics[width=\linewidth,clip=True]{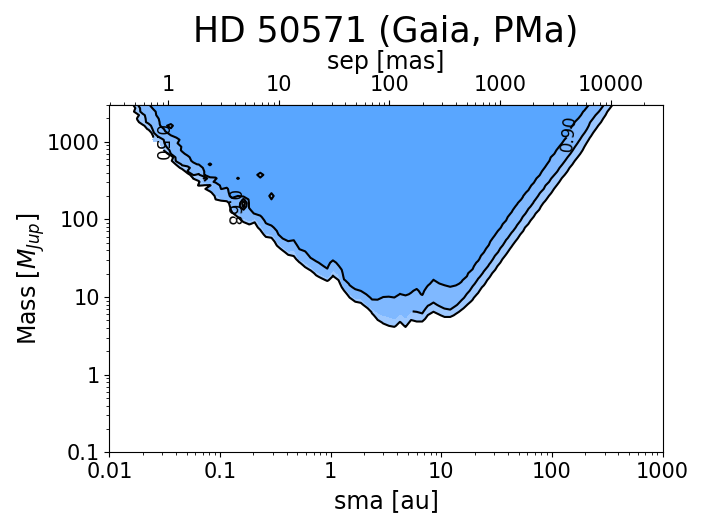}
    \end{subfigure}
    \par\medskip
    \begin{subfigure}{0.32\textwidth}
        \includegraphics[width=\linewidth,clip=True]{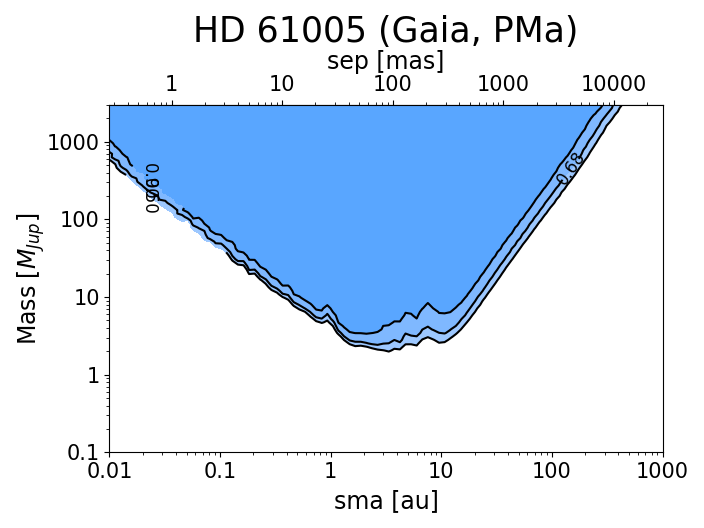}
    \end{subfigure}\hfill
    \begin{subfigure}{0.32\textwidth}
        \includegraphics[width=\linewidth,clip=True]{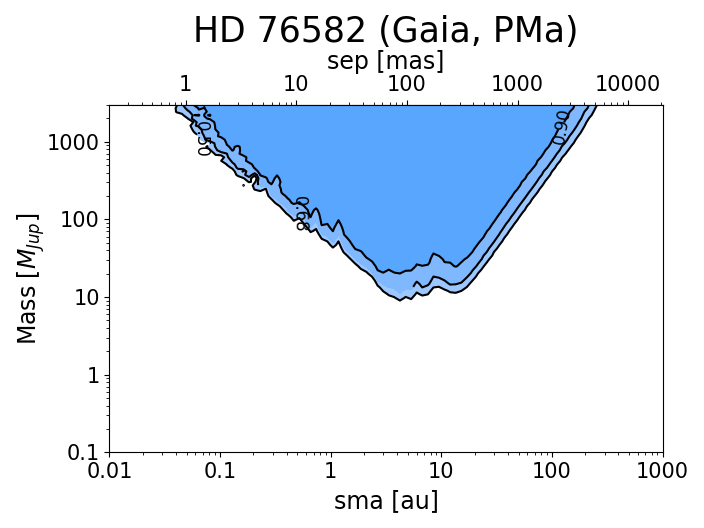}
    \end{subfigure}\hfill
    \begin{subfigure}{0.32\textwidth}
        \includegraphics[width=\linewidth,clip=True]{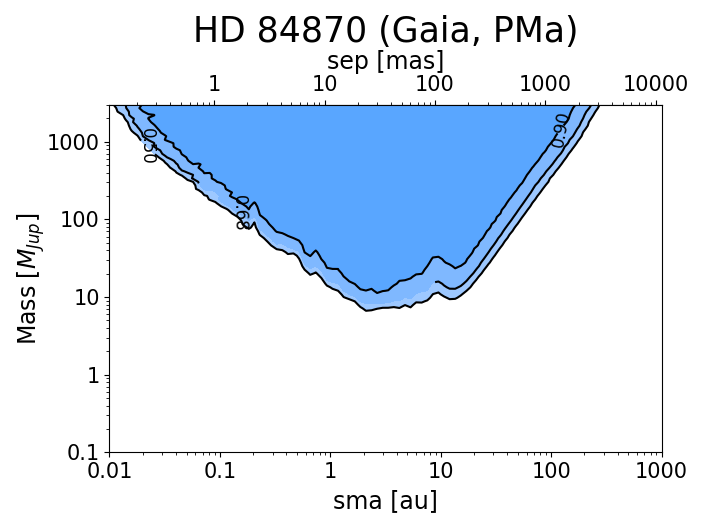}
    \end{subfigure}
    \par\medskip
    \begin{subfigure}{0.32\textwidth}
        \includegraphics[width=\linewidth,clip=True]{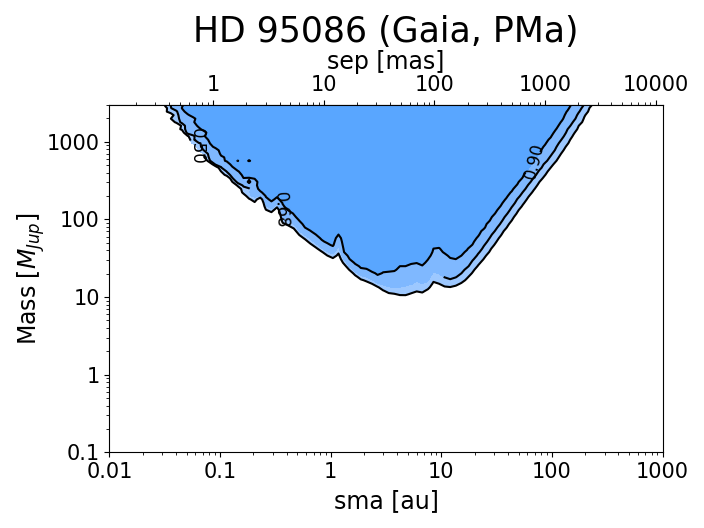}
    \end{subfigure}\hfill
    \begin{subfigure}{0.32\textwidth}
        \includegraphics[width=\linewidth,clip=True]{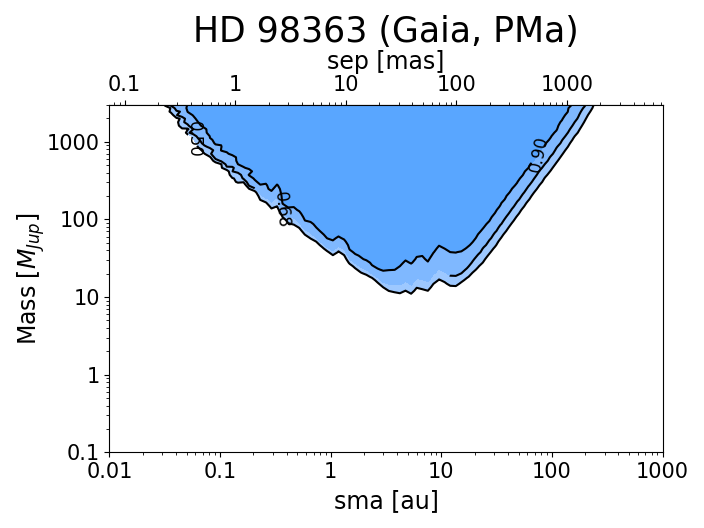}
    \end{subfigure}\hfill
    \begin{subfigure}{0.32\textwidth}
        \includegraphics[width=\linewidth,clip=True]{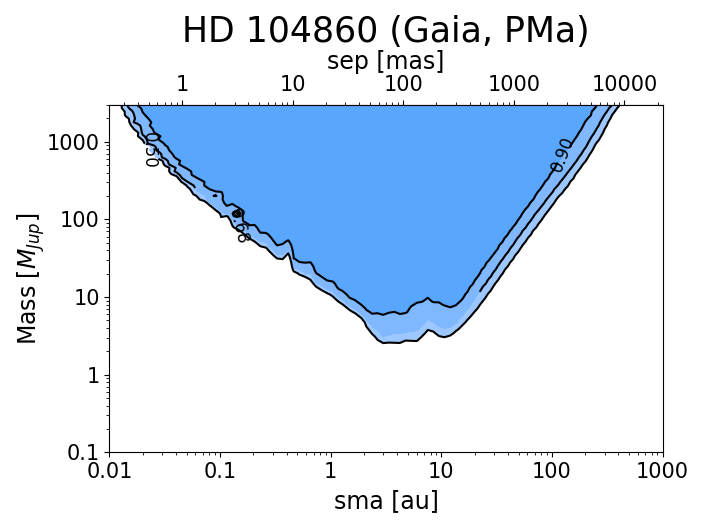}
    \end{subfigure}
    \par\medskip
    \begin{subfigure}{0.32\textwidth}
        \includegraphics[width=\linewidth,clip=True]{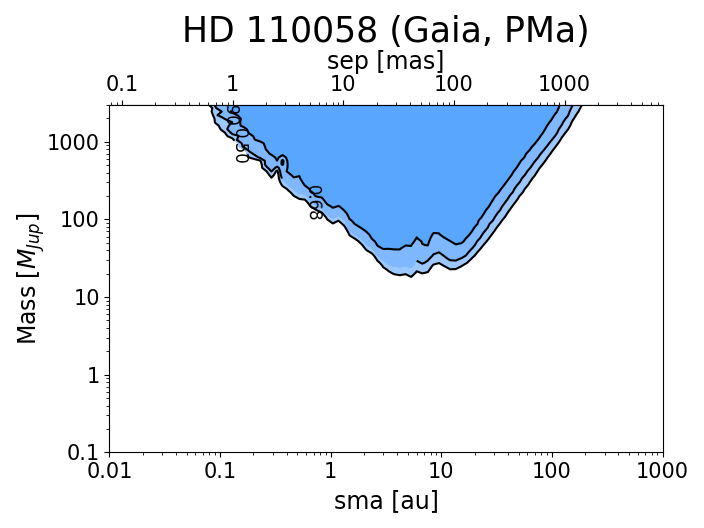}
    \end{subfigure}\hfill
    \begin{subfigure}{0.32\textwidth}
        \includegraphics[width=\linewidth,clip=True]{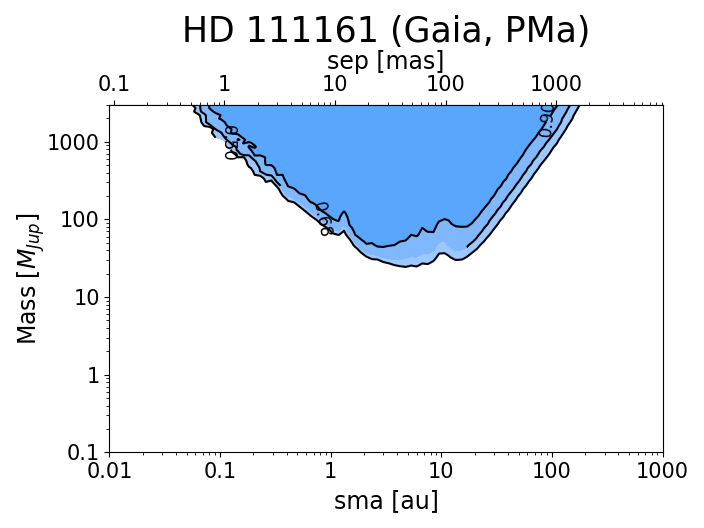}
    \end{subfigure}\hfill
    \begin{subfigure}{0.32\textwidth}
        \includegraphics[width=\linewidth,clip=True]{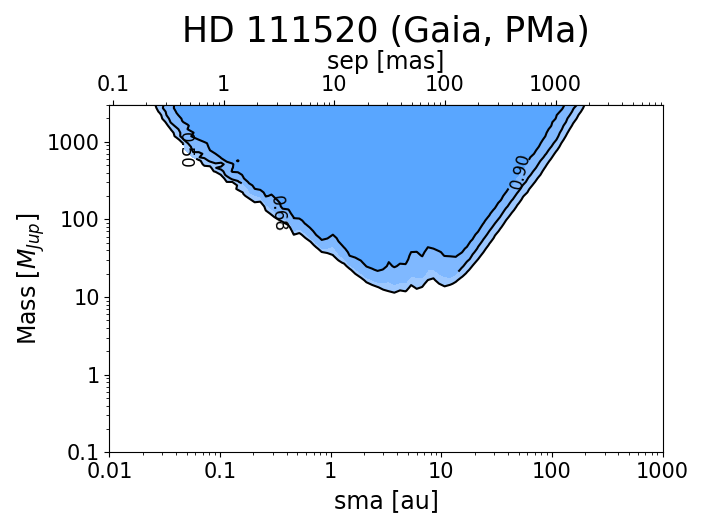}
    \end{subfigure}
    
    \caption*{continued.}
\end{figure*}

\begin{figure*}[p]\ContinuedFloat
    \centering
    \begin{subfigure}{0.32\textwidth}
        \includegraphics[width=\linewidth,clip=True]{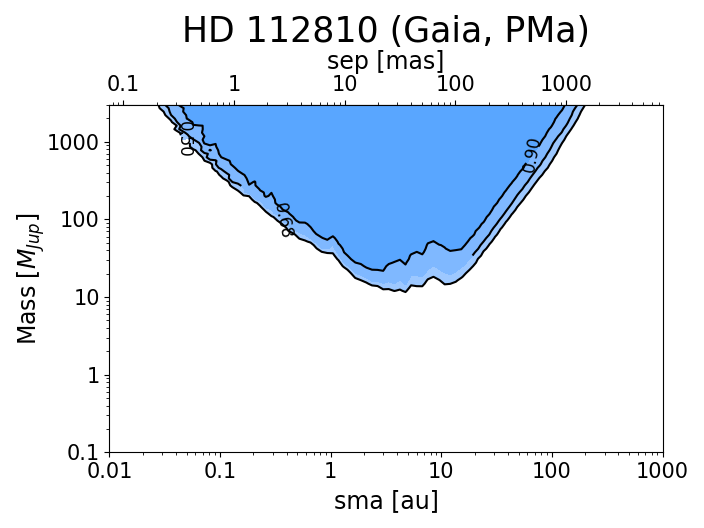}
    \end{subfigure}\hfill
    \begin{subfigure}{0.32\textwidth}
        \includegraphics[width=\linewidth,clip=True]{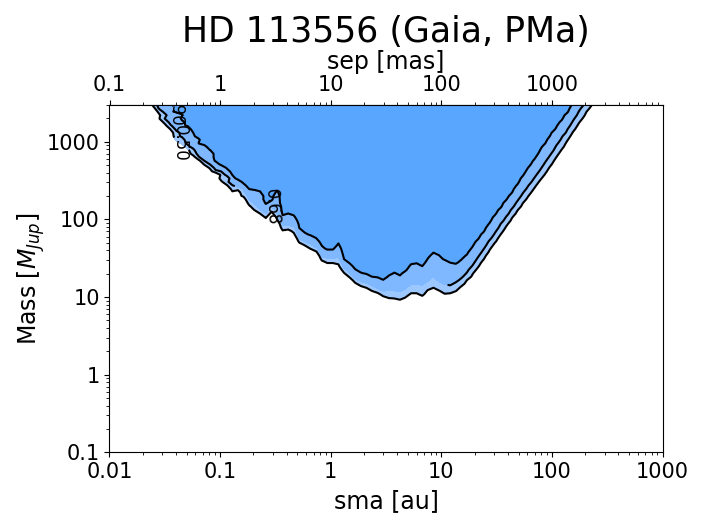}
    \end{subfigure}\hfill
    \begin{subfigure}{0.32\textwidth}
        \includegraphics[width=\linewidth,clip=True]{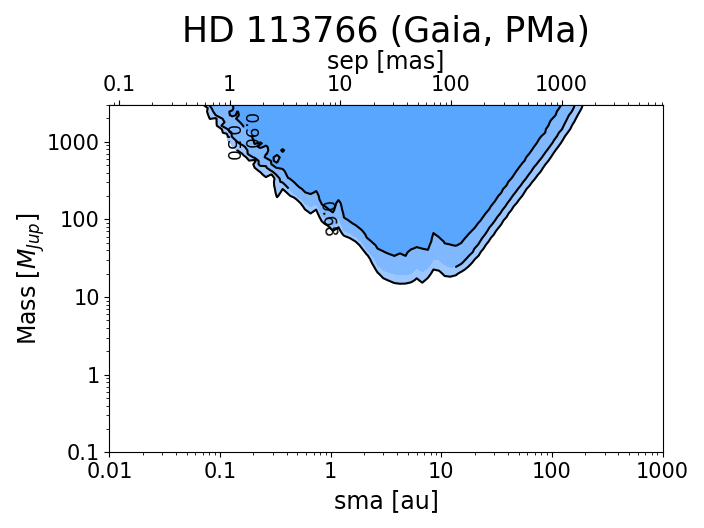}
    \end{subfigure}
    \par\medskip
    \begin{subfigure}{0.32\textwidth}
        \includegraphics[width=\linewidth,clip=True]{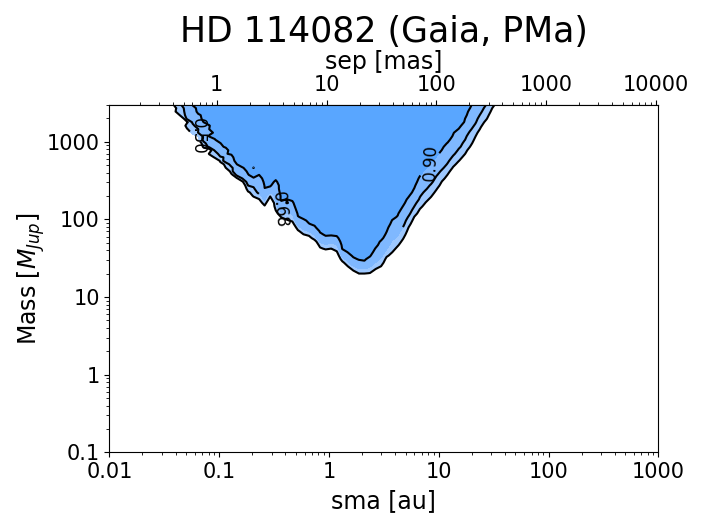}
    \end{subfigure}\hfill
    \begin{subfigure}{0.32\textwidth}
        \includegraphics[width=\linewidth,clip=True]{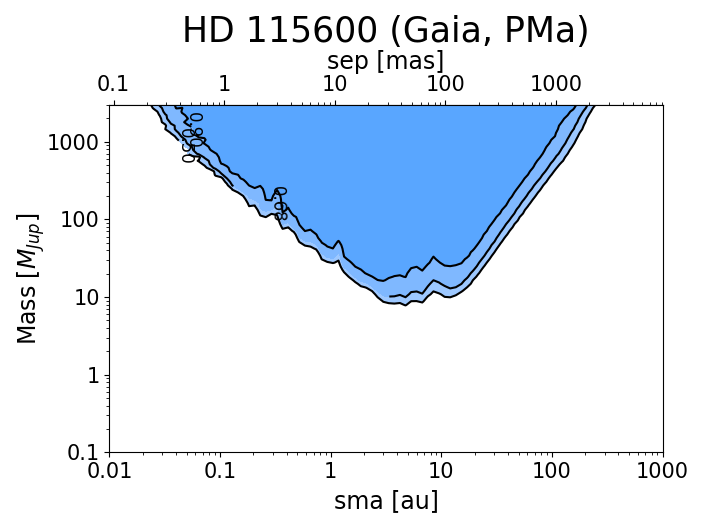}
    \end{subfigure}\hfill
    \begin{subfigure}{0.32\textwidth}
        \includegraphics[width=\linewidth,clip=True]{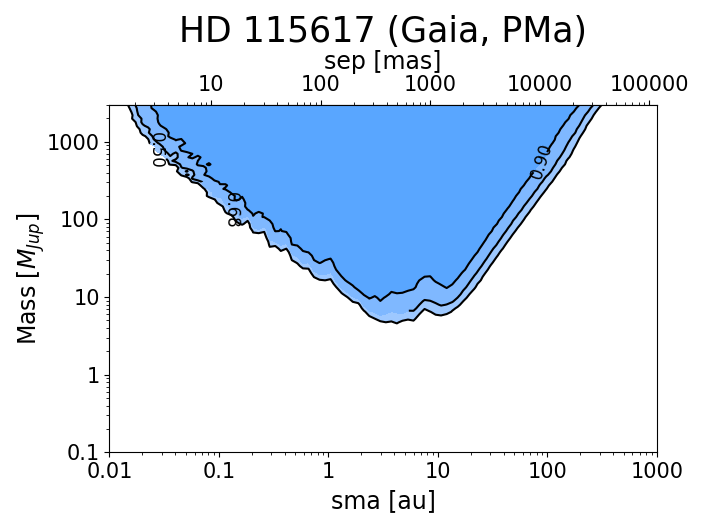}
    \end{subfigure}
    \par\medskip
    \begin{subfigure}{0.32\textwidth}
        \includegraphics[width=\linewidth,clip=True]{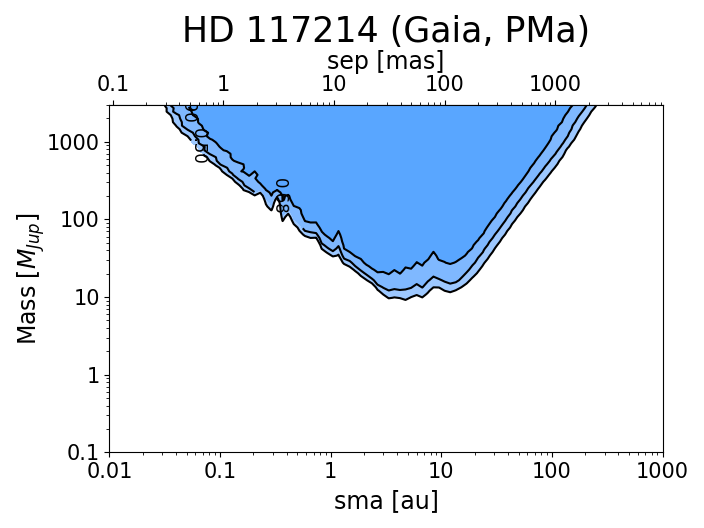}
    \end{subfigure}\hfill
    \begin{subfigure}{0.32\textwidth}
        \includegraphics[width=\linewidth,clip=True]{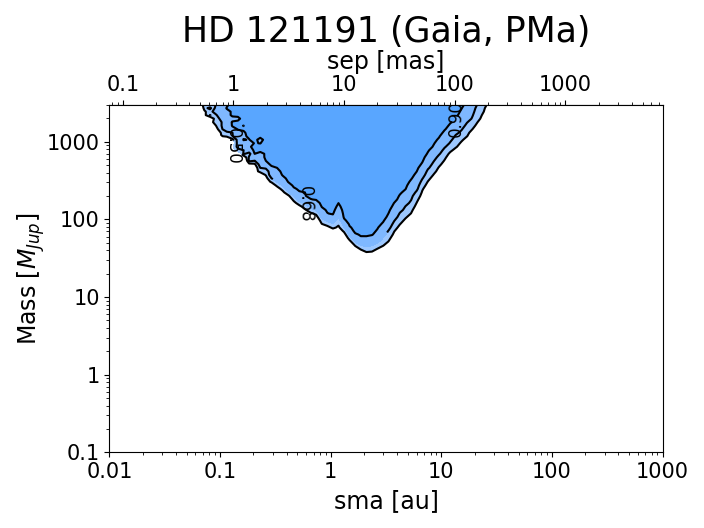}
    \end{subfigure}\hfill
    \begin{subfigure}{0.32\textwidth}
        \includegraphics[width=\linewidth,clip=True]{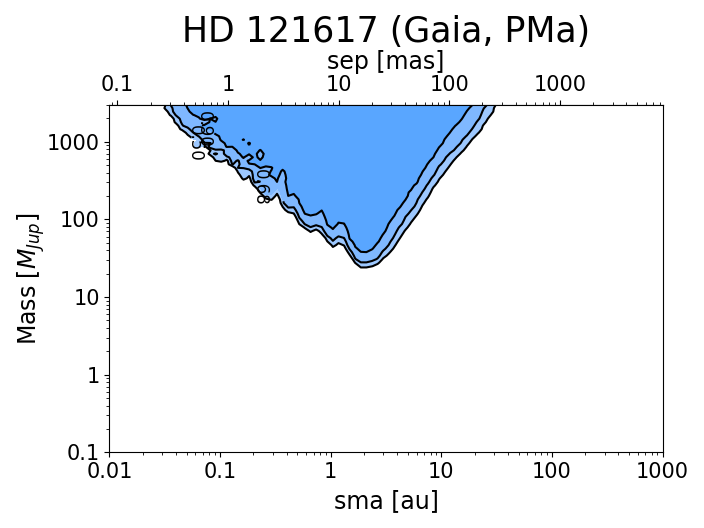}
    \end{subfigure}
    \par\medskip
    \begin{subfigure}{0.32\textwidth}
        \includegraphics[width=\linewidth,clip=True]{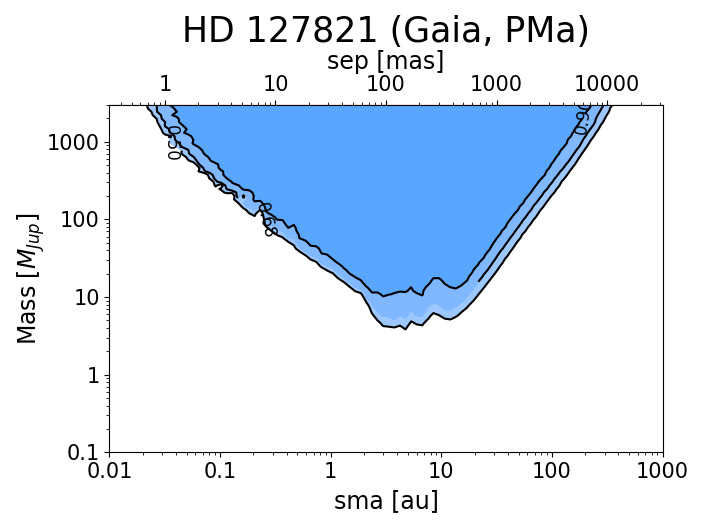}
    \end{subfigure}\hfill
    \begin{subfigure}{0.32\textwidth}
        \includegraphics[width=\linewidth,clip=True]{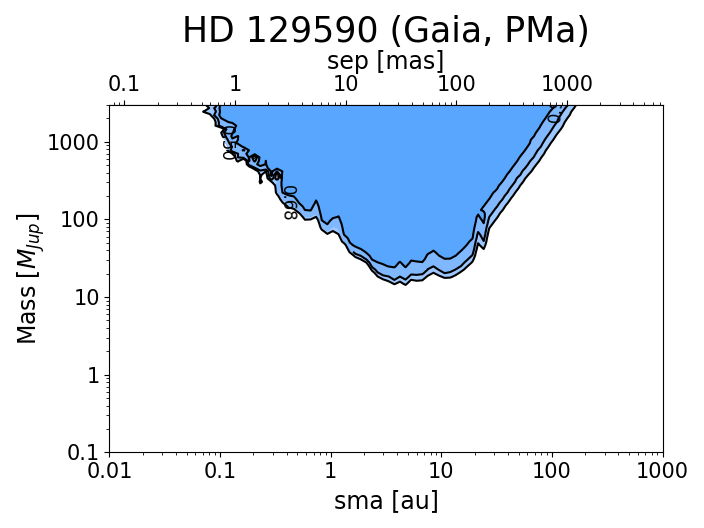}
    \end{subfigure}\hfill
    \begin{subfigure}{0.32\textwidth}
        \includegraphics[width=\linewidth,clip=True]{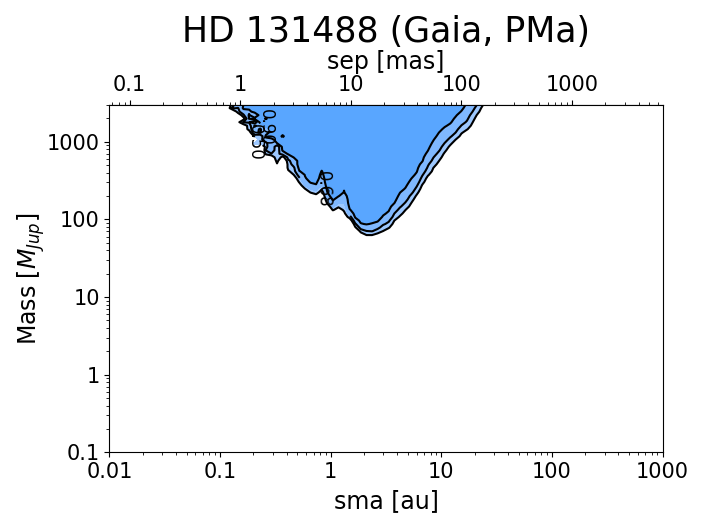}
    \end{subfigure}
    \par\medskip
    \begin{subfigure}{0.32\textwidth}
        \includegraphics[width=\linewidth,clip=True]{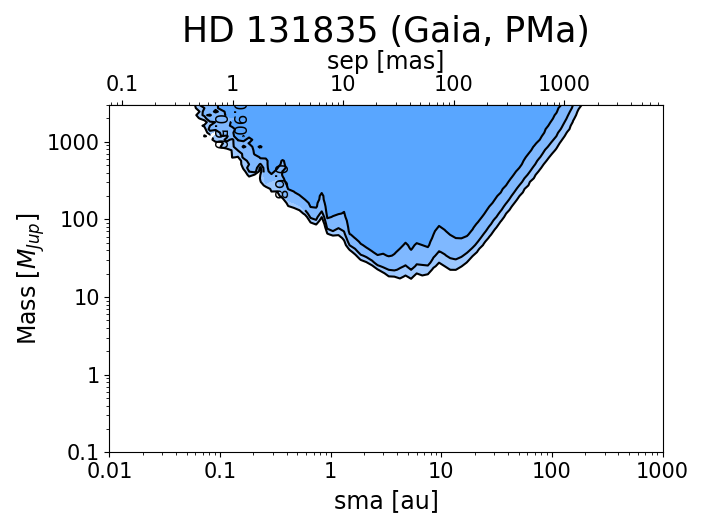}
    \end{subfigure}\hfill
    \begin{subfigure}{0.32\textwidth}
        \includegraphics[width=\linewidth,clip=True]{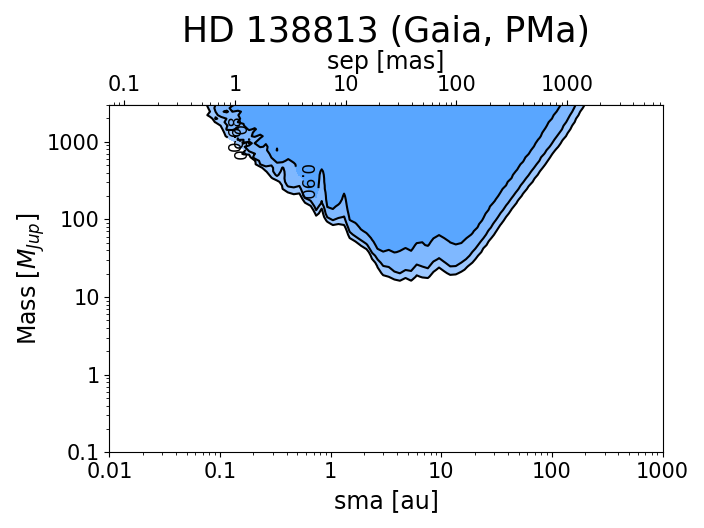}
    \end{subfigure}\hfill
    \begin{subfigure}{0.32\textwidth}
        \includegraphics[width=\linewidth,clip=True]{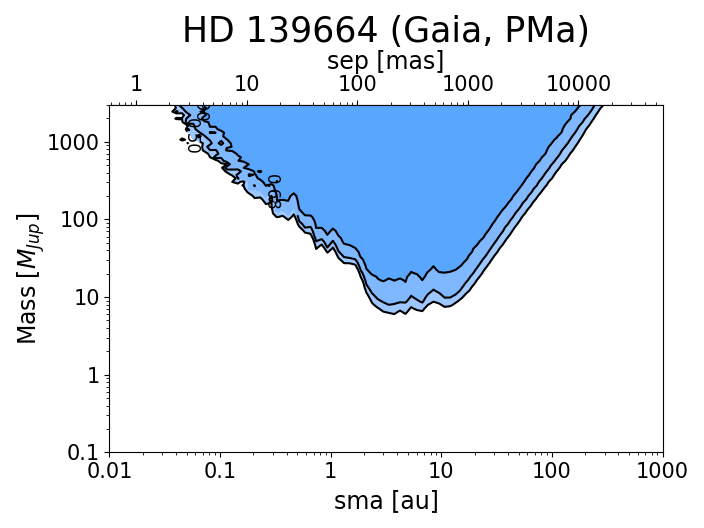}
    \end{subfigure}
    
    \caption*{continued.}
\end{figure*}

\begin{figure*}[p]\ContinuedFloat
    \centering
    \begin{subfigure}{0.32\textwidth}
        \includegraphics[width=\linewidth,clip=True]{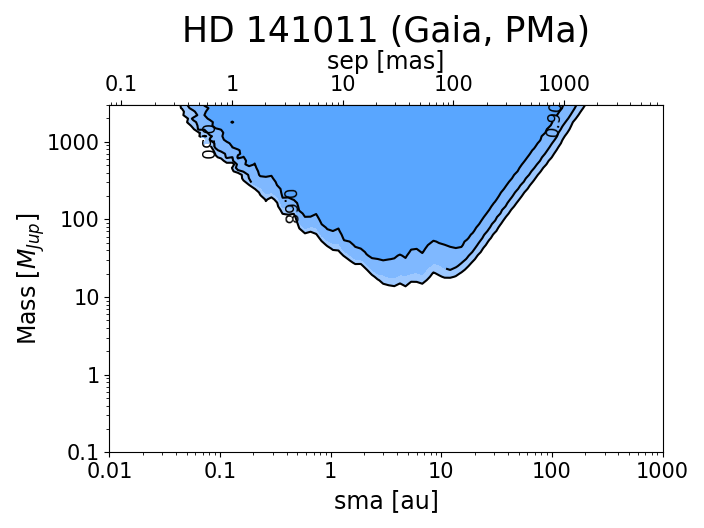}
    \end{subfigure}\hfill
    \begin{subfigure}{0.32\textwidth}
        \includegraphics[width=\linewidth,clip=True]{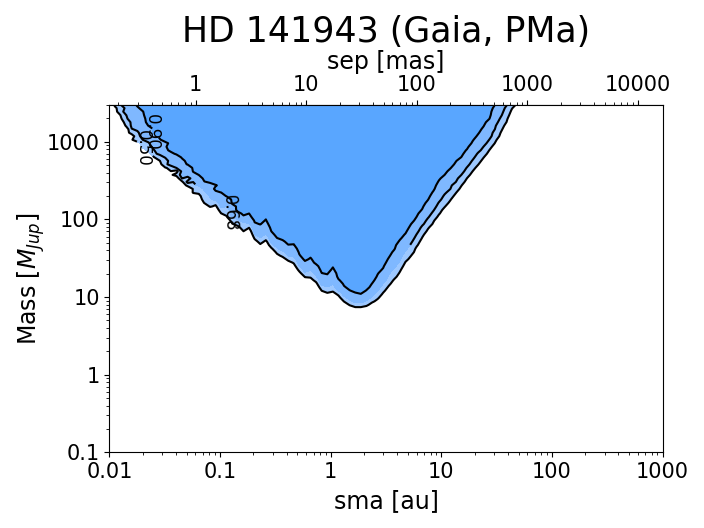}
    \end{subfigure}\hfill
    \begin{subfigure}{0.32\textwidth}
        \includegraphics[width=\linewidth,clip=True]{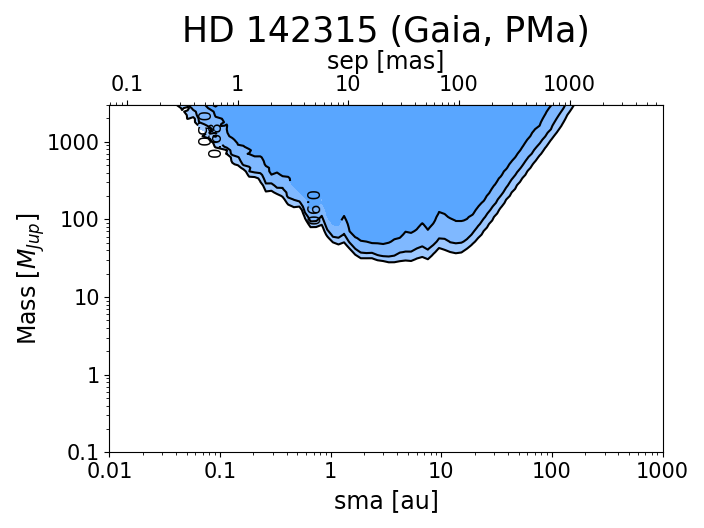}
    \end{subfigure}
    \par\medskip
    \begin{subfigure}{0.32\textwidth}
        \includegraphics[width=\linewidth,clip=True]{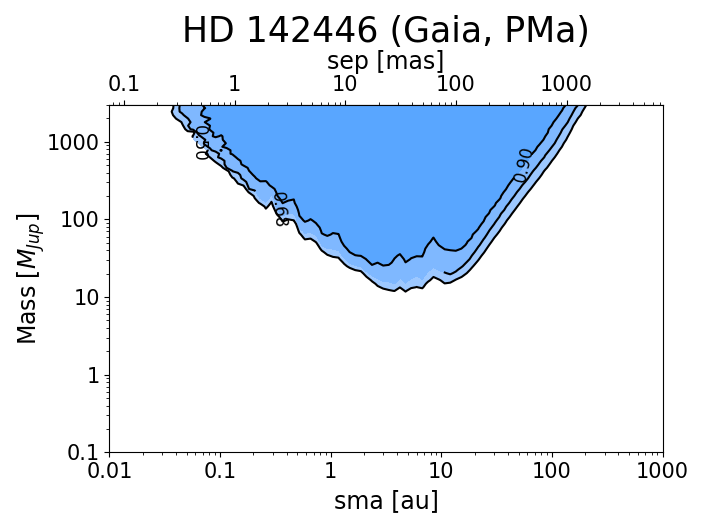}
    \end{subfigure}\hfill
    \begin{subfigure}{0.32\textwidth}
        \includegraphics[width=\linewidth,clip=True]{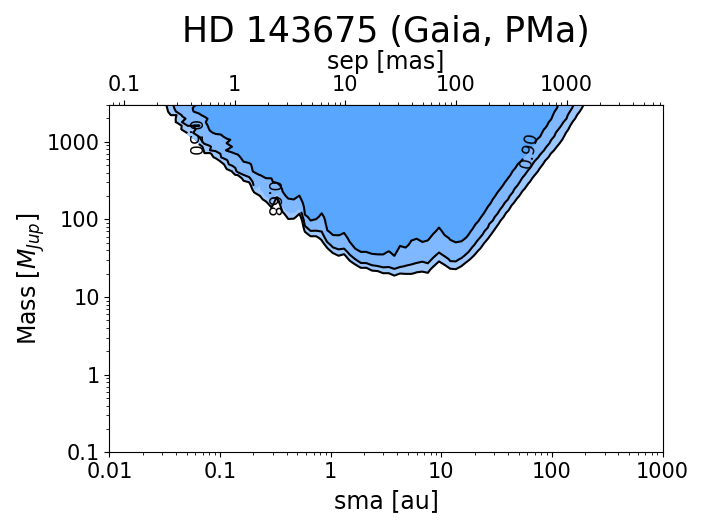}
    \end{subfigure}\hfill
    \begin{subfigure}{0.32\textwidth}
        \includegraphics[width=\linewidth,clip=True]{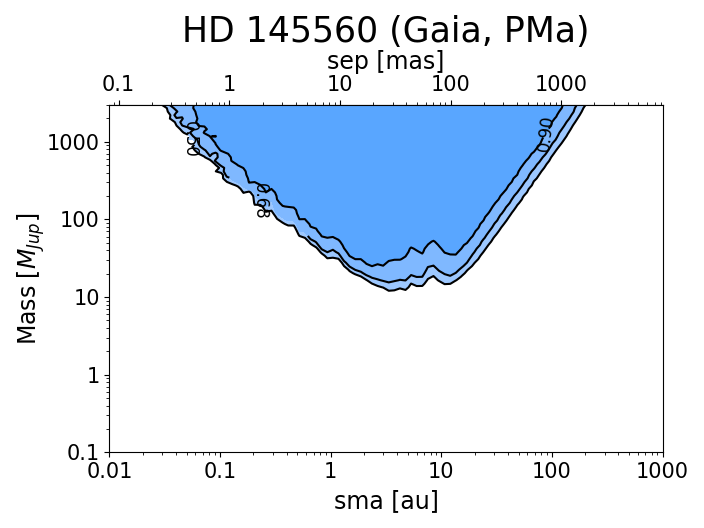}
    \end{subfigure}
    \par\medskip
    \begin{subfigure}{0.32\textwidth}
        \includegraphics[width=\linewidth,clip=True]{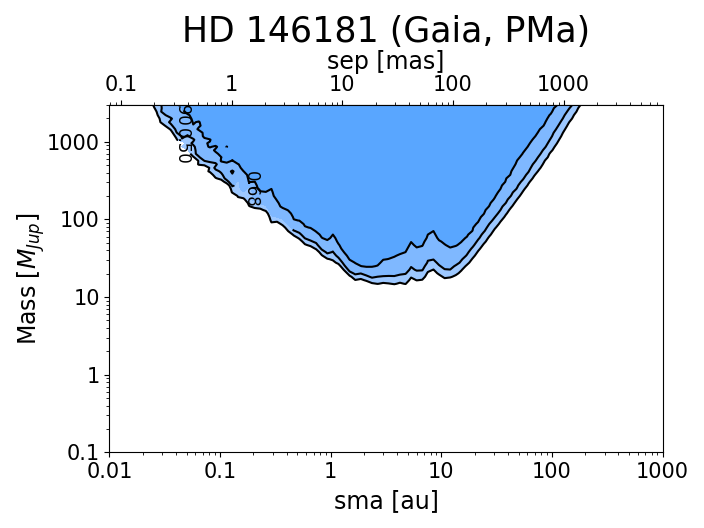}
    \end{subfigure}\hfill
    \begin{subfigure}{0.32\textwidth}
        \includegraphics[width=\linewidth,clip=True]{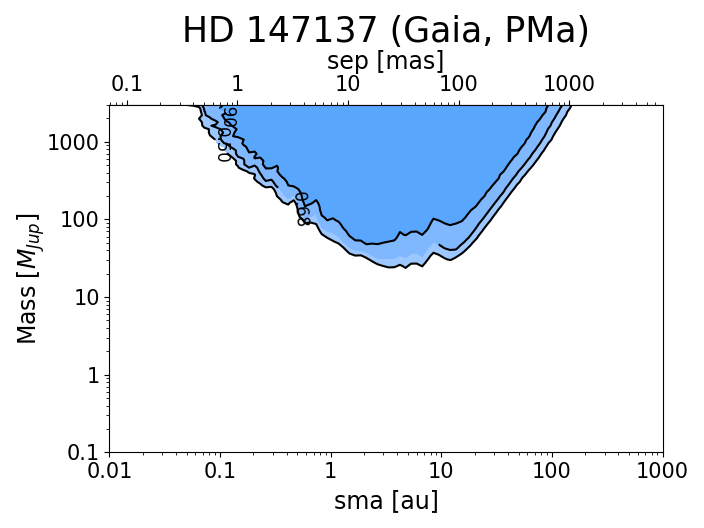}
    \end{subfigure}\hfill
    \begin{subfigure}{0.32\textwidth}
        \includegraphics[width=\linewidth,clip=True]{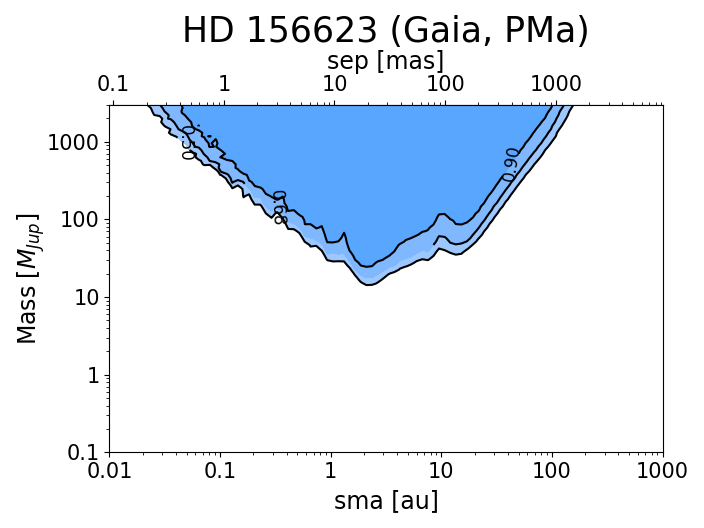}
    \end{subfigure}
    \par\medskip
    \begin{subfigure}{0.32\textwidth}
        \includegraphics[width=\linewidth,clip=True]{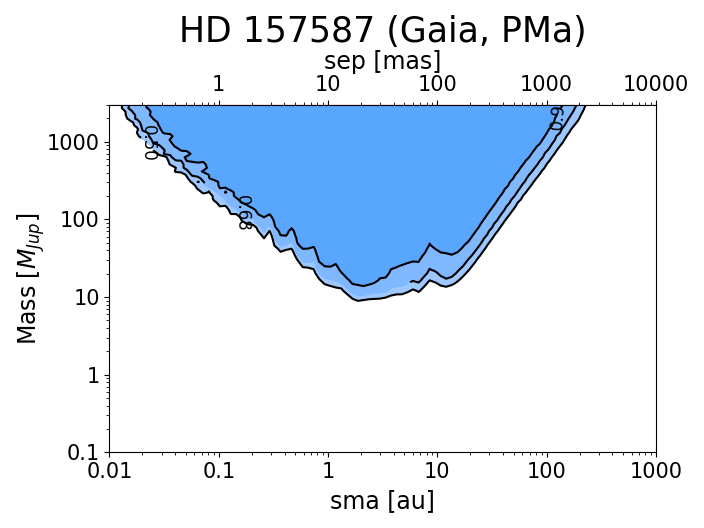}
    \end{subfigure}\hfill
    \begin{subfigure}{0.32\textwidth}
        \includegraphics[width=\linewidth,clip=True]{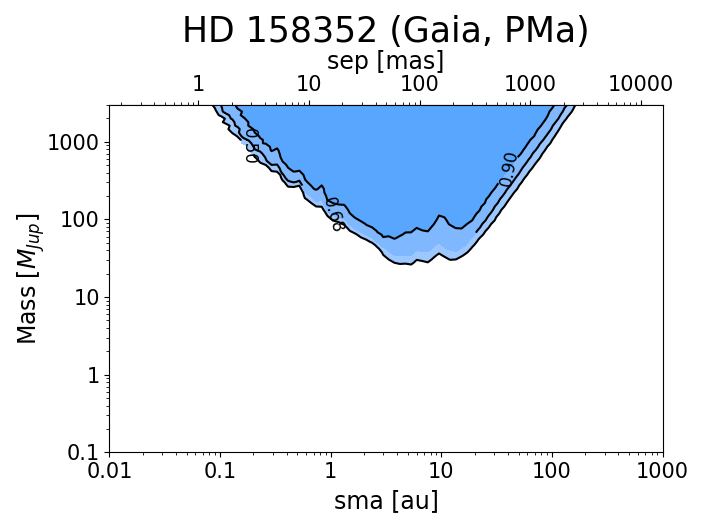}
    \end{subfigure}\hfill
    \begin{subfigure}{0.32\textwidth}
        \includegraphics[width=\linewidth,clip=True]{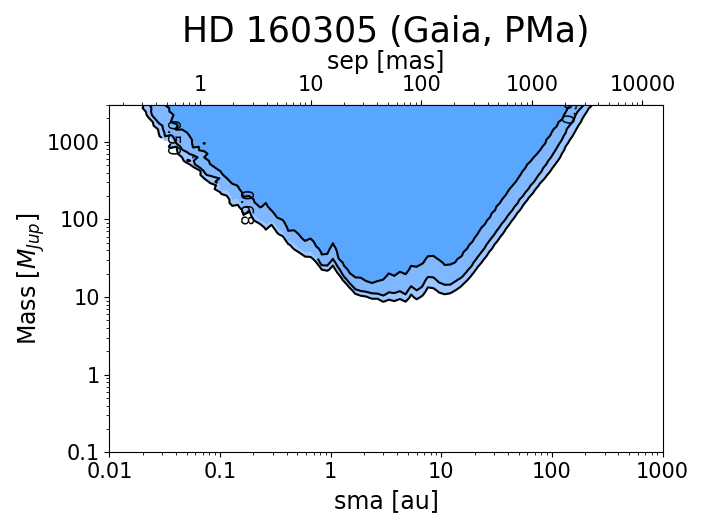}
    \end{subfigure}
    \par\medskip
    \begin{subfigure}{0.32\textwidth}
        \includegraphics[width=\linewidth,clip=True]{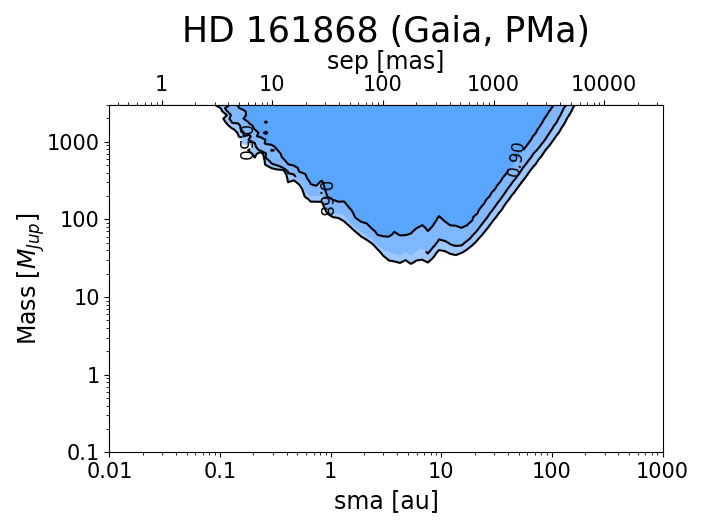}
    \end{subfigure}\hfill
    \begin{subfigure}{0.32\textwidth}
        \includegraphics[width=\linewidth,clip=True]{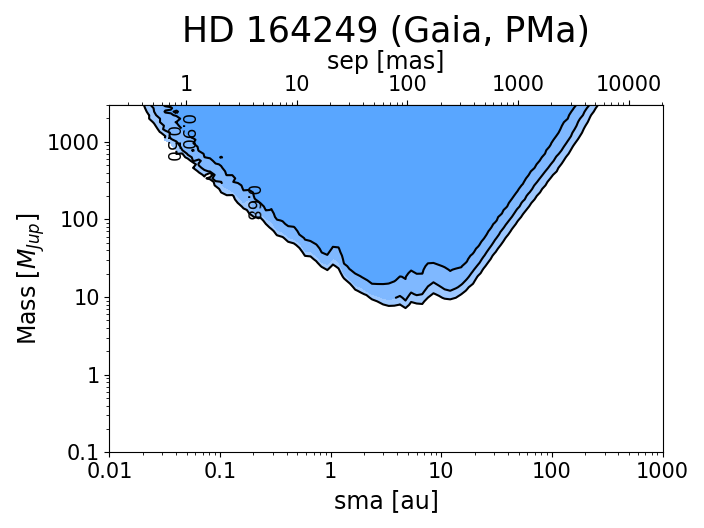}
    \end{subfigure}\hfill
    \begin{subfigure}{0.32\textwidth}
        \includegraphics[width=\linewidth,clip=True]{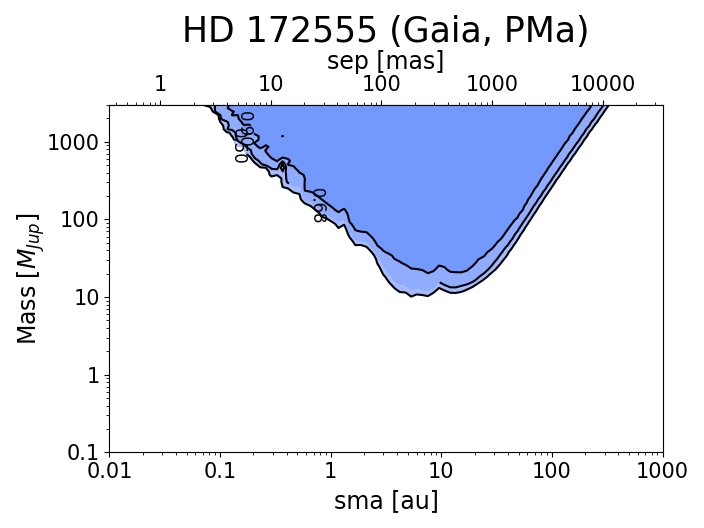}
    \end{subfigure}
    
    \caption*{continued.}
\end{figure*}

\begin{figure}[H]\ContinuedFloat
    \centering
    \begin{subfigure}{0.32\textwidth}
        \includegraphics[width=\linewidth,clip=True]{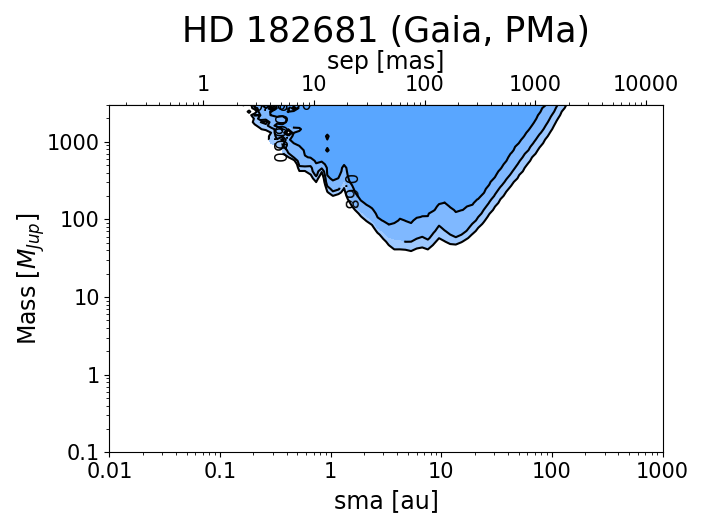}
    \end{subfigure}\hfill
    \begin{subfigure}{0.32\textwidth}
        \includegraphics[width=\linewidth,clip=True]{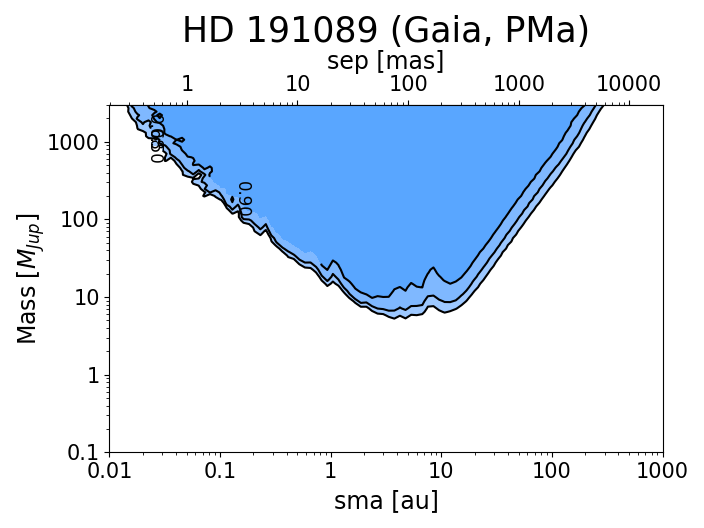}
    \end{subfigure}\hfill
    \begin{subfigure}{0.32\textwidth}
        \includegraphics[width=\linewidth,clip=True]{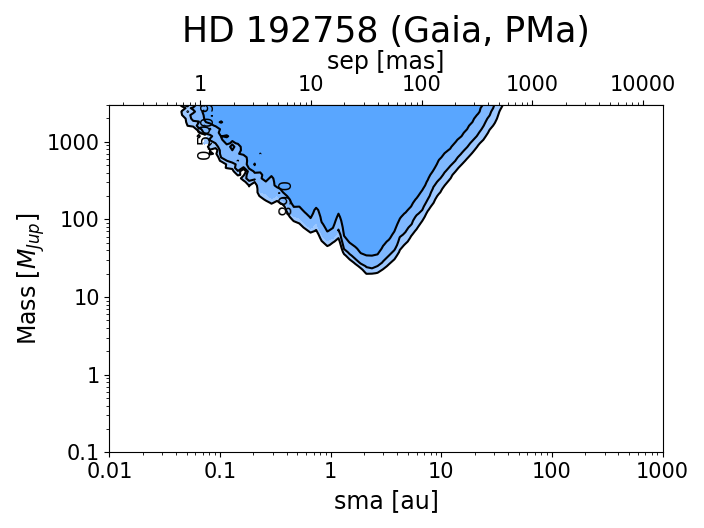}
    \end{subfigure}
    \par\medskip
    \begin{subfigure}{0.32\textwidth}
        \includegraphics[width=\linewidth,clip=True]{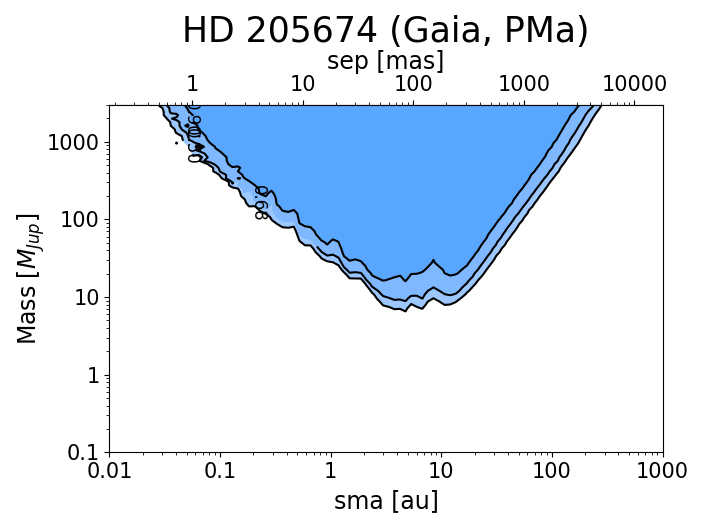}
    \end{subfigure}\hfill
    \begin{subfigure}{0.32\textwidth}
        \includegraphics[width=\linewidth,clip=True]{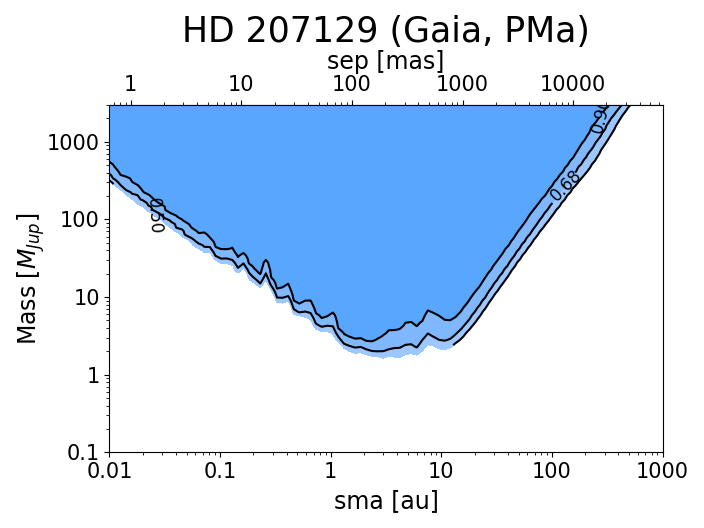}
    \end{subfigure}\hfill
    \begin{subfigure}{0.32\textwidth}
        \includegraphics[width=\linewidth,clip=True]{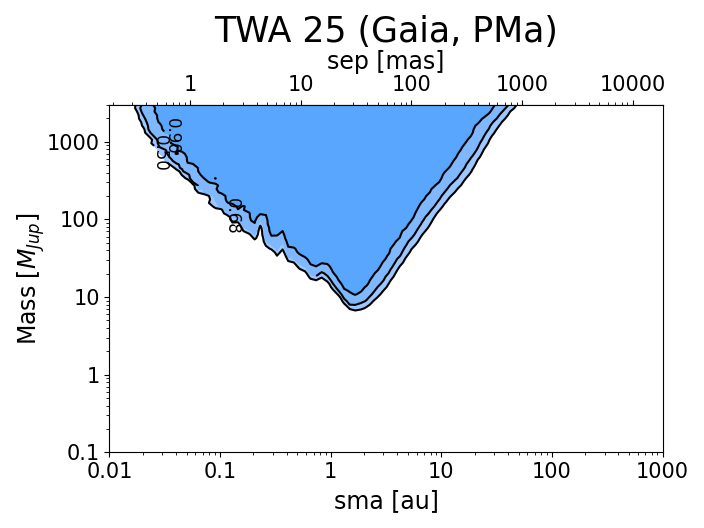}
    \end{subfigure}
    \par\medskip

    \begin{subfigure}{0.32\textwidth}
        \includegraphics[width=\linewidth,clip=True]{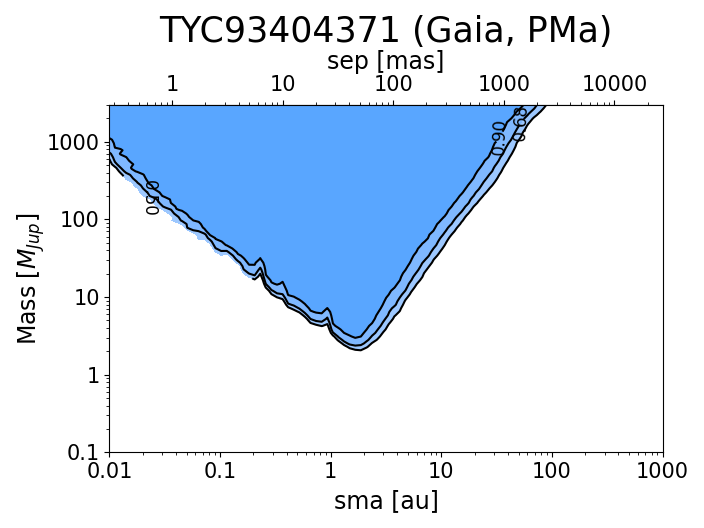}
    \end{subfigure}
    
    \caption*{continued.}
\end{figure}

\newpage
\section{Sample}\label{sec:Sample}

\begin{longtable}[c]{l@{~~}ccccccc}
\caption{Information on all systems. }
\label{tab:all_systems}\\
\hline
Star name               & spectral type    & ${M_*}$ & age & {age ref.} & distance & $L_{\text{disk}}/L_*$ \\
\endfirsthead

\multicolumn{7}{c}
{{ \textbf{Table \thetable}.\ continued.}} \\
\hline
Star name               & spectral type    & ${M_*}$ & age & {age ref.}  & distance & $L_{\text{disk}}/L_*$ \\

\endhead

\endfoot
\endlastfoot

    &   & (${M_{\odot}}$)  & (Myr)  & & (pc) & \\
\hline

49 Cet (HD 9672)  & A1V & 1.89 & $45\pm4$ & 1 & 57.23 & $3.0\times10^{-4}$ \\
\hline
$\epsilon$ Eri (HD 22049)   & K2V & 0.81 & $600\pm200$ & 26 & 3.220 & $3.0\times 10^{-5}$ \\
\hline
$\beta$ Pic (HD 39060)  & A6V & 1.78 & $23\pm3$ & 1 & 19.44 & $2.8\times10^{-3}$  \\
\hline
AU Mic (HD 197481) & M1VeBa1 & 0.64 & $23\pm3$ & 1 & 9.714 & $4.4\times10^{-3}$ \\
\hline
BD+45° 598 & K0 & - & $23\pm3$ & 27 & 73.27 & $6.0\times10^{-4}$  \\
\hline
Fomalhaut (HD 216956) & A4V & 1.87 & $440\pm40$ & 18 &  7.70 & $4.6\times10^{-5}$ \\
\hline
Fomalhaut C (HD 216956 C)  & M4.0Ve & 0.20 & $440\pm40$ & 18 & 7.68 & $1.7\times10^{-5}$ \\
\hline
GJ 14   & K7V & 0.65 & $500\pm300$ & 2 & 14.69 & $3.4\times10^{-5}$ \\
\hline
GSC07396-00759 & M1Ve & 0.69 & $23\pm3$ & 1 & 71.84 & $8\times10^{-5}$  \\
\hline
HD 105   & G0V & 1.04 & $45\pm4$ & 1 & 38.83 & $2.6\times10^{-4}$ \\
\hline
HD 377   & G2V & 1.11 & $220\pm50$ & 4 & 38.41 & $2.0\times10^{-4}$ \\
\hline
HD 8907 & F8 & 1.2 & $500\pm200$ & 4 & 33.40 & $6.2\times10^{-5}$ \\
\hline
HD 10638   & G0V & 1.59 & $100\pm50$ & 2 & 67.77 & $2.0\times10^{-4}$ \\
\hline
HD 10647   &  F9V  & 1.09 & $1700\pm600$ & 4 & 17.35& $6.2\times10^{-5}$ \\
\hline
HD 10700   &  G8V  & 0.9 & $6000\pm2000$ & 4 & 3.65& $2.5\times10^{-5}$ \\
\hline
HD 14055   &  A1Vnn  & 2.03 & $300\pm200$ & 5 & 35.74& $8.9\times10^{-5}$ \\
\hline
HD 15115   &  F4IV  & 1.33 & $45\pm5$ & 1 & 48.77& $8.0\times10^{-5}$ \\
\hline
HD 15257   &  F0III  & 1.92 & $1000\pm800$ & 3 & 48.99& $6.0\times10^{-5}$ \\
\hline
HD 15745   &  F0  & 1.38 & $23\pm3$ & 1 & 71.73& $1.9\times10^{-3}$ \\
\hline
HD 16743   &  F0/2III/IV  & 1.46 & $100^{+400}_{-60}$ & 24 & 57.81& $7.0\times10^{-5}$ \\
\hline
HD 21997   &  A3IV/V  & 1.81 & $42^{+6}_{-4}$ & 1 & 69.69& $5.8\times10^{-4}$ \\
\hline
HD 30447   &  F3V  & 1.46 & $42^{+6}_{-4}$ & 4 & 80.31& $8.8\times10^{-4}$ \\
\hline
HD 32297 &  A0V  & 1.59 & $30\pm10$ & 6 & 129.73& $6.1\times10^{-3}$ \\
\hline
HD 35650 &  K6V  & 0.66 & $150\pm50$ & 28 & 17.46& $1.8\times10^{-4}$ \\
\hline
HD 35841 &  F3V  & 1.20 & $42^{+6}_{-4}$ & 1 & 103.08& $1.4\times10^{-3}$ \\
\hline
HD 36546 &  B8  & 1.92 & $6\pm4$ & 20 & 100.18& $4.7\times10^{-3}$ \\
\hline
HD 38206 &  A0V  & 2.01 & $42^{+6}_{-4}$ & 21 & 70.69& $1.2\times10^{-4}$ \\
\hline
HD 38858 &  G2V  & 0.97 & $5000\pm5000$ & 7 & 15.21& $7.7\times10^{-5}$ \\
\hline
HD 48682 &  F9V  & 1.16 & $4000\pm2000$ & 8 & 16.61& $6.2\times10^{-5}$ \\
\hline
HD 50571 &  F5VFe+0.4  & 1.31 & $300\pm100$ & 22 & 33.93& $1.2\times10^{-4}$ \\
\hline
HD 53143   & G9V & 0.91 & $1010\pm130$ & 25 & 18.34 & $2.5\times 10^{-4}$ \\
\hline
HD 54341   & A0V & 1.97 & $200^{+100}_{-200}$ & 3 & 101.46 & $1.7\times 10^{-4}$ \\
\hline
HD 61005   & G8Vk & 0.92 & $45\pm5$ & 1 & 36.45 & $2.8\times 10^{-3}$ \\
\hline
HD 76582   & F0IV & 1.77 & $1200\pm900$ & 9 & 48.89 & $2.0\times 10^{-4}$ \\
\hline
HD 84870   & A3 & 1.69 & $300\pm200$ & 10 & 88.77 & $4.3\times 10^{-4}$ \\
\hline
HD 92945   & K1V & 0.82 & $29\pm23$ & 29 & 21.51 & $7.6\times10^{-4}$ \\
\hline
HD 95086   & A8III & 1.53 & $13.3^{+1.1}_{-0.6}$ & 11 & 86.46 & $7\times10^{-4}$ \\
\hline
HD 98363   & A2V & 1.9 & $500\pm200$ & 4 & 137.21 & $6.4\times10^{-4}$ \\
\hline
HD 104860  & G0/F9V & 1.03 & $190^{+110}_{-100}$ & 12 & 45.19 & $6.1\times10^{-4}$ \\
\hline
HD 105211   & F2V & 1.56 & $1000\pm1000$ & 13 & 19.69 & $7.3\times 10^{-5}$  \\
\hline
HD 106906       &  F5V &  1.53 & $13\pm2$ & 23 & 103.3 &  $1.4\times 10^{-3}$  \\
\hline
HD 107146  & G2V & 0.99 & $150^{+50}_{-30}$ & 30 & 27.47 & $1.2\times10^{-3}$  \\
\hline
HD 109085   & F2V & 1.46 & $1400\pm300$ & 31 & 18.24 & $3.6\times10^{-4}$  \\
\hline
HD 110058  & A0V & 1.75 & $15\pm3$ & 1 & 130.08 & $9\times 10^{-4}$  \\
\hline
HD 111161  & A3III/IV & 2.4 & $15\pm13$ & 32 & 109.37 & $4.2\times 10^{-3}$ \\
\hline
HD 111520  & F5/6V & 1.23 & $15\pm3$ & 1 & 108.05 & $2.3\times 10^{-3}$  \\
\hline
HD 112810  & F3/5IV/V & 1.29 & $15\pm3$ & 1 & 133.66 & $8.9\times 10^{-4}$  \\
\hline
HD 113556  & F2V & 1.41 & $15\pm3$ & 1 & 100.48 & $5.1\times 10^{-4}$  \\
\hline
HD 113766  & F3/5V & 1.54 & $15\pm3$ & 1 & 122.55 & $2.2\times 10^{-2}$  \\
\hline
HD 114082  & F3V & 1.34 & $15\pm3$ & 1 & 95.06 & $3.6\times 10^{-3}$  \\
\hline
HD 115600  & F2IV/V & 1.44 & $15\pm3$ & 1 & 109.04 & $2.0\times 10^{-3}$  \\
\hline
HD 115617  & G6.5V & 1.0 & $4000\pm2000$ & 4 & 8.53 & $1.0\times 10^{-5}$  \\
\hline
HD 117214  & F6V & 1.48 & $15\pm3$ & 1 & 107.35 & $2.7\times 10^{-3}$  \\
\hline
HD 120326  & F0V & 1.6 & $16\pm1$ & 33 & 113.27 & $1.1\times10^{-3}$  \\
\hline
HD 121191  & A5IV/V & 1.61 & $16\pm2$ & 1 & 132.29 & $3.2\times10^{-3}$  \\
\hline
HD 121617  & A1V & 1.92 & $16\pm2$ & 1 & 117.89 & $5.1\times10^{-3}$  \\
\hline
HD 127821  & F4IV & 1.27 & $1300\pm500$ & 4 & 31.69 & $1.9\times10^{-4}$  \\
\hline
HD 129590  & G3V & 1.27 & $16\pm2$ & 1 & 136.32 & $5.5\times10^{-3}$  \\
\hline
HD 131488  & A1V & 1.85 & $16\pm2$ & 1 & 152.24 & $2.9\times10^{-3}$  \\
\hline
HD 131835  & A2IV  & 1.80 & $16\pm2$ & 1 & 129.74 & $2.7\times10^{-3}$  \\
\hline
HD 138813  & A0V  & 1.89 & $10\pm3$ & 1 & 136.60 & $7.7\times10^{-4}$  \\
\hline
HD 139664  & F4V  & 1.31 & $200^{+200}_{-100}$ & 4 & 17.40 & $1.3\times10^{-4}$  \\
\hline
HD 141011  & F5V  & 1.4 & $16\pm2$ & 34 & 128.38 & $1.7\times10^{-4}$ \\
\hline
HD 141943  & G2  & 1.2 & $16\pm2$ & 4 & 60.14 & $1.1\times10^{-4}$ \\
\hline
HD 142315  & B9V  & 2.07 & $10\pm3$ & 1 & 144.74 & $6.6\times10^{-4}$  \\
\hline
HD 142446  & F3V  & 1.34 & $16\pm2$ & 1 & 135.61 & $1.8\times10^{-4}$  \\
\hline
HD 143675  & A5IV/V  & 2.0 & $16\pm2$ & 32 & 135.7 & $4.1\times10^{-4}$  \\
\hline
HD 145560  & F5V  & 1.30 & $16\pm2$ & 1 & 121.23 & $2.0\times10^{-3}$  \\
\hline
HD 146181  & F6V  & 1.26 & $16\pm2$ & 1 & 127.52 & $1.9\times10^{-3}$  \\
\hline
HD 146897  & F2/3V & 1.31 & $10\pm3$ & 1 & 132.19 & $1.0\times10^{-2}$  \\
\hline
HD 147137  & A9V & 1.40 & $10\pm3$ & 1 & 143.47 & $4.4\times10^{-4}$  \\
\hline
HD 156623  & A0V & 2.11 & $16\pm2$ & 4 & 108.33 & $4.3\times10^{-3}$  \\
\hline
HD 157587  & F5V & 1.47 & $3000\pm1500$ & 35 & 99.87 & $3.2\times10^{-3}$  \\
\hline
HD 158352  & A8Vp & 1.94 & $900\pm700$ & 5 & 63.81 & $7.1\times10^{-5}$  \\
\hline
HD 160305  & F8/G0V & 1.13 & $20.4\pm2.5$ & 36 & 65.80 & $1.4\times10^{-4}$  \\
\hline
HD 161868  & A1VnkA0mA0 & 1.99 & $300\pm200$ & 4 & 29.75 & $6.3\times10^{-5}$  \\
\hline
HD 164249  & F6V & 1.29 & $23\pm3$ & 1 & 49.30 & $9.4\times10^{-4}$  \\
\hline
HD 170773  & F5V & 1.32 & $1500^{+1200}_{-700}$ & 37 & 36.94 & $4.8\times 10^{-4}$ \\
\hline
HD 172555  & A7V & 1.87 & $24\pm3$ & 4 & 28.79 & $8.1\times 10^{-4}$ \\
\hline
HD 181327  & F6V & 1.26 & $18.5^{+2.0}_{-2.4}$ & 38 & 47.78 & $2.0\times10^{-3}$  \\
\hline
HD 182681  & B8.5V & 2.00 & $110\pm90$ & 15 & 70.69 & $2.8\times10^{-4}$  \\
\hline
HD 191089  & F5V & 1.25 & $23\pm3$ & 1 & 50.11 & $1.6\times10^{-3}$  \\
\hline
HD 192758  & F0V & 1.57 & $45\pm5$ & 4 & 66.50 & $4.1\times10^{-4}$  \\
\hline
HD 202628  & G5V & 0.99 & $1100\pm400$ & 16 & 23.79 & $1.4\times10^{-4}$  \\
\hline
HD 202917  & G7V & 0.9 & $45\pm4$ & 28 & 46.77 & $2.9\times10^{-4}$  \\
\hline
HD 205674  & F3/5IV & 1.31 & $800\pm600$ & 17 & 55.75 & $3.5\times10^{-4}$  \\
\hline
HD 206893  & F5V & 1.26 & $155\pm15$ & 39 & 40.77 & $2.7\times10^{-4}$ \\
\hline
HD 207129  & G2V & 1.03 & $3000^{+2000}_{-1000}$ & 4 & 15.56 & $9.3\times10^{-5}$ \\
\hline
HR 4796 (HD 109573)  & A0V & 1.99 & $10\pm3$ & 1 & 70.77 & $2.4\times 10^{-3}$  \\
\hline
HR 8799 (HD 218396) & F0+VkA5mA5 & 1.47 & $30\pm10$ & 19 & 40.88 & $3.7\times10^{-5}$ \\
\hline
TWA 7 & M2Ve & 0.46 & $6.4\pm1$ & 40 & 34.10 & $1.7\times10^{-3}$  \\
\hline
TWA 25 & M0.5 & 0.67 & $13\pm4$ & 28 & 53.60 & - \\
\hline
TYC93404371  & K7Ve & 0.72 & $23\pm3$ & 1 & 36.59 & $1.0\times10^{-3}$ \\
\hline
Vega (HD 172167)  & A0V & 2.72 & $460\pm10$ & 14 & 7.76 & $4.3\times 10^{-6}$ \\
\hline
\caption*{{{Notes:} the spectral types and distances are taken from SIMBAD. The disk brightness and stellar masses are taken from Ma25; \citet{Engler2025, Moor2025}.\\
{Age references}: 1. Ma25 - 2. \citet{Chen2014} - 3. \citet{Rhee2007} - 4. \citet{Pearce2022} - 5. \citet{ZorecRoyer2012} - 6. \citet{Kalas2005} - 7. \citet{Kennedy2015} - 8. \citet{Eiroa2013} - 9. \citet{Marshall2016} - 10. \citet{Vican2016} - 11. \citet{Booth2021} - 12. \citet{Brandt2014} - 13. \citet{Dodson-Robinson2016} - 14.\citet{Yoon2010} - 15. \citet{Gray2006} - 16. \citet{Faramaz2019} - 17. \citet{Casagrande2011} - 18. \citet{Mamajek2012} - 19. \citet{Baines2012} - 20. \citet{Currie2017} - 21. \citet{torres08} - 22. \citet{Moor2011} - 23. \citet{Bailey2014} - 24. \citet{Desidera2015} - 25. \citet{Barnes2007} - 26. \citet{Llop2021} - 27. \citet{Hinkley2021} - 28. \citet{Bell2015} - 29. \citet{Plavchan2009} - 30. \citet{Mesa2021} - 31. \citet{Lisse2012} - 32. \citet{Moor2025} - 33. \citet{Pecaut2012} - 34. \citet{Engler2025} - 35. \citet{Millar-Blanchaer2016} - 36. \citet{Couture2023} - 37. \citet{Sepulveda2019} - 38. \citet{Miret-Roig2020} - 39. \citet{Hinkley2023} - 40. \citet{Binks2020}}}
\end{longtable}

\begin{longtable}[c]{l@{~~}cccccccc}
\caption{Target sample. }
\label{tab:planet_catalog}\\
\hline
Star name                 & $i_{\text{disk}}$ & Rin 1 & Rout 1 & Rin 2 & Rout 2  & Planets sma/mass & $\sigma_{\rm RUWE}$ & $\sigma_{\rm PMa}$  \\
\endfirsthead

\multicolumn{8}{c}
{{\textbf{Table \thetable}.\ continued.}} \\
\hline
Star name                 & $i_{\text{disk}}$ &  Rin 1 & Rout 1 & Rin 2 & Rout 2  & Planets sma/mass & $\sigma_{RUWE}$ & $\sigma_{PMa}$  \\

\endhead

\endfoot
\endlastfoot

   & (°)   & (au) & (au)  & (au) & (au) &   & &   \\
\hline

AU Mic (HD 197481)   &  $88.4^{+0.1}_{-0.1}$  &  28.7  &  41.1  &    &    &  b: ${8.46321\pm0.00004}\text{ days}$  &  0.262  &  0.481  \\
   &    &    &    &    &    &  $ 0.053\pm0.016 M_J$   &    &    \\
   &    &    &    &    &    &  c: ${18.859019\pm0.000016} \text{ days}$   &    &    \\
     &    &    &    &    &    &  $0.0428^{+0.0359}_{-0.0359}M_J$   &    &    \\
   &    &   &    &    &    &  d: ${12.73596\pm0.00793} \text{ days}$  &    &    \\
     &   &    &    &    &    &  $0.00331^{+0.0016}_{-0.0016}M_J $  &    &    \\
   &    &   &    &    &    &  {e: ${33.11\pm0.06\text{ days}}$}  &    &    \\
     &   &    &    &    &    &  {${0.0664^{+0.0170}_{-0.0135}M_J }$}  &    &    \\
\hline
49 Cet (HD 9672)   &  $79.1^{+0.4}_{-0.4}$  &  4  &  60  &  65  &  250  &    &  0.138  &  0.306 \\
\hline
$\epsilon$ Eri (HD 22049)    &  $30^{+2}_{-2}$  &  12  &  16  &  62.6  &  74.6  &  b: $3.53^{+0.03}_{-0.04}\text{au}$  &  1.053  &  1.563  \\
     &   &    &    &    &    &  $0.66^{+0.12}_{-0.09}M_J$  &    &   \\
\hline
$\beta$ Pic (HD 39060)   &  $86.6^{+0.4}_{-0.3}$  &  63.2  &  155.2  &    &    &  b: $9.0^{+0.4}_{-0.4}\text{au}$  &  0.393  &  0.158  \\
 &    &   &    &    &    &  $9.9^{+0.5}_{-0.5}M_J$  &    &    \\
   &   &   &    &    &    &  c: $2.69^{+0.003}_{-0.003}\text{au}$  &    &   \\
 &   &   &    &    &    &  $8.93^{+0.14}_{-0.14}M_J$  &    &   \\
\hline
BD+45° 598    &  $86.1^{+2.2}_{-2.0}$  & 50.5  &  85.5  &    &    &    &  1.371   &  - \\
\hline
GJ 14    &  $64^{+1}_{-2}$  &  35.9  &  73.1  &    &    &    &  0.474  &  0.289 \\
\hline
HD 105    &  $49^{+3}_{-2}$  &  55  &  85  &    &    &    &  0.288  &  1.143 \\
\hline
HD 53143    &  $56.1^{+0.6}_{-0.5}$  &  5.5  &  55  &  76.9  &  99.9  &    &  1.156  &  0.410  \\
\hline
HD 54341    &  $50^{+10}_{-10}$  &  4.9  &  13.6  &  106.7  &  128.1  &    &  ${>10}$  &  3.314  \\
\hline
HD 92945    &  $66.8^{+1.0}_{-1.3}$  &  52  &  63.4  &  83.4  &  121  &    &  0.474  &  1.840  \\
\hline
HD 105211    &  $66.4^{+0.8}_{-0.7}$  &  120  &  143  &    &    &    &  0.015  &  0.671  \\
\hline
HD 106906        &  $85.3^{+0.1}_{-0.1}$  &  35  &  110  &    &     &  B: $0.4\text{au}$  &  0.180  &  3.600     \\
 &   &    &    &    &    &  $1309\pm52M_J$  &    &     \\
 &   &    &    &    &    &  b: $650\text{au}$  &    &     \\
 &   &    &    &    &    &  $11^{+2}_{-2}M_J$  &    &     \\
\hline
HD 107146   &  $22.2^{+0.6}_{-0.9}$  &  46.6  &  ~60  &  100  &  135.6  &    &  0.987  &  3.856 \\
\hline
HD 109085    &  $39^{+1}_{-1}$  &  129  &  182  &    &    &    &  0.061  &  0.486 \\
\hline
HD 109573 (HR 4796)   &  $76.5^{+0.2}_{-0.2}$  &  70.4  &  85.2  &    &    &    &  0.560  &  0.396 \\
\hline
HD 120326   &  $80^{+1}_{-1}$  &  -  &  $\sim$59  &  -  &  $\sim$130  &    &  2.508  &  0.521 \\
\hline
HD 146897   &  $84.0^{+3.0}_{-3.0}$  &  60  &  80  &    &    &    &  4.102  &  1.343 \\
\hline
HD 170773   &  $33^{+2}_{-2}$  &  160  &  228  &    &    &    &  1.061  &  0.677 \\
\hline
HD 181327   &  $29.9^{+0.5}_{-0.5}$  &  73.3  &  89.3  &    &    &    &  0.547  &  1.024 \\
\hline
HD 202628   &  $57.9^{+0.4}_{-0.4}$  &  163.4  &  179.4  &    &    &    &  0.283  &  1.918 \\
\hline
HD 202917   &  $68.6^{+1.5}_{-1.5}$  &  55.4  &  68.6  &    &    &    &  0.297  &  1.721 \\
\hline
HD 206893   &  $24^{+9}_{-11}$  &  <28  &  60.5  &  87.5  &  180  &  b: $9.6^{+0.4}_{-0.3}\text{au}$  &  1.677  &  6.765  \\
 &   &    &    &    &    &  $28.0^{+2.2}_{-2.1}M_J$  &    &     \\
   &   &    &    &    &    &  c: $3.53^{+0.08}_{-0.06}\text{au}$  &    &     \\
     &    &    &    &    &    &  $12.7^{+1.2}_{-1.0}M_J$  &    &   \\
\hline
TWA7   &  $13.1^{+3.1}_{-2.6}$  &  <28  &  >100  &  48.5  &  55.5  &  b: $52\text{au}$  &  1.790  &  -  \\
 &    &    &    &    &    &  $0.34\pm0.06 M_J$  &    &     \\
 \hline
 \caption*{{{Notes:} the references for the disk inclination and separations are given in the paragraphs dedicated to each system in the present paper. For the vast majority, they are taken from Ma25. The planet sma/mass and associated uncertainties are retrieved from the NASA Exoplanet Archive. Note that AU Mic e is to be confirmed. The $\sigma_{\rm RUWE}$ and $\sigma_{\rm PMa}$ columns correspond to the significance of the RUWE and PMa signal obtained with GaiaPMEX.}}
\end{longtable}

\twocolumn
\end{appendix}
\end{document}